%% file: main.tex
\documentclass{sn-jnl}
\usepackage{graphicx}%
\usepackage{multirow}%
\usepackage{amsmath,amssymb,amsfonts}%
\usepackage{amsthm}%
\usepackage{mathrsfs}%
\usepackage[title]{appendix}%
\usepackage{xcolor}%
\usepackage{textcomp}%
\usepackage{manyfoot}%
\usepackage{booktabs}%
\usepackage{algorithm}%
\usepackage{algorithmicx}%
\usepackage{algpseudocode}%
\usepackage{listings}%
\usepackage{longtable}
\usepackage{lscape}
\usepackage{natbib}

\theoremstyle{thmstyleone}%
\theoremstyle{thmstyletwo}%

\theoremstyle{thmstylethree}%

\begin{document}

\title[Carbon-Chain Chemistry in the Interstellar Medium]{Carbon-Chain Chemistry in the Interstellar Medium}


\author*[1]{\fnm{Kotomi} \sur{Taniguchi}}\email{kotomi.taniguchi@nao.ac.jp}
\equalcont{These authors contributed equally to this work.}

\author[2,3,4]{\fnm{Prasanta} \sur{Gorai}}\email{prasanta.astro@gmail.com}
\equalcont{These authors contributed equally to this work.}

\author[2,5]{\fnm{Jonathan C.} \sur{Tan}} 

\affil*[1]{\orgdiv{Division of Science}, \orgname{National Astronomical Observatory of Japan}, \orgaddress{\street{2-21-1 Osawa, Mitaka}, \city{Tokyo}, \postcode{181-8588}, \country{Japan}}}

\affil[2]{\orgdiv{Department of Space, Earth and Environment}, \orgname{Chalmers University of Technology}, \orgaddress{\street{SE-412 96}, \city{Gothenburg}, \country{Sweden}}} 

\affil[3]{\orgdiv{Rosseland Centre for Solar Physics}, \orgname{University of Oslo}, \orgaddress{\street{PO Box 1029 Blindern, 0315}, \city{Oslo}, \country{Norway}}} 

\affil[4]{\orgdiv{Institute of Theoretical Astrophysics}, \orgname{University of Oslo}, \orgaddress{\street{PO Box 1029 Blindern, 0315}, \city{Oslo}, \country{Norway}}} 

\affil[5]{\orgdiv{Department of Astronomy}, \orgname{University of Virginia}, \orgaddress{\city{Charlottesville}, \postcode{22904-4325}, \state{Virginia}, \country{USA}}}


\abstract{
The presence of carbon-chain molecules in the interstellar medium (ISM) has been known since the early 1970s and $>130$ such species have been identified to date, making up $\sim 43$\% of the total of detected ISM molecules. 
They are prevalent not only in star-forming regions in our Galaxy but also in other galaxies. 
These molecules provide important information on physical conditions, gas dynamics, and evolutionary stages of star-forming regions. 
Larger species of polycyclic aromatic hydrocarbons (PAHs) and fullerenes (C$_{60}$ and C$_{70}$), which may be related to the formation of the carbon-chain molecules, have been detected in circumstellar envelopes around carbon-rich Asymptotic Giant Branch (AGB) stars and planetary nebulae, while PAHs are also known to be a widespread component of the ISM in most galaxies. 
Recently, two line survey projects toward Taurus Molecular Cloud-1 with large single-dish telescopes have detected many new carbon-chain species, including molecules containing benzene rings. 
These new findings raise fresh questions about carbon-bearing species in the Universe.
This article reviews various aspects of carbon-chain molecules, including observational studies, chemical simulations, quantum calculations, and laboratory experiments, and discusses open questions and how future facilities may answer them.
}




\maketitle

\section{General Introduction}
\subsection{Brief Overview of Astrochemistry} \label{sec:1_1}

Astrochemistry is an interdisciplinary research field concerning ``study of the formation, destruction, and excitation of molecules in astronomical environments and their influence on the structure, dynamics, and evolution of astronomical objects'' \citep{dalgarno08}. Astrochemical studies can involve various approaches: astronomical observations; laboratory experiments on reactions and diagnostic spectroscopy; chemical simulations; and quantum chemical calculations. Collaborative studies among these approaches have been crucial in revealing the great variety of chemical pathways that operate in space.

More than 300 molecules have been discovered in the interstellar medium (ISM) or circumstellar envelopes (CSEs) to date (the Cologne Database for Molecular Spectroscopy (CDMS)\footnote{\url{https://cdms.astro.uni-koeln.de/classic/molecules}}; \citet{mcgu22}).
Technical innovations and advances in observational facilities have boosted the detection of new, rarer interstellar molecules including isotopologues. 
These molecules have been detected in various physical conditions of the ISM; diffuse atomic H clouds ($n_{\rm {H}}\approx100$ cm$^{-3}$, $T\approx70-80$ K), molecular clouds ($n_{\rm {H}}\approx10^{4}$ cm$^{-3}$, $T\approx10$ K), prestellar cores\footnote{We use this term for gravitationally bound objects with central number densities above $10^{5}$ cm$^{-3}$ \citep{caselli2022}.} ($n_{\rm {H}}\approx10^{5}-10^{6}$ cm$^{-3}$, $T\approx10$ K), protostellar cores ($n_{\rm {H}}\approx10^{7}$ cm$^{-3}$, $T\approx100-300$ K), protoplanetary disks ($n_{\rm {H}}\approx10^{4}-10^{10}$ cm$^{-3}$, $T\approx10-500$ K), and envelopes of evolved stars ($n_{\rm {H}}\approx10^{10}$ cm$^{-3}$, $T\approx2000-3500$ K).
Beyond our Galaxy, about 73 molecules have been detected in extragalactic sources.

Although 98\% of the total mass of baryons consists of hydrogen (H) and helium (He), heavier trace elements such as carbon (C), oxygen (O), and nitrogen (N) are important constituent elements of interstellar molecules. These elements can make interstellar molecules complex and chemically rich. 
In particular, carbon composes the backbones of many molecules and is a prerequisite for organic chemistry.

Astrochemical studies of star-forming regions in our Galaxy have progressed rapidly in recent years, including in both nearby low-mass and more distant high-mass star-forming regions.
The interstellar molecules in these regions provide information on both macroscopic aspects and microscopic processes that help us to understand physical conditions and star formation histories. 
In the Universe, including our Galaxy, most stars form from self-gravitating molecular clouds, i.e., where hydrogen exists predominantly in the form of $\rm H_2$, mediated via formation on dust grain surfaces \citep{Hollenbach1971}.
In the gas phase, ion-molecule reactions, which can proceed even at cold temperatures as low as $\sim10$ K, synthesize many molecules \citep[e.g.,][]{herbst83,Herbst1983ApJ269.329,herbst1984C3O}.
At the same time, complex organic molecules (COMs)\footnote{Molecules consisting of more than 6 atoms \citep{herbst2009}.} begin to form mainly by hydrogenation reactions on dust grain surfaces (e.g., CH$_3$OH formation by successive hydrogenation reactions of CO; \citet{Watanabe2002}). 
During the protostellar and protoplanetary disk stage, chemical processes and chemical composition become much more complex because of stellar feedbacks, such as protostellar radiative heating via dust reprocessed infrared radiation, direct impact of energetic UV and X-ray photons and relativistic cosmic ray particles, and shock heating produced by protostellar outflows and stellar winds. 

In addition to chemical composition, isotopic fractionation in some molecules can be an indicator of chemical inheritance or {\it{in situ}} chemical changes and has been subject to a variety of astrochemical studies.
Especially deuterium fractionation (D/H) and nitrogen fractionation ($^{14}$N/$^{15}$N) are essential for helping to trace the journey of materials during star and planet formation \citep[for reviews][]{caselli2012,jorgensen2020,oberg2021}.
These are particularly important for revealing the formation of our Solar System \citep[e.g.,][]{Jensen2019}, one of the most fundamental questions of astronomy.

This review focuses on ``carbon-chain molecules'', one of the major groups of molecules in the Universe.
They are abundant in the ISM and known to be useful tracers of current physical conditions and past evolutionary history. In particular, as we will see, they can be used to probe the kinematics of chemically young gas and of the early stage of molecular clouds \citep[e.g.,][]{dobashi2018,pineda2020}.
This means that line emission from rotational transitions of carbon-chain species is unique probes of gas kinematics related to star formation.
Some carbon-chain species have been suggested to possess the potential to form COMs or more complex molecules that are related to biologically relevant molecules including amino acids.
For example, cyanoacetylene (HC$_3$N) has been suggested to be a candidate for the precursor of Cytosine, Uracil, and Thymine \citep{choe2021}.
These aspects further motivate us to study their chemical characteristics in the Universe.

\subsection{History of Studies of Carbon-Chain Molecules}\label{sec:1_2}

Carbon-chain molecules are one of the major constituents of molecules detected in the Universe (see section \ref{sec:2_1}).
After the discovery of the first carbon-chain molecules in the ISM in the 1970s, many efforts to explain their formation routes were made by laboratory experiments and chemical simulations in the 1980s.
In the beginning, the focus was on gas-phase chemical reactions of small species \citep{prasad80I,prasad80II,graedel82}.
\citet{herbst83} was able to reproduce the observed abundance of a larger species, C$_4$H, in Taurus Molecular Cloud-1 (TMC-1; $d\approx140$ pc), which is one of the most carbon-chain-rich sources.
\citet{herbst83} found that ion-molecule reactions with a large amount of atomic carbon (with its abundance of $\sim 10^{-5}$) are necessary to explain the observed C$_4$H abundance.
\citet{suzuki83} found that reactions including C$^+$ can also play essential roles in carbon-chain growth.
These results suggested that carbon-chain species could efficiently form in young molecular clouds before carbon is locked into CO molecules.

\begin{figure*}
\begin{center}
\includegraphics[bb = 0 0 650 400, width=0.95\textwidth]{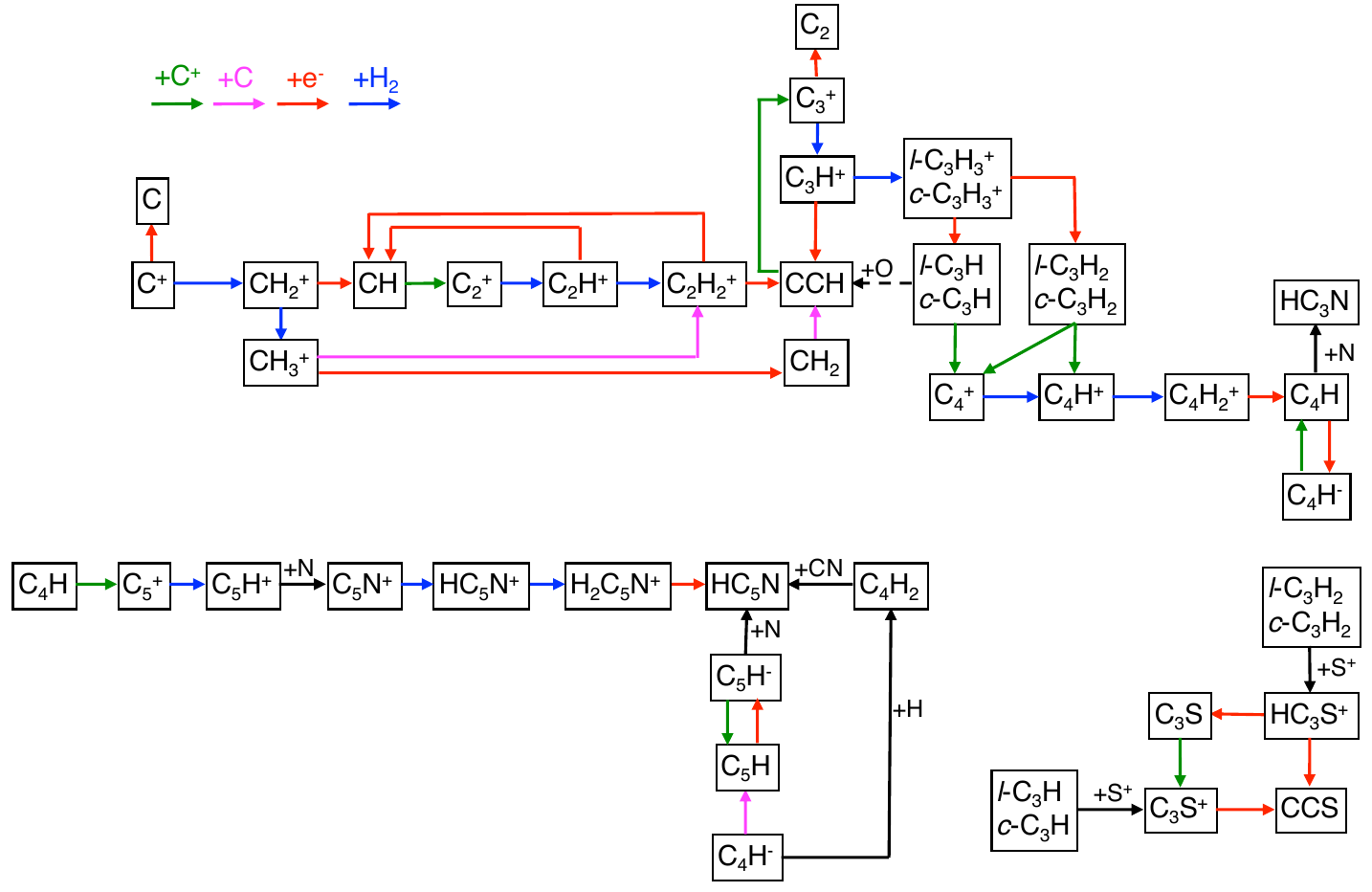}
\end{center}
\caption{Carbon-chain growth (for number of C~$\leq4$) and formation processes of HC$_5$N and CCS in early stages of molecular clouds ($t<10^{4}$ yr). 
The dashed arrow from $l,c-$C$_3$H to CCH indicates that this reaction is important around $10^{5}$ yr.}
\label{fig:route}
\end{figure*}

In the 1990s, carbon-chain molecules were detected in many molecular clouds, beyond the previously well-studied examples such as TMC-1.
Survey observations revealed that carbon-chain molecules are evolutionary indicators of starless and star-forming cores in low-mass star-forming regions; they are abundant in starless cores \citep{suzuki92,benson98}.
These studies reinforced the view that these molecules are formed from ionic (C$^+$) or atomic (C) carbon in the early stages of molecular clouds, before CO formation, as predicted by chemical simulations.
Fig. \ref{fig:route} shows the carbon-chain growth and formation pathways of nitrogen- and sulfur-bearing species (HC$_3$N, HC$_5$N, and CCS), which have been frequently detected in low-mass starless cores \citep{suzuki92}. 
These reaction schemes are constructed from results of the latest chemical simulations with a constant temperature of 10 K and a constant density of $n_{\rm{H}}=10^4$ cm$^{-3}$ in $t<10^4$ yr \citep{tani2019}.
Hydrocarbons can form efficiently from C$^+$ and C via ion-molecule reactions involving H$_{2}$ and dissociative recombination reactions leading to the formation of neutral hydrocarbons.
Such carbon-chain formation in cold molecular clouds is basically consistent with that proposed in the 1980s.

During the later stages of starless cores, carbon-chain species are adsorbed onto dust grains or destroyed by reactions with ions such as He$^+$ and H$^+$, and reactions with atomic oxygen (O).
These processes result in the depletion of carbon-chain species by the later stages of prestellar cores and protostellar cores.
Thus, carbon-chain molecules were classically known as ``early-type species''.

Subsequent studies found that carbon-chain species exist in lukewarm regions ($T\approx25-35$ K) around low-mass protostars, and a new carbon-chain formation mechanism starting from CH$_4$ was proposed (see section \ref{sec:2_2}). 
This was named Warm Carbon-Chain Chemistry \citep[WCCC; for a review][]{sakai2013}.
More recently, carbon-chain chemistry around massive young stellar objects (MYSOs) has been explored (section \ref{sec:2_3}), and Hot Carbon-Chain Chemistry (HCCC) has been proposed \citep{tani2023}.

Furthermore, very complex carbon-chain species, branched-chain molecules, and molecules including benzene rings have been discovered in the ISM in the last few years (section \ref{sec:3_11}). 
These new findings bring fresh challenges and excite our curiosity for a deeper understanding of carbon-chain chemistry in the ISM.

\subsection{Outline of This Review} \label{sec:1_3}

In this review article, we summarize results from studies of carbon-chain molecules by astronomical observations, chemical simulations, quantum chemical calculations, and laboratory experiments.
In section \ref{sec:2}, we overview the current status of the detected carbon-chain species in the ISM and the main concepts of carbon-chain chemistry around protostars, which are different from the chemistry in cold starless cores ($i.e.,$ Fig.\,\ref{fig:route}).
We review observational studies (section \ref{sec:3}), chemical models (section \ref{sec:4}), quantum chemical calculations (section \ref{sec:5}), and laboratory experiments (section \ref{sec:Experiment}). 
Finally, we list current open and key questions regarding carbon-chain species and summarise points of this review in section \ref{sec:7}.

Here, we set a definition of ``carbon-chain molecules'' for this review article, as recent detections of new interstellar species complicate this categorization. 
We include linear carbon-chain species with more than two carbon atoms and cyclic species with more than three carbon atoms containing at least one double (-=-) or triple (-$\equiv$-) bond as carbon-chain molecules.
Even if molecules meet the above criteria, species containing functional groups related to organic chemistry (e.g., -OH, -NH$_2$) are excluded from carbon-chain molecules because they are generally categorized as COMs. 
As an exception, $cyclic$-C$_2$Si, which consists of a cyclic structure with two carbon atoms and one Si atom, is treated as a carbon-chain species. 

In addition, we include molecules containing the structure of benzene, polycyclic aromatic hydrocarbons (PAHs), and fullerenes (C$_{60}$ or C$_{70}$) in this review.
Currently, it is unclear whether these species are directly or indirectly related to the above-defined ``standard'' carbon-chain species.
However, for instance, PAHs are one of the larger carbon reservoirs in the ISM; they account for up to $\sim 15$\% of the interstellar carbon and their IR luminosity is up to 20\% of the total IR power emitted by the Milky Way and star-forming galaxies \citep[e.g.,][and references therin]{Li2020NatAs...4..339L}. 
These species could be the origin of some carbon-chain species in early-stage star-forming regions via the top-down mechanism operating in harsh, diffuse ISM environments \citep[e.g.,][]{2005A&A...435..885P}. Addressing the importance of the top-down scenario and connection of carbon-chain species with PAHs is a major topic of study, which can be addressed by sensitive mm, e.g., with the Atacama Large Millimeter/submillimeter Array (ALMA), and infrared, e.g., with the James Webb Space Telescope (JWST), observations.

In the following sections, we abbreviate $linear$- and $cyclic$- as $l$- and $c$-, if necessary to indicate the molecular structure (e.g., $l$-C$_3$H$_2$ and $c$-C$_3$H$_2$).
This review article summarizes literature results until the end of December 2023.

\section{Development of Carbon-Chain Chemistry} \label{sec:2}

\subsection{Different carbon-chain families and their present status} \label{sec:2_0}

All of the carbon-chain species belonging to the various groups, C$_n$, C$_n$H, C$_n$H$^{-}$, C$_n$O, C$_n$N, C$_n$N$^{-}$,  C$_n$S, C$_n$P, HC$_{2n+1}$N, HC$_{2n}$N, HC$_{n}$O, HC$_{n}$S, H$_2$C$_{n}$, HC$_{n}$H, MgC$_{n}$H, MgC$_{n}$N, are summarized in Table \ref{tab:quant}. 
Information is given on their electronic ground state, electric polarizability, electric dipole moment, and present astronomical status.

Regarding the present astronomical detection status provided in Table \ref{tab:quant}, we have organized the data into three columns. The first column contains information on carbon-chain species detected in the ISM or CSEs.
The second column pertains to detections in TMC-1 and the third one is specific to detections in IRC+10216 only.
We have made these distinctions because the majority of known carbon-chain species have been discovered in these two sources.
Note that the absence of a check mark ($\surd$) does not necessarily mean that a species does not exist in TMC-1 or IRC+10216 because some species have not yet been searched for in each source or may not have been detected due to low abundance.
We will summarize each group in the following subsections.

\input{table1.tex}

\subsubsection{Pure carbon chains - C$_n$}

All pure linear carbon-chain species are indicated as C$_n$ ($n >1$). 
The electronic ground state of all these species is singlet, and they do not have a permanent dipole moment (see Table \ref{tab:quant}). Hence, they do not show rotational transitions, and so are not detectable via rotational transitions using radio observations. 
Instead, they show emission in the infrared domain through their rotational-vibrational transitions, and so far three chains ($n=2, 3, 5$) are astronomically detected from this group (see Table \ref{tab:quant}). 
In diffuse and translucent environments, C$_2$ formation starts with the reaction of $\rm{C^{+} + CH \rightarrow C_2^{+}+ H}$ (see Fig.\,\ref{fig:route}), followed by a series of hydrogen abstraction reactions and dissociative recombination reactions that yield C$_2$ via several channels 
\citep[][and reference therein]{welty13}.  
C$_3$ is formed via a dissociative recombination reaction of C$_3$H$^+$ that can also produce CCH (Fig.\,\ref{fig:route}), though neutral–neutral reactions (e.g., C + C$_2$H$_2$) may also contribute to the formation of C$_3$ \citep{roueff2002c3}. 

\subsubsection{Hydrocarbons - C$_n$H}

The C$_n$H group represents the simplest hydrocarbons and carbon-chain radicals. 
All carbon chains from this group have permanent dipole moments and show strong rotational transitions. 
To date, seven neutral ($n=2-8$) carbon-chain species have been detected from this group.
All of them have been identified towards both TMC-1 and IRC+10216. 
In starless cores, carbon-chain species are generally considered to form mainly by the electron recombination reactions of protonated ions such as C$_2$H$_2^+$ and $c,l$-C$_3$H$_3^+$, as shown Fig.\,\ref{fig:route}.

Apart from neutrals, four anions ($\rm{C_4H^{-}}$, $\rm{C_6H^{-}}$, $\rm{C_8H^{-}}$, and C$_{10}$H$^-$) and two cations ($\rm{C_3H^{+}}$ and $\rm{C_5H^{+}}$) have also been identified in TMC-1. 
The anions could form by the electron attachment for the neutral species; the cases of C$_4$H$^-$ and C$_5$H$^-$ are shown in Fig.\,\ref{fig:route}.
The anions belong to the even series ($n=2, 4, 6$), while the cations belong to the odd series ($n=3, 5$). 
In addition, two cyclic chains, $c$-C$_3$H, and $c$-C$_5$H, have been identified. 
All neutral species have a doublet ground state and show a trend of increasing dipole moment with the number of carbon atoms ($n$), especially for neutrals and anions (see Table \ref{tab:quant}). 

The C$_n$H family is mainly formed through the atomic reactions in the following channel, $\rm{C + C_{n-1}H_2 \rightarrow C_nH + H}$ \citep{Remijan2023}. 
This reaction scheme is indicated in Fig.\,\ref{fig:route}, from CH$_2$ to CCH.
Another two channels, which involve atomic and their related anions, can also form C$_n$H family species efficiently: $\rm{C + C_{n-1}H^{-} \rightarrow C_nH + e^{-}}$ and $\rm{H + C_n^{-} \rightarrow C_nH + e^{-}}$.

\subsubsection{Oxygen-bearing carbon chains - C$_n$O \label{sec:2_2_3}}

To date, three oxygen (O)-bearing carbon chains, C$_n$O ($n=2, 3, 5$) have been detected in the ISM. In this series, C$_3$O was the first, detected toward TMC-1 in 1984, while C$_2$O was identified in the same source in 1991. 
It took around three decades to detect the higher-order chain, C$_5$O, in TMC-1. C$_4$O is yet to be detected. Carbon chains in this group have alternate ground states, i.e., triplet and singlet, and show a trend of increasing dipole moment with the number of carbon atoms (see Table \ref{tab:quant}). 
A protonated species, HC$_3$O$^+$ with singlet ground state, has been detected in TMC-1 \citep{cernicharo2020HC3O}. 
The observed trend toward TMC-1 shows that C$_3$O is the most abundant, followed by C$_2$O and C$_5$O, with C$_5$O about 50 times less abundant than C$_2$O and about 80 times less abundant than C$_3$O \citep{Cernicharo2021C5O}. 
In addition, all these species have been identified towards the circumstellar envelope of IRC+10216. 
The formation of C$_n$O and HC$_n$O chains follows similar formation mechanisms as discussed above. The first step involves the radiative association of $\rm{C_{n-1}H^{+}}$, $\rm{C_{n-1}H_2^{+}}$, and $\rm{C_{n-1}H_3^{+}}$ ions with CO, which is then followed by dissociative electron recombination reactions \citep{adams89,Cernicharo2021C5O}. 

\subsubsection{Sulfur-bearing carbon chains - C$_n$S}

Similar to C$_n$O, several sulfur (S)-bearing carbon chains, C$_n$S ($n=2, 3, 4, 5$), have been identified in the ISM. 
Two protonated species, HCCS$^{+}$ and HC$_3$S$^+$, have only been detected towards TMC-1 so far. 
The discovery of HC$_3$S$^+$ supports the formation route of CCS and C$_3$S via the electron recombination reactions of HC$_3$S$^+$ indicated in Fig.\,\ref{fig:route}. 
Carbon chains in this group have alternative ground states, i.e., triplet and singlet, and show a trend of increasing dipole moment with the number of carbon atoms, similar to the C$_n$O group (see Table \ref{tab:quant}). 
The abundances of C$_2$S and C$_3$S are almost three orders of magnitude higher than C$_4$S and C$_5$S toward TMC-1 \citep{Cernicharo2021S}. On the other hand, the C$_5$S column density is slightly less than those of C$_2$S and C$_3$S, with differences less than one order of magnitude, toward IRC+10216 \citep{Agundez2014}. 
C$_2$S and C$_3$S are mainly produced via several ion-neutral reactions followed by electron recombination reactions and via several neutral-neutral reactions \citep{sakai2007}. 
Higher order chains of this family, such as C$_4$S and C$_5$S, are thought to be formed via reactions of S + C$_4$H and C + HC$_3$S, and C$_4$H + CS and S + C$_5$H, respectively.
However, the kinetics and product distribution of these reactions are poorly known \citep{Cernicharo2021S}.

\subsubsection{Nitrogen-bearing carbon chains - C$_n$N}

In this group, C$_3$N was the first detected species, done so tentatively toward IRC+10216 in 1977 and more robustly toward TMC-1 in 1980. The next higher order chain in this series, C$_5$N was detected toward TMC-1 and tentatively detected toward IRC+10216 in 1998. The lower order chain, CCN, was found in 2014.
Their anions, $\rm{C_3N^{-}}$ and $\rm{C_5N^{-}}$, were discovered in the circumstellar envelope of the carbon-rich star IRC+10216 \citep{Thaddeus08,Cernicharo2008C5Nm}. 
They have also been identified toward TMC-1 by the QUIJOTE group, including their neutral analogs (C$_3$N, C$_5$N) \citep{Cernicharo2020C3Nm}. 
They measured similar abundance ratios of $\rm{C_3N^{-}}$/C$_3$N = 140 and 194, and $\rm{C_5N^{-}}$/C$_5$N = 2 and 2.3 in TMC-1 and IRC+10216, respectively, even though physical conditions are completely different for TMC-1 and IRC+10216. 
It might be a coincidence that there are similar abundance ratios of anion and neutral forms of C$_{n}$N ($n = 3, 5$).
All carbon chains from this group have doublet ground state, and the two anionic forms have singlet state. 
Since the dipole moment of CCN is low compared to those of C$_3$N, C$_5$N, and their anionic forms, the detection of CCN is more challenging, even though it is of lower order in the chain (see Table \ref{tab:quant}). 
C$_4$N, C$_6$N, and C$_7$N show even smaller values of their dipole moments, which suggests that much high sensitivity observations are required for their identification. 
C$_2$N is produced through the reactions of N + C$_2$ and C + CN. 
Similarly, C$_3$N is produced in reactions of N + C$_3$ and C + CCN, and C$_5$N is produced through N + C$_5$ on dust surfaces\footnote{\url{https://kida.astrochem-tools.org/}}. 
The production of $\rm{C_3N^{-}}$ mainly comes from the reaction between N atoms and bare carbon-chain anions $\rm{C_n^{-}}$ \citep{Cernicharo2020C3Nm}, whereas $\rm{C_5N^{-}}$ is produced via the electron radiative attachment to C$_5$N \citep{walsh09}. 

\subsubsection{Phosphorus-bearing carbon chains - C$_n$P}

Although phosphorus (P) has a relatively small elemental abundance, it plays a crucial role in the development of life \citep{Chantzos2020P}. Among the known eight phosphorus-bearing molecules, C$_2$P (or CCP) is the only P-bearing carbon-chain species and detected toward IRC+10216 \citep{Halfen08}. 
All carbon chains in this group have doublet ground states (Table \ref{tab:quant}). Higher order chains, C$_3$P and C$_4$P, show a higher value of dipole moments, but they are yet to be detected in the ISM or circumstellar environments. Since the overall elemental abundance of phosphorous is small, higher-order phosphorous chains are expected to have very low abundances. 
CCP may be produced by radical-radical reactions, between CP and hydrocarbons (CCH and C$_3$H), or ion-molecule chemistry involving P$^{+}$ and HCCH followed by the dissociative electron recombination reaction \citep{Halfen08}. 

\subsubsection{HC$_n$O family}

Four neutral HC$_n$O ($n=2, 3, 5, 7$) chains have been identified toward TMC-1 \citep{Cernicharo2021C5O,McGuire2017,Cordiner2017}. 
The detection summary of this group indicates odd $n$ chains are more abundant compared to their even $n$ counterparts. This trend is the same as in the C$_n$O family.  
All neutral chains have doublet ground states and dipole moment values are less than 3 Debye. The observed cation HC$_3$O$^+$ has a singlet ground state and a dipole moment of 3.41 Debye (see Table \ref{tab:quant}). As mentioned before, C$_n$O and HC$_n$O are linked through their formation routes (sec Sec. \ref{sec:2_2_3}).  

\subsubsection{HC$_n$S family}

This family is similar to HC$_n$O but contains sulfur instead of oxygen. Only two neutral species, HCCS and HC$_4$S have been identified toward TMC-1 \citep{Cernicharo2021S,Fuentetaja2022HCCCHCCC}. 
Observed statistics suggest chains with even $n$ have higher abundance than odd $n$ species. 
All neutral chains of this group have doublet ground states (see Table \ref{tab:quant}). The dipole moments of neutral species are less than 2.2 Debye. HCCS is mainly formed through the reaction, C + H$_2$CS \citep{Cernicharo2021S}. 
HC$_4$S is produced through the reaction between C and $\rm{H_2C_3S}$ and by the dissociative recombination reaction of $\rm{H_2C_4S^{+}}$, which is formed via reactions of S + $\rm{C_4H_3^{+}}$ and S$^{+}$ + $\rm{C_4H_3}$ \citep{fuentetaja2022}.
This group has protonated ions, and HC$_2$S$^+$ and HC$_3$S$^+$ have been detected in TMC-1 \citep{cabezas2022,cernicharo2021HC3S}.
For $\rm{HC_3S^{+}}$, proton transfer to C$_3$S from HCO$^{+}$ and $\rm{H_3O^{+}}$ is the main formation route. 
The reactions of S$^+$ + $c,l-\rm{C_3H_2}$ (see Fig.\,\ref{fig:route}) and S + $c,l-\rm{C_3H_3}$ are also equally important and efficient \citep{cernicharo2021HC3S}.

\subsubsection{Cyanopolyynes - HC$_{2n+1}$N}

Cyanopolyynes are the most important, interesting, and ubiquitous organic carbon chains ($n=1-5$) detected in the ISM so far. As mentioned above, HC$_3$N was the first detected carbon-chain molecule in space. In this series, five species, starting from HC$_3$N to HC$_{11}$N, have been found in TMC-1.
All these species have also been detected toward IRC+10216, except HC$_{11}$N \citep{Morris76,Winnewisser1978,Matthews1985}. 
In this series, especially HC$_3$N and HC$_5$N, have been identified in various star-forming environments (see section \ref{sec:3_32}). 
Three cations, HC$_3$NH$^+$, HC$_5$NH$^+$, and HC$_7$NH$^+$, have also been identified toward TMC-1 \citep{Kawaguchi94,Marcelino20,cabezas2022HC7NH}. 
The detection of these protonated species in TMC-1 supports the formation route of the neutral species via electron recombination reactions (see Fig.\,\ref{fig:route}).
Their main formation pathways have been studied by observations of their $^{13}$C isotopic fractionation, as we discuss in section \ref{sec:3_12}.
All neutral cynaopolyynes have a singlet ground state and show a trend of increasing dipole moment with length of the chain (see Table \ref{tab:quant}). 
Unlike other carbon-chain species, the cyanopolyyne family could form on dust surfaces through reactions N + C$_{2n+1}$H ($n=1-4$) and H + C$_{2n+1}$N ($n=1-4$) (see section \ref{sec:3_12} for more detail regarding the formation of cyanopolyynes in the gas phase). Protonated cyanopolyynes (e.g., $\rm{HC_3NH^{+}}$, $\rm{HC_5NH^{+}}$) are mainly formed via a proton donor (e.g., HCO$^+$) to cyanopolyynes (e.g., HC$_3$N, HC$_5$N). 
Protonated cyanopolyynes are destroyed by dissociative electron recombination reactions \citep{Marcelino20}.

\subsubsection{Allenic chain family - HC$_{2n}$N}

HCCN was the first member of the allenic chain family, HC$_{2n}$N, observed in space \citep{Guelin91}, and HC$_4$N was the second. These species have been identified toward IRC+10216. The allenic chain family has a triplet ground state and shows increasing dipole moment with size, similar to cyanopolyynes and other families (Table \ref{tab:quant}). HC$_4$N may form through the reactions of C$_3$N + CH$_2$ and C$_3$H + HCN. For this family, ion-molecule paths are relatively slow \citep{Cernicharo04}. However, HCCN is formed by the reactions between atomic nitrogen and $\rm{H_nCCH^{+}}$.

\subsection{Statistics of Detected Species} \label{sec:2_1}

\begin{sidewaysfigure*}
\includegraphics[bb = 0 0 850 300, width=\columnwidth]{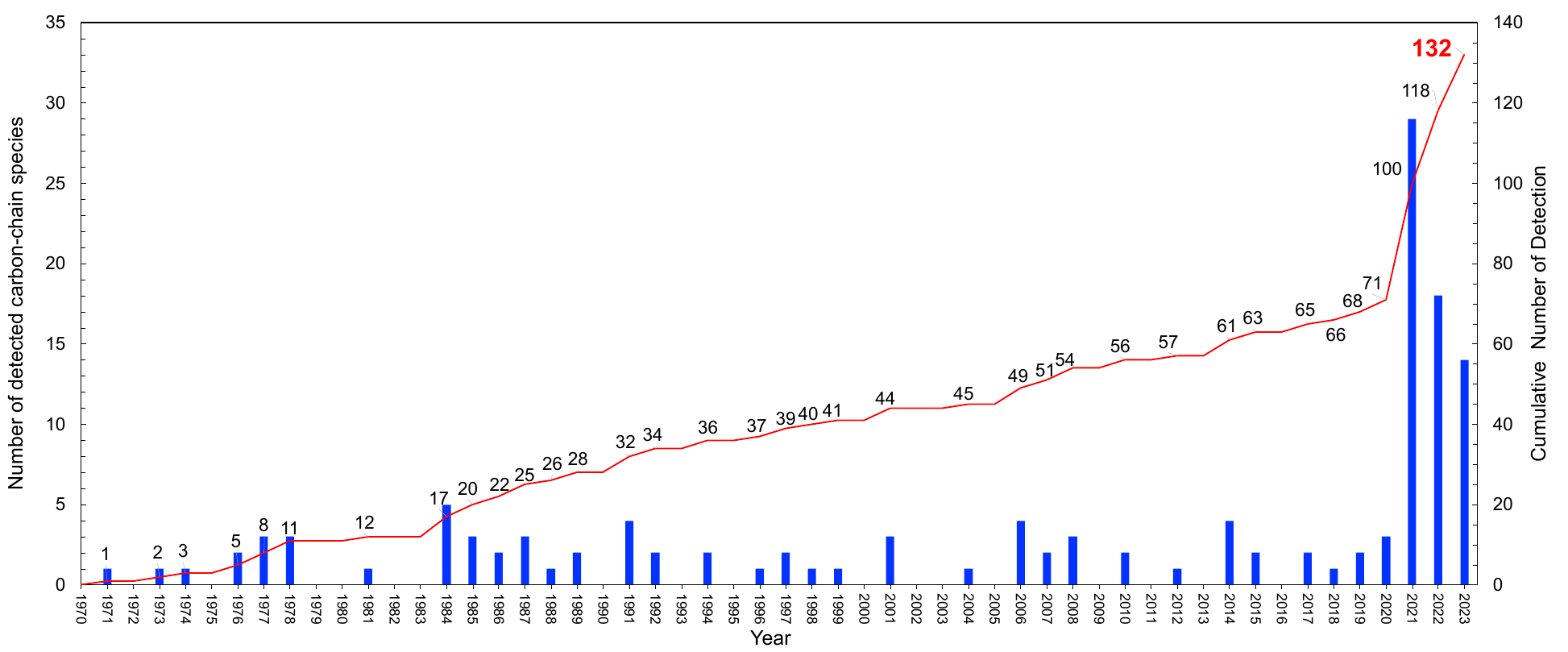}
\caption{The number of detected carbon-chain species in each year (blue bars) and its cumulative plot (red curves and numbers).}
\label{fig:detectionsummary}
\end{sidewaysfigure*}

Fig.\,\ref{fig:detectionsummary} shows the cumulative plot of carbon-chain detection together with the histogram plot in each year starting from 1971, the first carbon-chain detection year, until 2023. 
Following the definition mentioned earlier in section \ref{sec:1_3}, 132 carbon-chain species have been discovered until the end of 2023.  
This accounts for approximately 43\% of all the known 305 molecules. 
Carbon-chain species that have been detected so far and candidates for future detection are summarized in Table \ref{tab:quant}.

Following the discovery of HC$_3$N in 1971 \citep{Turner71}, new carbon-chain species continued to be found at a rate of just over one species per year for the next 50 years. 
A steep increase is seen after 2021 in Fig.\,\ref{fig:detectionsummary}, thanks to the two deep line survey projects toward TMC-1 CP (section \ref{sec:3_11}).
29, 18, and 14 carbon-chain species have been discovered in 2021, 2022, and 2023, respectively.
Before that, the maximum number was five in 1984. 
Hence, the number of discovered new carbon-chain species in the last three years is larger than those in the previous era by a factor of three to five.

Note that most of the achievements in the last three years are at TMC-1 CP.
We cannot conclude that these new carbon-chain species are prevalent in other starless cores and star-forming regions. In addition, 
this also raises the question as to why TMC-1 appears to be particularly rich in carbon-chain species compared to other starless cores.
Line survey observations toward a large sample of starless and star-forming cores in various environments are needed to answer this question.

\subsection{Warm Carbon-Chain Chemistry (WCCC)} \label{sec:2_2}

In section \ref{sec:1_2}, we mentioned that carbon-chain molecules have been classically known as early-type species, because they are abundant in young starless cores and deficient in evolved star-forming cores. 
Against this classical picture, \citet{sakai2008} detected various carbon-chain molecules toward IRAS\,04368+2557 in the low-mass star-forming region L1527 in Taurus.
The derived rotational temperature from the C$_{4}$H$_{2}$ lines is $12.3\pm0.8$ K, which is higher than excitation temperatures of carbon-chain species in the starless core TMC-1 ($\approx 4-8$ K).
They proposed that evaporation of CH$_{4}$ from ice mantles could be the trigger of formation of carbon-chain molecules in the lukewarm envelopes around low-mass protostars, and named such a carbon-chain formation mechanism ``Warm Carbon-Chain Chemistry (WCCC)''.
A second WCCC source, IRAS\,15398-3359 in the Lupus star-forming region, was discovered soon after \citep{sakai2009}.
This suggested that the WCCC mechanism may be a common feature around low-mass protostars, and WCCC has been widely accepted in the astrochemical field.

Later studies using chemical simulations support the formation mechanism of carbon-chain molecules starting from CH$_4$ around temperatures of 25--30 K \citep{hassel2008}.
The CH$_4$ molecules react with C$^+$ in the gas phase to produce C$_2$H$_3$$^+$ or C$_2$H$_2$$^+$.
The C$_2$H$_2$$^+$ ion reacts with H$_2$ leading to C$_2$H$_4$$^+$.
Then, electron recombination reactions of C$_2$H$_3$$^+$ and C$_2$H$_4$$^+$ produce C$_2$H$_2$.
Regarding WCCC, the review article by \citet{sakai2013} summarized related studies in detail.
We thus avoid duplication here.

However, an important question has been raised since the review of \citet{sakai2013}, namely, the origin(s) of WCCC sources, which is still controversial.
The focus is on how CH$_4$-rich ice is formed.
This means that carbon atoms need to be adsorbed onto dust grains without being locked up in CO molecules.
\citet{sakai2008} proposed a possible solution involving short collapse times of prestellar cores in order to produce conditions needed for WCCC sources. However, there is limited evidence for such short collapse times, e.g., based on observed infall velocities \citep{2010MNRAS.402.1625K}, levels of deuteration \citep{2016ApJ...821...94K} or demographics of prestellar versus protostellar cores \citep{2015A&A...584A..91K}.

As an alternative scenario, \citet{spezzano2016} suggested that variations in the far ultraviolet (FUV) interstellar radiation field (ISRF) could produce carbon-chain-rich or COM-rich conditions, based on their observational results toward the prestellar core L1544. They found a spatial distribution of $c$-C$_3$H$_2$ in a region relatively exposed to the ISRF, while CH$_3$OH was found in a relatively shielded region. In this scenario, the FUV ISRF destroys CO, a precursor of CH$_3$OH, leading to formation of C and/or C$^+$, precursors of carbon-chain species.
\citet{spezzano2020} also found similar trends with observations of $c$-C$_3$H$_2$ and CH$_3$OH toward six other starless cores.
They concluded that the large-scale effects have a direct impact on the chemical segregation; $c$-C$_3$H$_2$ is enhanced in the region more illuminated by the ISRF, whereas CH$_3$OH tends to reside in the region shielded from the ISRF.
Such chemical segregation observed in starless cores may be inherited to the protostellar stage and recognized as the different classes of WCCC protostars and COM-rich hot corinos. 
The feasibility of this scenario is boosted by its ability to explain the data of multiple independent cores.

In addition, several authors have since presented chemical simulations of the effects of these factors on the abundances of carbon-chain species and COMs  \citep{aikawa2020, kalvans2021}. We discuss their modeling results in section \ref{sec:4}.

\subsection{Concept of Hot Carbon-Chain Chemistry} \label{sec:2_3}

The discovery of the WCCC mechanism around low-mass protostars naturally raised a question: are carbon-chain molecules formed around high-mass ($m_{\ast}>8 M_{\odot}$) protostars?
With such a motivation, observations and chemical simulations focusing on carbon-chain species around massive young stellar objects (MYSOs) have proceeded since the late 2010s. 
Observational studies show that carbon-chain species are abundant around some MYSOs, but their abundances, especially cyanopolyynes, cannot be reproduced by WCCC.
Then, a different carbon-chain chemistry has been proposed \citep{tani2023}.
Here, we briefly explain the concept of ``Hot Carbon-Chain Chemistry (HCCC)'' proposed to reproduce the observational results around MYSOs. 
Details of these observational and simulation studies are summarized in sections \ref{sec:3_3} and \ref{sec:4}, respectively.

\begin{figure*}[ht]
\begin{center}
\includegraphics[bb = 0 0 700 300, width=\textwidth]{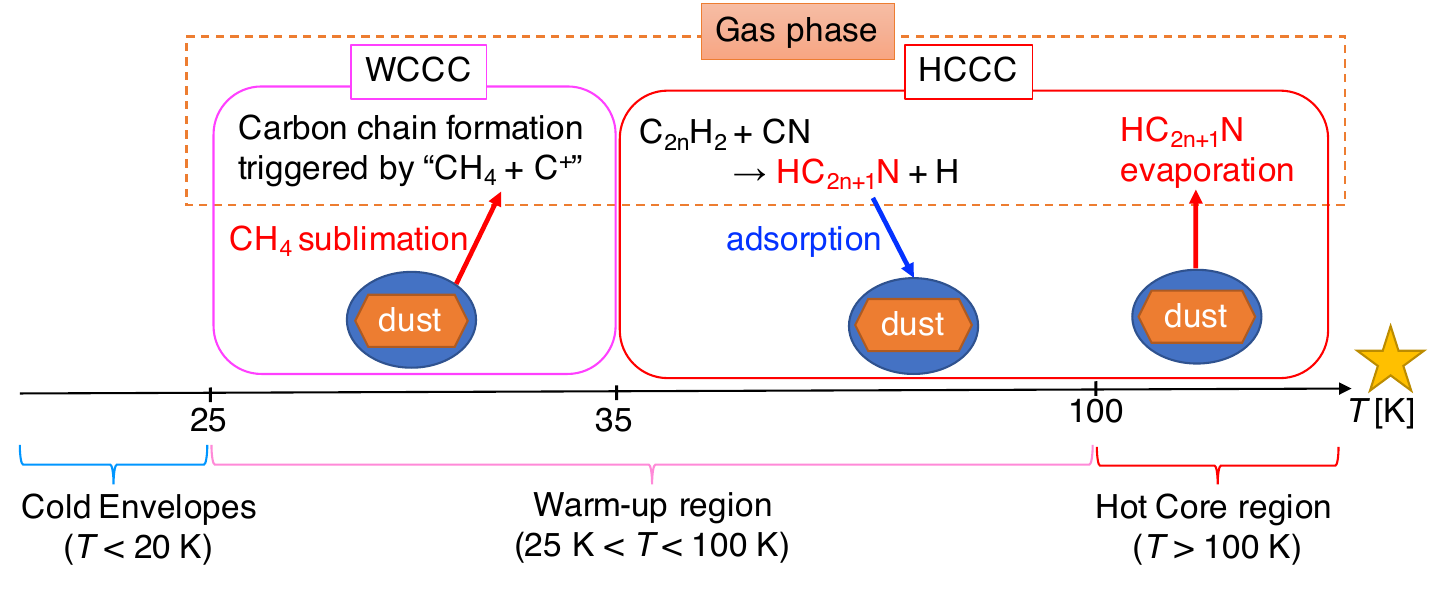}
\end{center}
\caption{Temperature dependence of carbon-chain chemistry. WCCC occurs around 25--35 K, while HCCC occurs in higher-temperature regions.}
\label{fig:HCCCconcept}
\end{figure*}

Fig.\,\ref{fig:HCCCconcept} shows a schematic view of carbon-chain chemistry around MYSOs based on the result of the chemical simulation by \citet{tani2019model}. 
We distinguish HCCC from WCCC depending on the temperature.
The HCCC mechanism refers to carbon-chain formation in the gas phase, adsorption onto dust grains, and accumulation in ice mantles during the warm-up phase (25 K $<T<$ 100 K), followed by evaporation into the gas phase in the hot-core stage ($T>$ 100 K).

This mechanism is particularly important for cyanopolyynes (HC$_{2n+1}$N).
Cyanopolyynes are relatively stable species because of no dangling bond, and they are not destroyed in the gas phase by reactions with O or H$_2$, which are major destroyers of other carbon-chain species with dangling bonds in the warm gas. 
Instead, cyanopolyynes are consumed by reactions with protonated ions such as HCO$^+$ in the gas phase.
Thus, the gas-phase destruction rates of cyanopolyynes are lower than those of the other carbon-chain species, which enables cyanopolyynes to be adsorbed onto dust grains.
During the warm-up stage, cyanopolyynes are efficiently formed by the gas-phase reactions between C$_{2n}$H$_{2}$ and CN. The formed cyanopolyynes are adsorbed onto dust grains and accumulated in ice mantles.
When conditions reach their sublimation temperatures above 100 K, the species sublimate into the gas phase and their gas-phase abundances show peaks.

Radical-type species, such as CCH and CCS, would not behave as cyanopolyynes do, because they are efficiently destroyed by the gas-phase reactions with O or H$_2$ \citep{tani2019model}. 
Their gas-phase peak abundances are reached just after WCCC starts to form them around 25 K, and decrease as the temperature increases.
Thus, we expect that radical-type carbon-chain species are abundant in the lukewarm regions and deficient in the hot-core regions, whereas the emission of cyanopolyynes is expected to show their peaks at the hot-core regions, similar to the emission of COMs.

Basically, HCCC can operate even around low-mass YSOs, because higher temperature is the only important factor to distinguish from WCCC and low-mass YSOs will have localized regions reaching the required temperatures of $\sim 100\:$K.
However, these regions are much smaller than the equivalent ones around MYSOs, making it more difficult to resolve the relevant temperature structures and detect the presence of HCCC around low-mass YSOs.

The main points of this section are summarized below:
\begin{enumerate}
\item Carbon-chain molecules account for around 44\% of the interstellar molecules. These molecules have been detected in the ISM since the 1970s, and an increased number of reported detections made by recent Green Bank 100m telescope (GBT) and Yebes 40m telescope observations are astonishing.
\item WCCC refers to the carbon-chain formation mechanism in the lukewarm gas ($T\approx25-35$ K) starting from CH$_4$ desorbing from dust grains around 25 K. The gas-phase reaction between CH$_4$ and C$^+$ is the trigger of the WCCC mechanism.
\item HCCC refers to the gas-phase carbon-chain formation and adsorption and accumulation in ice mantles during the warm-up phase ($T< 100$ K), and their sublimation in the hot-core phase ($T>100$ K).
\end{enumerate}

\section{Observations} \label{sec:3}

\subsection{Carbon-Chain Species in TMC-1} \label{sec:3_11}

\begin{sidewaysfigure*}
\includegraphics[bb = 0 0 700 400, width=\columnwidth]{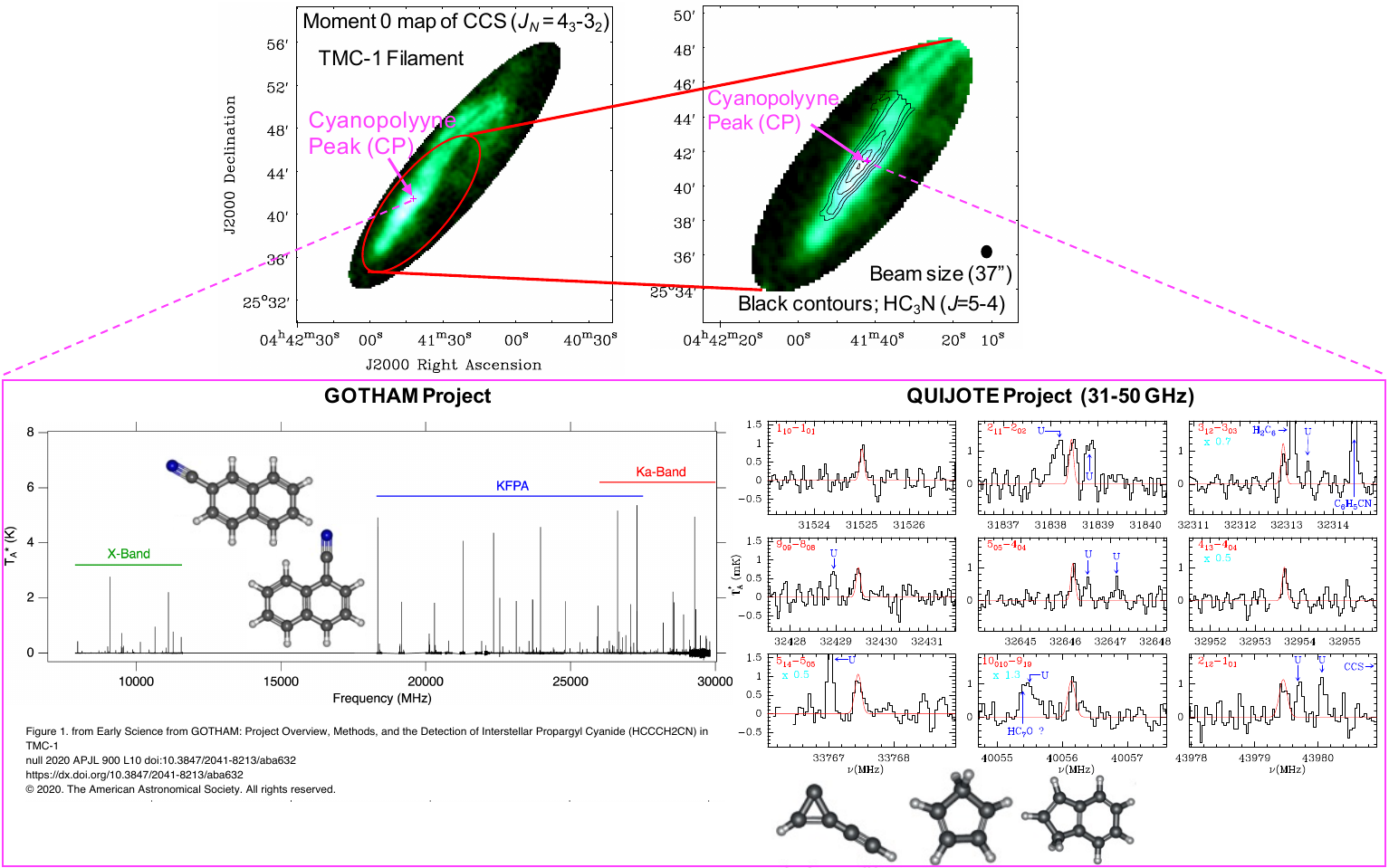}
\caption{Top panels: Moment 0 map of the CCS ($J_N=4_3-3_2$) line overlaid by black contours indicating moment 0 map of the HC$_3$N ($J=5-4$) line. These data were obtained with the Nobeyama 45m radio telescope (beam size=$37^{\prime\prime}$ at 45 GHz). The original data were provided by Dr. Fumitaka Nakamura (NAOJ) and Dr. Kazuhito Dobashi (Tokyo Gakugei University). The magenta cross shows the position of the Cyanopolyyne Peak (TMC-1 CP) observed by the two line survey projects, GOTHAM and QUIJOTE projects. The spectral figures are from \citet{McGuire2020summary} and \citet{cernicharo2021hydrocarboncycle}.}
\label{fig:tmc1}
\end{sidewaysfigure*}

Taurus Molecular Cloud-1 (TMC-1) is one of the best-studied filament \citep[e.g.,][]{kaifu2004}.
Its ``Cyanopolyyne Peak'' (hereafter TMC-1 CP), which is located southeast of the filament, is the famous site where carbon-chain molecules are particularly abundant (Fig.\,\ref{fig:tmc1}).
A number of deep-integration observations of carbon-chain species have been conducted toward this position.

\citet{dobashi2018} identified four velocity components ($v_{\rm {LSR}}$ = 5.727, 5.901, 6.064, and 6.160 km\,s$^{-1}$) at TMC-1 CP with very high velocity resolution (0.0004 km\,s$^{-1}$) spectra of the CCS and HC$_3$N lines in the 45 GHz band obtained by the Z45 receiver \citep{Nakamura2015PASJ} installed on the Nobeyama 45m radio telescope.
They revealed the gas kinematics of the TMC-1 filament and found that these four velocity components indicate moving inward toward the center of the TMC-1 filament.
\citet{dobashi2019} identified 21 subfilaments in the TMC-1 filament using CCS ($J_{N}=4_{3}-3_{2}$; 45.379033 GHz) line data. 
They found that the subfilaments have line densities that are close to the critical line density for dynamical equilibrium ($\sim 17 M_{\odot}$\,pc$^{-1}$).
These results indicate that self-gravity is important in the dynamics of the subfilaments.

The CCS ($J_{N}=4_{3}-3_{2}$) line was also used for measurement of the line-of-sight magnetic field strength by its Zeeman splitting \citep{nakamura2019}.
The derived magnetic field strength is $\sim117\pm21$ $\mu$G, which implies that the TMC-1 filament is magnetically supercritical.
As these studies show, rotational-transition lines of carbon-chain species are useful to investigate physical conditions of starless cores.

Two research groups have been carrying out line survey observations toward TMC-1 CP and have reported the detection of new interstellar molecules. Their research programs are still ongoing as of the writing of this review article.
Here we highlight groundbreaking results done by each project. The sample of detected species from these projects is also summarized in section \ref{sec:2_0}.

One project is GOTHAM (GBT Observations of TMC-1: Hunting Aromatic Molecules\footnote{\url{https://greenbankobservatory.org/science/gbt-surveys/gotham-survey/}}) using the Green Bank 100m telescope. 
This project is a high sensitivity (2 mK) and high velocity resolution (0.02 km\,s$^{-1}$) spectral line survey in the X, K, and Ka bands (see Fig.\,\ref{fig:tmc1}).
The beam sizes (FWHM) are $1.4^{\prime}$, $32^{\prime\prime}$, and $26.8^{\prime\prime}$ for the X-Band receiver, KFPA, and Ka-Band (MM-F1) receiver, respectively\footnote{\url{https://www.gb.nrao.edu/scienceDocs/GBTpg.pdf}}.
They have analyzed spectra using the spectral line stacking and matched filter methods \citep{loomis2018,loomis2021} utilizing the velocity information derived by \citet{dobashi2018}, and achieved detection of new, rare interstellar molecules.

\citet{mcguire2021} detected two polycyclic aromatic hydrocarbons (PAHs), 1- and 2-cyanonaphthalene, containing two rings of benzene via spectral matched filtering.
Some strong lines of 1-cyanonaphthalene can be identified in the smoothed spectra with a 14-kHz resolution.
Their molecular structures are shown in Fig.\,\ref{fig:tmc1}.
The nitrile bond (-CN) makes the dipole moment larger, thus aiding the detection of benzonitrile ($c$-C$_6$H$_5$CN), the first detected species with a benzene ring at TMC-1 CP \citep{McGuire2018}.
Benzonitrile has also been detected in other sources: Serpens 1A, Serpens 1B, Serpens 2, and L1521F \citep{burkhardt2021natureastronomy}.
Although the detection of pure hydrocarbon rings was considered to be difficult due to their small permanent dipole moments, \citet{burkhardt2021} achieved the detection of indene ($c$-C$_9$H$_8$).
\citet{mccarthy2021} detected 1-cyano-1,3-cyclopentadiene ($c$-C$_5$H$_5$CN), a five-membered ring, and \citet{lee2021} detected 2-cyano-1,3-cyclopentadiene, which is an isomer with a little higher energy (5 kJ\,mol$^{-1}$ or 600 K).
Thus, not only molecules with benzene structure but also molecules with five-membered rings with a nitrile bond have been detected.
\citet{loomis2021} reported the detection of HC$_{11}$N.
HC$_4$NC, an isomer of HC$_5$N, has been detected by \citet{Xue2020}, and soon after \citet{cernicharo2020HC4NC} also reported its detection using the Yebes 40m telescope.
\citet{Xue2020} ran chemical simulations with formation pathways of electron recombination reactions of HC$_5$NH$^+$ and HC$_4$NCH$^+$, and reproduced the observed abundance of HC$_4$NC.

The other project is QUIJOTE (Q-band Ultrasensitive Inspection Journey to the Obscure TMC-1 Environment) line survey using the Yebes 40m telescope. 
This line survey project covers the frequency range 31.0--50.3 GHz, with beam sizes of $56^{\prime\prime}$ and $31^{\prime\prime}$ at 31 GHz and 50 GHz, respectively \citep{Fuentetaja2022HCCCHCCC}. The frequency resolution is 38.15 kHz, corresponding to $\sim0.29$ km\,s$^{-1}$ at 40 GHz. We note that this velocity resolution is not ideal to observe lines in TMC-1 CP, which have a typical line width of $\sim 0.5$ km\,s$^{-1}$. However,
this survey has achieved very high sensitivities of 0.1 -- 0.3 mK and various molecules have been successfully detected without resort to stacking analysis. 

The QUIJOTE project has reported the detection of many pure hydrocarbons consisting of only carbon and hydrogen: e.g., 1- and 2-ethynyl-1,3-cyclopentadiene \citep[$c$-C$_5$H$_5$CCH,][]{cernicharo2021cyclopentadiene}; benzyne \citep[$ortho$-C$_6$H$_4$,][]{cernicharo2021benzyne}; $c$-C$_3$HCCH, $c$-C$_5$H$_6$, $c$-C$_9$H$_8$ \citep{cernicharo2021hydrocarboncycle}; fulvenallene \citep[$c$-C$_5$H$_4$CCH$_2$,][]{cernicharo2022fulvenallene}; and CH$_2$CCHC$_4$H \citep{fuentetaja2022}.
The detection of such pure hydrocarbons is astonishing because their dipole moments are very small.
In addition to pure hydrocarbons, the QUIJOTE project has also detected carbon-chain ions: e.g., HC$_3$O$^+$ \citep{cernicharo2020HC3O}; HC$_7$NH$^+$ \citep{cabezas2022HC7NH}; HC$_3$S$^+$ \citep{cernicharo2021HC3S}; HCCS$^+$ \citep{cabezas2022}; C$_5$H$^+$ \citep{cernicharo2022C5H}. It has also detected five cyano derivatives \citep[$trans$-CH$_3$CHCHCN, $cis$-CH$_3$CHCHCN, CH$_2$C(CH$_3$)CN, $gauche$-CH$_2$CHCH$_2$CN, $cis$-CH$_2$CHCH$_2$CN;][]{cernicharo2022cyano}.
The very high sensitivity line survey observations achieved by the QUIJOTE project reveal a wide variety of carbon-chain species.
At the same time, these results raise new questions because the abundances of some of the newly detected molecules cannot be explained by chemical models.

\begin{figure}
    \centering
    \includegraphics[bb = 0 20 900 750, width=0.95\textwidth]{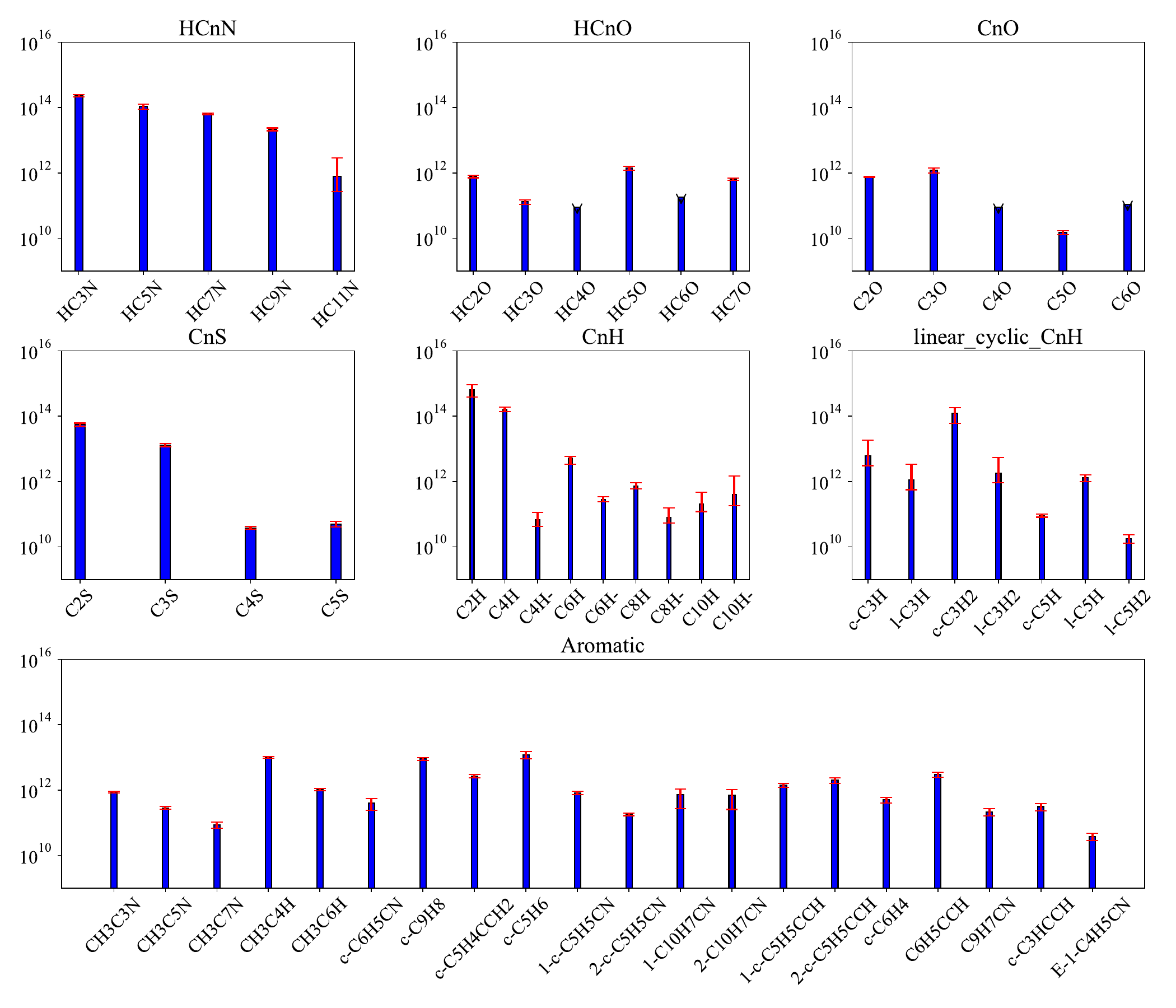}
    \caption{Column density variation of carbon chains toward TMC-1 CP.}
    \label{fig:tmc_column}
\end{figure}

Fig.\,\ref{fig:tmc_column} shows column densities of several carbon-chain series at TMC-1 CP
(see also Table \ref{tab:colden-tmc} in Appendix \ref{sec:appendix2}).
We note that several factors could produce artificial (systematic) differences in derived column densities; 
\begin{enumerate}
    \item variation in telescope beam sizes and pointing relative to the source (i.e., leading to different levels of beam dilution) 
    \item variation of excitation conditions probed by the different transitions
    \item variation of analytical methods (e.g., rotational diagram vs. Markov Chain Monte Carlo (MCMC); single-velocity components vs. multi-velocity components)
    \item variation of data selection criteria and data quality (e.g., system noise temperatures) 
\end{enumerate}
For example, regarding the column densities of cyanopolyynes (HC$_{2n+1}$N) obtained by GBT, the column density of HC$_9$N \citep[$2.15^{+0.23}_{-0.20} \times 10^{13}$ cm$^{-2}$;][]{loomis2021} is higher than that of HC$_7$N \citep[($1.39\pm0.36$)$\times 10^{13}$ cm$^{-2}$;][]{burkhardt2018}, which is an unexpected result. However, we consider that this may be affected by the use of different analytical methods.
Column densities smaller than HC$_7$N have been derived assuming a single velocity component, whereas HC$_9$N and HC$_{11}$N have been analyzed with the application of four-velocity components \citep{loomis2021}. 
The plotted column densities of HC$_9$N and HC$_{11}$N are summations of these four velocity components.
This illustrates the need to compare results among different methods carefully.
In the following discussion, we will mainly compare carbon-chain species whose column densities have been obtained from the same papers using the same methods in order to avoid such potential systematic uncertainties.

The C$_n$H series shows an odd-even fluctuation, with the even-numbered species having higher abundances compared to the odd-numbered. 
The anions have been detected only for the even-numbered members. 
The neutral/anion abundance ratios are derived to be $\sim 2386$, 18, 9, and 0.5 for C$_4$H, C$_6$H, C$_8$H, and C$_{10}$H, respectively.
Note that the identification of C$_{10}$H is tentative \citep{Remijan2023}, and the ratio is just a reference value level.
Thus, the longer members have higher relative abundances of anion forms to their neutrals (Table \ref{tab:colden-tmc} in Appendix \ref{sec:appendix2}). 

The cyclic-to-linear ratios of C$_3$H$_2$ and C$_3$H have been discussed \citep[e.g.,][]{loison2017,Olli2016}.
Such ratios were derived to be 67 and 5.5 for C$_3$H$_2$ and C$_3$H at TMC-1 CP \citep{loison2017}, and these ratios likely depend on the density.
As seen in Fig.\,\ref{fig:route}, these four species are chemically close to each other; electron recombination reactions of $c/l-$C$_3$H$_3^+$ could produce these molecules.
It is important to fully understand these ratios in various astronomical environments to measure branching ratios of dissociative recombination reactions of their parent hydrocarbon ions ($c.f.,$ Fig.\,\ref{fig:route}).
The cyclic forms may not be necessarily more abundant than the linear form. 
In the case of C$_5$H, $l-$C$_5$H is the most stable isomer, and observations show that $c-$C$_5$H is less abundant than $l-$C$_5$H by around one order of magnitude \citep{Cabezas2022cyclicC5H}.
Thus, the stability of the molecules is likely one of the key factors in determining which isomer(s) is the most abundant. 
Quantum calculation is essential to investigate the stability.
The number of isomers increases as the molecule becomes larger, and sophisticated calculations are necessary (section \ref{sec:5_2}).

The question of why TMC-1 CP is rich in carbon-chain species may be related to the formation and evolution of the TMC-1 filament.
If gravitational collapse is impeded leading to a longer duration starless core phase, then this may allow more time for the growth of longer carbon chains.
Magnetic field support is the most likely cause of delayed gravitational collapse. However, precise measurements of magnetic field strengths, e.g., by Zeeman measurements, are difficult to obtain and such information is generally lacking for most starless cores.
We note that the extended Q-band (eQ) receiver, recently installed on the Nobeyama 45m radio telescope, will enable improved Zeeman measurements of line of sight magnetic field strengths in starless and star-forming cores with a much shorter observing time \citep{eQ2022}. Such measurements are need to study the potential relationships between core dynamics and carbon-chain chemistry.

Another possibility to explain elevated abundances of long carbon chains in TMC-1 is enhanced efficiency of the top-down mechanism, in which large carbon-bearing species such as PAHs are destroyed to form small hydrocarbons.
This scenario may be consistent with the detection of molecules including benzene, such as benzonitrile and 1- and 2-cyanonaphthalene. If the enhancement of PAH destruction is due to the global environment, e.g., of Taurus, then one would expect generally elevated abundances of larger carbon-chain species in this region.
Revealing the origin of the chemical characteristics of TMC-1 CP, likely requires a better understanding of the chemical heritage from the diffuse ISM to molecular gas conditions across the wider region.

\subsection{Dilution of $^{13}$C species in Starless Cores}\label{sec:3_13}

It has been found that the $^{12}$C/$^{13}$C ratios of carbon-chain species show higher values compared to the local isotopic abundance ratio of about 60 to 70 \citep{milam2005}.
This so called dilution of $^{13}$C species was first discovered in TMC-1 CP, but has since been found in other starless cores.

Variation of the $^{12}$C/$^{13}$C ratio of carbon-chain species give further constraints on the processes regulating carbon-chain chemistry in dark clouds.
One important cause of $^{12}$C/$^{13}$C isotopologue abundance variation is the reaction \citep[e.g.,][]{langer1984}:
\begin{equation} \label{equ:13C}
^{13}{\rm {C}}^{+} + {\rm {CO}} \rightleftharpoons {\rm {C}}^{+} + {\rm{^{13}CO}} + 35 {\rm {K}}.
\end{equation}
The backward reaction is ineffective in cold-temperature conditions ($\sim10$ K), so that the abundance of $^{13}$C$^{+}$ should decrease.
Ionic and atomic carbons (C$^{+}$ and C) are the main parent species of carbon-chain molecules and a low abundance of $^{13}$C$^{+}$ is then expected to result in a deficit of $^{13}$C isotopologues of carbon-chain molecules.  
Table \ref{tab:13C} summarizes the $^{12}$C/$^{13}$C ratios of carbon-chain molecules derived in three starless cores: TMC-1 CP; L1521B; and L134N.
From Table~\ref{tab:13C}, the following trends can be inferred:
\begin{enumerate}
    \item Cyanopolyynes (HC$_{2n+1}$N) have relatively lower $^{12}$C/$^{13}$C ratios compared to the other hydrocarbons. Especially, the $^{13}$C isotopomers of CCH have high values.
    \item The $^{12}$C/$^{13}$C ratios are different among the dark clouds. The ratios in L134N are relatively low compared to the others.   
\end{enumerate}
An explanation for the first point was proposed by \citet{tani2019}; the high $^{12}$C/$^{13}$C ratios of CCH seem to be caused by reactions between hydrocarbons (CCH, C$_2$H$_2$, $l,c$-C$_3$H) and C$^+$.
The formed ions will go back to CCH through reactions including electrons and H$_2$.
If $^{13}$C$^{+}$ is diluted by reaction (\ref{equ:13C}), these reactions will produce hydrocarbons with high $^{12}$C/$^{13}$C ratios.
On the other hand, the $^{12}$C/$^{13}$C ratios of cyanopolyynes do not likely change after their formation.
The second point may be related to the evolution of the starless cores; TMC-1 CP and L1521B are considered to be chemically younger than L134N \citep[e.g.,][]{dickens2000}.

Currently, the available data are limited, and such studies have been conducted mainly at TMC-1 CP.
Thus, it is difficult to reach firm conclusions.
Future high-sensitivity survey observations are needed to reveal the detailed mechanisms causing the dilution of $^{13}$C species, which would give information about the chemical relationships among carbon-chain molecules in cold dark clouds.
In addition, chemical simulations including the $^{13}$C isotopomers are necessary for an improved quantitative understanding of $^{13}$C species dilution, including its evolution.

\begin{center}
\begin{table}[t]
\caption{The $^{12}$C/$^{13}$C ratios of carbon-chain molecules in dark clouds} \label{tab:13C}
\begin{tabular}{lccc}
\hline
Species & TMC-1 CP & L1521B & L134N \\
\hline
H$^{13}$CCCN & $79\pm11$$^{(a)}$ & $117\pm16$$^{(b)}$ & $61\pm9$$^{(b)}$ \\
HC$^{13}$CCN & $75\pm10$$^{(a)}$ & $115\pm16$$^{(b)}$ & $94\pm26$$^{(b)}$ \\
HCC$^{13}$CN & $55\pm7$$^{(a)}$ & $76\pm6$$^{(b)}$ & $46\pm9$$^{(b)}$ \\
H$^{13}$CCCCCN & $98\pm14$$^{(c)}$ & & \\
HC$^{13}$CCCCN & $101\pm14$$^{(c)}$ & \\
HCC$^{13}$CCCN & $95\pm12$$^{(c)}$ & \\
HCCC$^{13}$CCN & $93\pm13$$^{(c)}$ & \\
HCCCC$^{13}$CN & $85\pm11$$^{(c)}$ & \\
HC$_{7}$N & $73\pm21$$^{(d)}$ &\\
$^{13}$CCH & $>250$$^{(e)}$ & $>271$$^{(f)}$ & $>142$$^{(f)}$ \\
C$^{13}$CH & $>170$$^{(e)}$ & $252^{+77}_{-48}$$^{(f)}$ & $101^{+24}_{-16}$$^{(f)}$ \\
$^{13}$CCCCH & $141\pm15$$^{(g)}$ &  \\
C$^{13}$CCCH & $97\pm9$$^{(g)}$ & \\
CC$^{13}$CCH & $82\pm5$$^{(g)}$ & \\
CCC$^{13}$CH & $118\pm8$$^{(g)}$ & \\
$^{13}$CCS & $230\pm43$$^{(h)}$ & \\
C$^{13}$CS & $54\pm2$$^{(h)}$ & \\
$^{13}$CCCS & $>206$$^{(g)}$ & \\
C$^{13}$CCS & $48\pm5$$^{(g)}$ &  \\
CC$^{13}$CS & $30-206$$^{(g)}$ & \\
\hline
\end{tabular}
Errors indicate the standard deviation.
References: (a) \citet{takano1998}, (b) \citet{tani2017}, (c) \citet{tani2016}, (d) \citet{burkhardt2018} (average value), (e) \citet{sakai2010}, (f) \citet{tani2019}, (g) \citet{sakai2013JPCA}, (h) \citet{sakai2007}.
\end{table}
\end{center}

\subsection{Carbon-Chain Species around Low-Mass YSOs} \label{sec:3_2}

Carbon-chain chemistry around low-mass young stellar objects (YSOs), namely WCCC, has been reviewed in \citet{sakai2013}, and we do not discuss WCCC in detail. 
Instead, we summarize observational results published after the review article.

Several survey observations with single-dish telescopes targeting carbon-chain molecules and COMs have been conducted. 
\citet{grani2016} carried out survey observations of CH$_3$OH and C$_4$H toward 16 embedded low-mass protostars using the IRAM 30m telescope.
A tentative correlation between the gas-phase C$_4$H/CH$_3$OH abundance ratio and the CH$_4$/CH$_3$OH abundance ratio in ice was found.
At the protostellar stage, the gas-phase C$_4$H is considered to form from CH$_4$ sublimated from dust grains (WCCC), whereas the gas-phase CH$_3$OH is mainly originated from ice mantle (hot corino chemistry).
Thus, the suggested tentative correlation between the gas phase and ice mantle supports the scenario of WCCC; sublimation of CH$_4$ is a trigger of carbon-chain formation in lukewarm gas \citep[e.g.,][]{hassel2008}.

\citet{higuchi2018} conducted survey observations of two carbon-chain species (CCH and $c$-C$_3$H$_2$) and CH$_3$OH toward 36 Class 0/I protostars in the Perseus molecular cloud using the IRAM 30m and Nobeyama 45m radio telescopes. 
They found that the column density ratio of CCH/CH$_3$OH varies by two orders of magnitudes among the target sources, and the majority of the sources show intermediate characters between hot corino and WCCC.
In other words, hot corino and WCCC are at opposite ends of a spectrum and both carbon-chain species and COMs are present around most low-mass protostars.
In addition, they found a possible trend that sources with higher CCH/CH$_3$OH ratios are located near cloud edges or in isolated clouds.

However, we need to treat their results carefully because CCH and $c$-C$_3$H$_2$ can be enhanced in photodissociation regions (PDRs) as described in section \ref{sec:3_6}. 
Hence, the observed trend in the CCH/CH$_3$OH ratio could be a result of PDR chemistry; the ISRF promotes the PDR chemistry at the edge of the molecular cloud rather than WCCC.
In addition, CCH is enhanced in outflow cavity walls, as discussed below.
Single-dish observations cannot spatially resolve warm envelopes and outflow cavity walls, and thus we cannot distinguish between these components.
To investigate the characteristics of WCCC, we should observe large carbon-chain species (e.g., HC$_5$N) that are not predicted to be enhanced in PDRs. 

The IRAM Large Program ``Astrochemical Surveys At IRAM (ASAI)'' is an unbiased line survey from 80 to 272 GHz toward 10 sources with various evolutionary stages.
Using the ASAI data, \citet{lefloch2018} found a difference in environmental conditions between hot corino and WCCC sources: i.e., inside and outside dense filamentary cloud regions, respectively.
Their results are likely to be more secure than those of \citet{higuchi2018} because \citet{lefloch2018} included long carbon-chain species that cannot be abundant in PDRs due to destruction by the UV radiation.

High-angular-resolution observations with interferometers, such as ALMA and NOEMA, have revealed spatial distributions of carbon-chain molecules around low-mass YSOs. 
Such observations are essential to distinguish the emission regions of each species (e.g., warm envelopes, outer cold envelopes, cavity walls).
\citet{oya2017} detected both a carbon-chain molecule (CCH) and several COMs toward the low-mass Class 0 protostar L483.
They found that the spatial distribution of CCH is different from those of COMs; the CCH emission shows a hole at the protostar position, whereas emission from COMs is concentrated near the protostar.
Such emission features are attributed to WCCC and hot corino chemistry, respectively.
Their results present an example of the hybrid-type source (c.f., Fig.\,\ref{fig:chem_div}).

\citet{zhang2018} found that CCH emission traces the outflow cavity with signatures of rotation with respect to the outflow axis toward the NGC\,1333 IRAS 4C outflow in the Perseus molecular cloud using ALMA.
\citet{tychoniec2021} analyzed ALMA data set toward 16 protostars and investigated spatial distributions of each molecule.
They also found that CCH and $c$-C$_3$H$_2$ trace the cavity wall.
This could be explained by the fact that the chemistry of the cavity wall is similar to PDR chemistry. 
The cavity walls are created by the illuminated by the stellar FUV and X-ray radiation field, and the PDR-like chemistry dominates cavity walls.
The photodissociation of molecules by UV radiation keeps high gas-phase abundances of atomic carbon (C), which is a preferable condition for the formation of hydrocarbons.

\citet{pineda2020} found a streamer toward the Class 0 YSO IRAS\,03292+3039 (or Per-emb-2) in the Perseus star-forming region with NOEMA. 
Such a streamer may be well traced by carbon-chain molecules such as CCS and HC$_3$N, if it is considered to be chemically young.
The streamer in this source seems to bring fresh gas from outside of the dense core ($>10,500$ au) down to the central protostar where the disk forms.
Thus the properties of such streamers are potentially important for the formation and evolution of protoplanetary disks.
However, these NOEMA observations did not cover the origin of the streamer.
Follow-up single-dish mapping observations of carbon-chain species (HC$_3$N, HC$_5$N, CCS, and CCH) have revealed the reservoir of the streamer \citep{Taniguchi2024}.
The reservoir and streamer are found to be chemically young, and their chemical ages are similar to those of early starless cores. 
Overall, it can be concluded that a chemically young streamer has the potential to change the chemical composition close to the YSO.  

\begin{figure}[th]
\begin{center}
\includegraphics[bb = 0 20 600 420, width=0.8\textwidth]{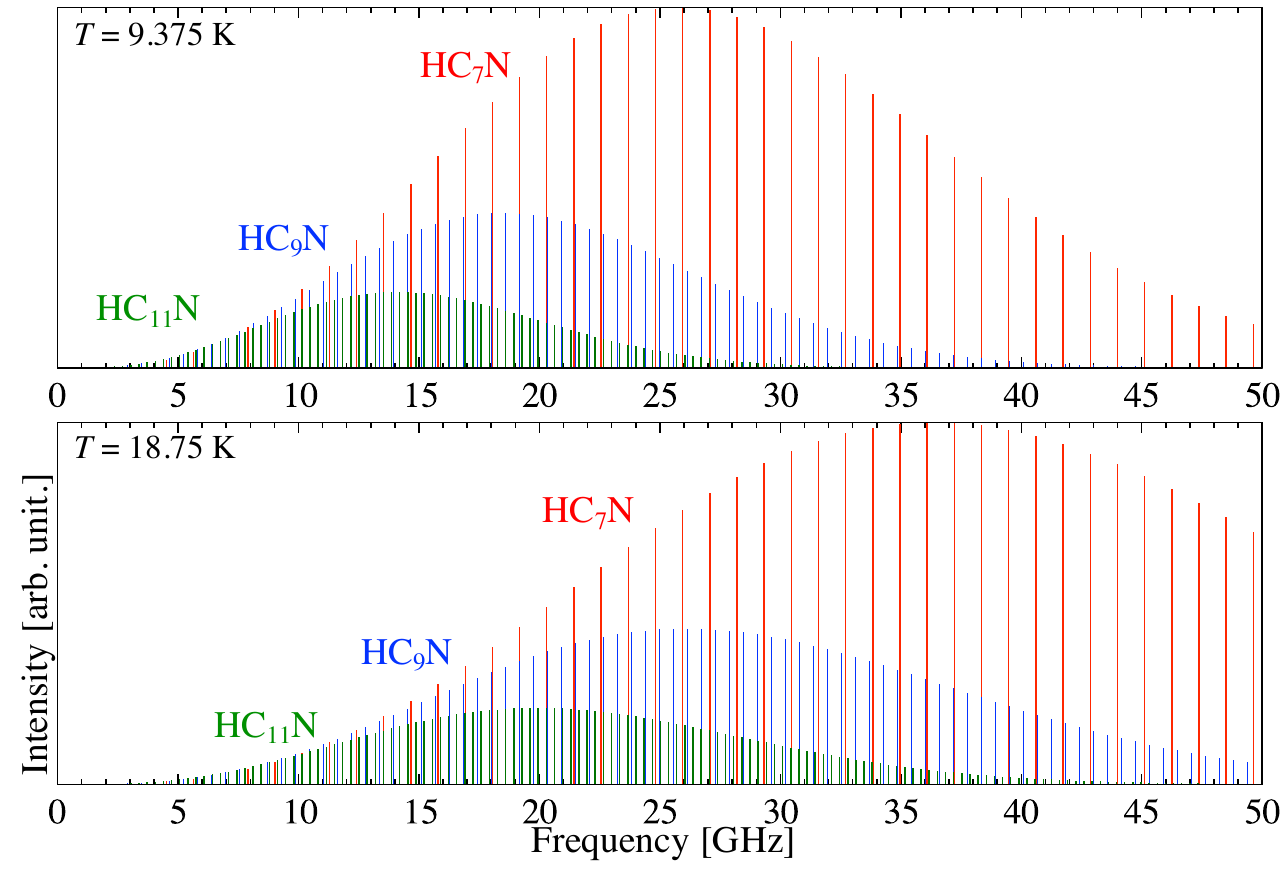}
\end{center}
\caption{Predicted line intensities of long cyanopolyynes (red; HC$_7$N, blue; HC$_9$N, and green; HC$_{11}$N) in the frequency range of $0-50$ GHz. The upper and lower panels show predictions at temperatures of 9.375 K and 18.75 K, respectively. The data on rest frequency and line intensity are taken from the CDMS database. Longer carbon-chain species show intensity peaks at lower frequencies.}
\label{fig:cyanopred}
\end{figure}

Taking advantage of the characteristics of carbon-chain molecules, we can trace unique features around low-mass YSOs.
Rotational-transition lines of carbon-chain molecules are now found to be useful tracers not only in early starless clouds but also around star-forming cores.
ALMA Band 1 and the next generation Very Large Array (ngVLA) will cover the 7 mm band or lower frequency bands, which are suitable for observations of carbon-chain molecules, especially longer ones (see Fig.\,\ref{fig:cyanopred}).
Future observations using such facilities will offer new insights into the carbon-chain chemistry around protostars, including low-, intermediate-, and high-mass systems.

\subsection{Carbon-Chain Species in High-Mass Star-Forming Regions} \label{sec:3_3}

\subsubsection{Chemical Evolutionary Indicators}\label{sec:3_31}

Carbon-chain molecules classically have been known to be abundant in young starless cores and good chemical evolutionary indicators in low-mass star-forming regions (section \ref{sec:1_2}).
However, it was unclear whether carbon-chain species can be used as chemical evolutionary indicators in high-mass star-forming regions and behave similarly as in the case of low-mass regions.

\begin{figure}[th]
\begin{center}
\includegraphics[bb = 0 10 480 330, width=0.8\textwidth]{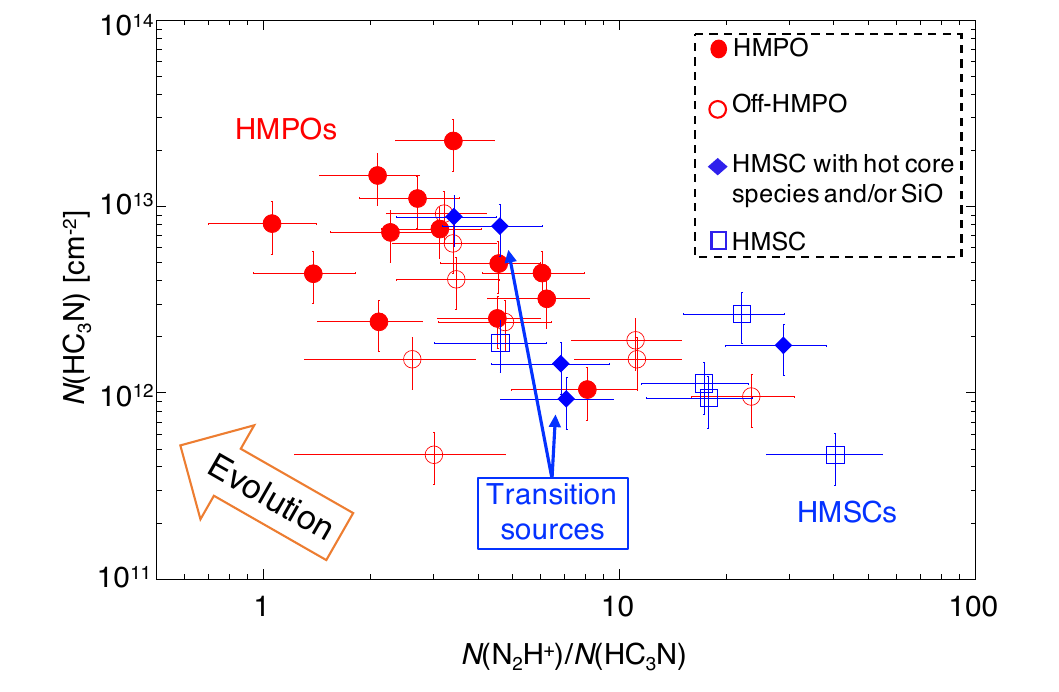}
\end{center}
\caption{A chemical evolutionary indicator in high-mass star-forming regions \citep{tani2019survey}. Off-HMPO means that IRAS-observed positions were not at exact continuum peak positions, but the beam covered the continuum core in the beam edge. Blue diamond plots are sources that were identified as HMSCs based on the IR observations, but which are associated with molecular emission lines of COMs (CH$_3$OH and/or CH$_3$CN) or SiO.}
\label{fig:evolution}
\end{figure}

Survey observations of several carbon-chain species (HC$_3$N, HC$_5$N, CCS, and $c$-C$_3$H$_2$) and N$_2$H$^+$ were carried out using the Nobeyama 45m radio telescope \citep{tani2018survey,tani2019survey}.
\citet{tani2018survey} observed the HC$_3$N and HC$_5$N lines in the 42--45 GHz band toward 17 high-mass starless cores (HMSCs) and 35 high-mass protostellar objects (HMPOs), and \citet{tani2019survey} observed HC$_3$N, N$_2$H$^+$, CCS, and $c$-C$_3$H$_2$ in the 81--94 GHz band toward 17 HMSCs and 28 HMPOs.
They proposed the column density ratio of $N$(N$_2$H$^+$)/$N$(HC$_3$N) as a chemical evolutionary indicator in high-mass star-forming regions (Fig.\,\ref{fig:evolution}). 
This column density ratio decreases as cores evolve from the starless (HMSC) to the protostellar (HMPOs) stage. 
Sources that were categorized as HMSCs based on the infrared (IR) observations but that are associated with molecular lines of COMs (CH$_3$OH or CH$_3$CN) and/or SiO (plotted as the blue diamond in Fig.\,\ref{fig:evolution}) tend to fall between HMSCs and HMPOs.
These sources are considered to contain early-stage protostars in the dense, dusty cores, which are not easily detected with IR observations.
Thus, these sources appear to be at an intermediate evolutionary stage between HMSCs and HMPOs.
It is essential to study the physical and chemical features of these sources in detail because they possess information on the initial conditions of high-mass protostars.

The decrease of the $N$(N$_2$H$^+$)/$N$(HC$_3$N) ratio means that HC$_3$N is efficiently formed and N$_2$H$^+$ is destroyed, as cores evolve.
It is a notable point that the tendency of this column density ratio is opposite to that in low-mass star-forming regions \citep{suzuki92,benson98}.
This tendency could be explained by higher temperatures and extended warm regions around HMPOs compared to low-mass protostars.
N$_2$H$^+$ is destroyed by a reaction with CO, i.e., abundant in the gas phase after being desorbed from dust grains, and HC$_3$N can be formed by CH$_4$ via the WCCC mechanism or via C$_2$H$_2$ desorbed from dust grains.
The desorption of CO, CH$_4$, and C$_2$H$_2$ from dust grains occurs when temperatures reach around 20 K, 25 K, and 50 K, respectively.
In summary, the gas-phase chemical composition is affected by the local heating from young massive protostars, and a chemical evolutionary indicator apparently shows an opposite trend compared to that of the low-mass case. Thus, carbon-chain species likely have the potential to be utilized as chemical evolutionary indicators even for high-mass protostars.

Note that the above observations were conducted using a single-dish telescope and the beam sizes are large ($\sim0.5-0.9$ pc at 5 kpc, which is a typical distance of high-mass star-forming regions).
Hence, higher angular-resolution observations with interferometers will be important to confirm such a chemical evolutionary trend.

\subsubsection{Cyanopolyynes around High-Mass Protostars}\label{sec:3_32}

From the survey observations mentioned in section \ref{sec:3_31}, the detection rates of HC$_3$N, HC$_5$N, $c$-C$_3$H$_2$, CCS are derived to be 93\%, 50\%, 68\%, and 46\%, respectively, in high-mass star-forming regions \citep{tani2018survey,tani2019survey}.
\citet{law2018} conducted survey observations toward 16 Class 0/I low-mass protostars with the IRAM 30m telescope and reported that the detection rates of HC$_3$N, HC$_5$N, $l$-C$_3$H, C$_4$H, CCS, and C$_3$S are 75\%, 31\%, 81\%, 88\%, 88\%, and 38\%, respectively.
Thus, cyanopolyynes (HC$_3$N and HC$_5$N) show higher detection rates, while CCS is relatively deficient in high-mass star-forming regions compared to low-mass regions.
These results imply that carbon-chain chemistry around MYSOs is different from WCCC found around low-mass YSOs.

\citet{tani2017MYSO} conducted observations of the multi-transition lines of HC$_5$N using the Green Bank 100m and Nobeyama 45m radio telescopes toward four MYSOs.
The derived rotational temperatures are $\sim20-25$ K, which are similar to the temperature regimes of the WCCC mechanism.
However, the derived rotational temperatures are lower limits due to contamination from extended cold components covered by the single-dish telescopes.
The MYSO G28.28-0.36 shows a particularly unique chemical character: carbon-chain species are rich, but COMs are deficient \citep{tani2018MYSO}.
This source may be analogous to the WCCC source L1527.
These results are suggestive of the chemical diversity around MYSOs, as similar to low-mass cases (hot corino and WCCC).

Since the above HC$_5$N excitation temperatures derived with single-dish data are lower limits due to contamination of cold outer envelopes, it could not be concluded that carbon-chain molecules exist in higher temperature regions around MYSOs compared to the WCCC sources.
\citet{tani2021carbon} derived the CCH/HC$_5$N abundance ratios toward three MYSOs and compared the observed ratio with the results of their chemical simulations to constrain temperature regimes where carbon-chain species exist.
The CCH/HC$_5$N abundance ratio is predicted to decrease as the temperature increases, because CCH shows a peak abundance in the gas phase around 30 K, while the gas-phase HC$_5$N abundance increases as the temperature rises up to $\sim100$ K \citep{tani2019model}.
Details about the chemical simulations are presented in section \ref{sec:4}.
The observed CCH/HC$_5$N abundance ratios toward all of the three MYSOs are $\sim15$, which is much lower than that toward the low-mass WCCC source L1527 ($625^{+3041}_{-339}$).
The observed abundance ratios around MYSOs agree with the simulations around 85 K, while the ratio in L1527 matches with the simulations around 35 K.
Therefore, carbon-chain species, at least HC$_5$N, around MYSOs appear to exist in higher temperature regions than the locations where the WCCC mechanism occurs.
Such results indicate that carbon-chain chemistry around MYSOs is different from the WCCC mechanism.
It is necessary to reveal spatial distributions of carbon-chain molecules around MYSOs to confirm that the carbon-chain chemistry around MYSOs is different from WCCC.

\begin{figure*}
\includegraphics[bb = 0 0 780 270, width=0.95\textwidth]{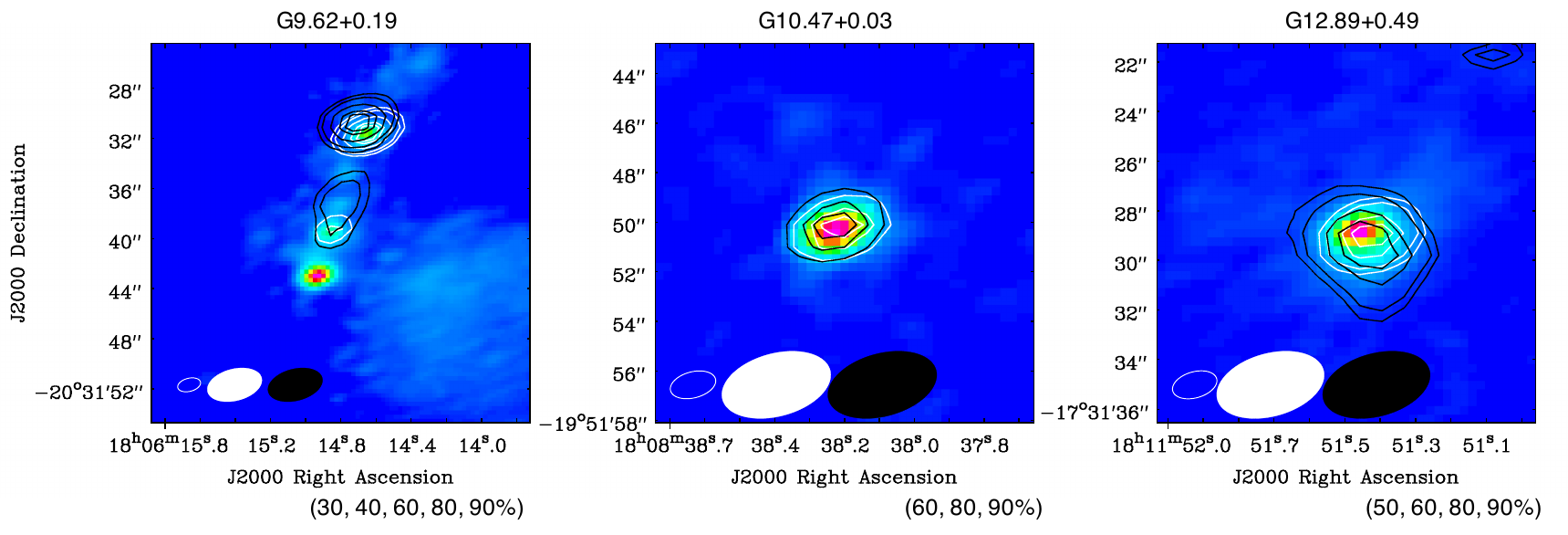}
\caption{Comparison of spatial distributions around MYSOs obtained by ALMA (color scale; continuum image, white contours; the CH$_3$OH line ($1_{0,1}-2_{1,2}$, $v_{t}=1$; $E_{\rm {up}}=302.9$ K), black lines; the HC$_5$N line ($J=35-34$; $E_{\rm {up}}=80.5$ K). 
This figure is a modified version of \citet{tani2023}. 
The contour levels are relative values of the peak intensities, and the contour levels are indicated below each panel. The ellipses at the bottom of each panel indicate the angular resolutions; open one corresponds to the continuum images, and white and black ones correspond to the moment 0 maps of CH$_3$OH and HC$_5$N, respectively.}
\label{fig:HC5Nmap}
\end{figure*}

More recently, spatial distributions of the HC$_5$N line ($J=35-34$; $E_{\rm {up}}=80.5$ K) around MYSOs have been revealed by ALMA Band 3 data \citep{tani2023}. 
This line has been detected from three sources among five target sources. Fig.\,\ref{fig:HC5Nmap} shows the comparison of spatial distributions among HC$_5$N, CH$_3$OH, and continuum emission in Band 3 toward the three MYSOs.
The HC$_5$N emission shows single peaks associated with the continuum peaks and is consistent with the emission of the CH$_3$OH line ($1_{0,1}-2_{1,2}$, $v_{t}=1$; $E_{\rm {up}}=302.9$ K) which should trace hot core regions with temperatures above 100 K. 
These results also support the ``Hot Carbon-Chain Chemistry (HCCC)'' scenario; HC$_5$N desorbs from dust grains with temperatures above 100 K and shows gas-phase peak abundances similar to those of COMs.

\begin{figure*}[hbt]
\begin{center}
\includegraphics[bb = 0 10 700 280, scale=0.5]{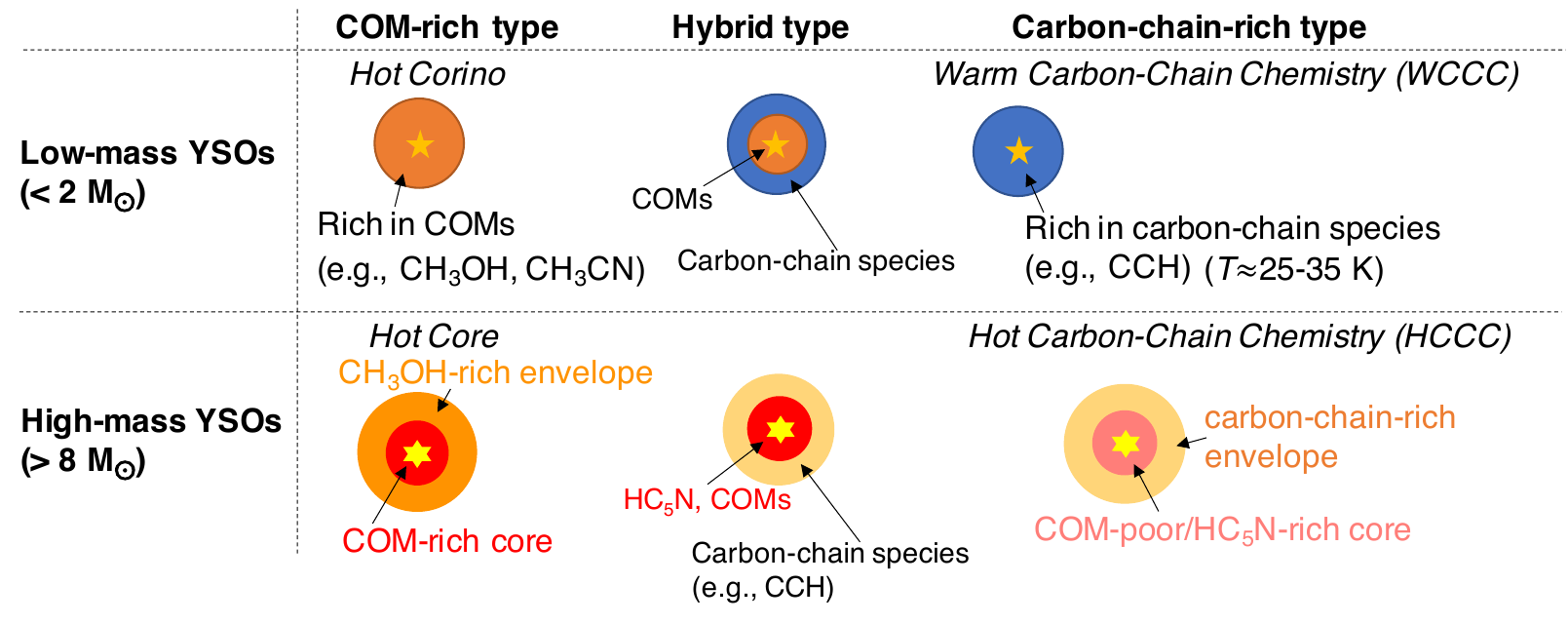}
\end{center}
\caption{Chemical properties found around low-mass and high-mass YSO.}
\label{fig:chem_div}
\end{figure*}

In summary, carbon-chain molecules are formed even around MYSOs by the HCCC mechanism.
Fig.\,\ref{fig:chem_div} shows a summary of chemical properties found around low-mass and high-mass YSOs, respectively.
Currently, a candidate of pure HCCC sources is the MYSO G28.28-0.36, in which COMs are deficient but HC$_5$N is abundant \citep{tani2018MYSO}. 
We note that high angular-resolution observations of COMs toward this MYSO, e.g., with ALMA, are still needed because COMs may be diluted in the large single-dish beams.
Only if COMs were found to be deficient in such high resolution data, it could be concluded that MYSO G28.28-0.36 is a pure HCCC source.
In addition, to statistically study the chemical diversity around MYSOs, larger source samples observed at high-angular-resolution to trace carbon-chain and COM species are needed.

As seen in Fig.\,\ref{fig:chem_div}, chemical differentiation emerges around both low-mass YSOs and high-mass YSOs.
However, at present it is still uncertain (1) what physical factor(s) is important for the chemical differentiation, and (2) whether a common factor(s) dominates the chemical differentiation around low-mass and high-mass YSOs.
For instance, temperature, radiation field of energetic photons (e.g., UV photons), and cosmic rays, are likely to be important for causing chemical differentiation \citep[e.g.,][]{fontani2017,lefloch2018,tani2019model}.
It is currently difficult to evaluate quantitatively the relative importance of such physical parameters between low-mass and high-mass YSOs. A study of the carbon-chain chemistry around intermediate-mass YSOs (i.e., $2 M_{\odot} < m_* < 8 M_{\odot}$), which bridge the gap between the low- and high-mass regimes, may help to solve these open questions.

\subsection{Revealing Main Formation Pathways of Carbon-Chain Molecules by $^{13}$C Isotopic Fractionation} \label{sec:3_12}

Since carbon-chain molecules are unstable even in vacuum chambers on the Earth, it is difficult to conduct laboratory experiments about their reactivity.
Consequently, their formation and destruction reactions remain unclear.
Instead, the observed $^{13}$C isotopic fractionation of carbon-chain molecules, which refers to the relative differences in abundance among the $^{13}$C isotopologues, has been proposed to be a key method for revealing their main formation mechanisms \citep{takano1998}.
In this subsection, we summarize observational results of the $^{13}$C isotopic fractionation of several carbon-chain species toward TMC-1 CP and studies on HC$_3$N toward various types of sources, from low-mass starless cores to MYSOs.

The local elemental $^{12}$C/$^{13}$C abundance ratio in the local ISM is around 60--70 \citep[e.g.,][]{milam2005}.
Hence, high-sensitivity observations are necessary to detect the $^{13}$C isotopologues of carbon-chain molecules with high enough signal-to-noise (S/N) ratios to compare their abundances.
With this constraint, TMC-1 CP is the most promising target source because of the high abundances of carbon-chain molecules (section \ref{sec:3_11}).
In fact, TMC-1 CP has the largest number of carbon-chain species investigated for $^{13}$C isotopic fractionation: seven species have been studied, HC$_3$N \citep{takano1998}, CCS \citep{sakai2007}, CCH \citep{sakai2010}, C$_3$S and C$_4$H \citep{sakai2013JPCA}, HC$_5$N \citep{tani2016}, HC$_7$N \citep{burkhardt2018}.
Thanks to the advent of new observational facilities, larger molecules can be investigated within reasonable observing times.

\citet{takano1998} observed the three $^{13}$C isotopologues of HC$_3$N (H$^{13}$CCCN, HC$^{13}$CCN, and HCC$^{13}$CN) at TMC-1 CP using the Nobeyama 45m radio telescope.
The relative abundance ratios of the three $^{13}$C isotopologues were derived to be $1.0:1.0:1.4$ ($\pm0.2$) ($1\sigma$) for [H$^{13}$CCCN]:[HC$^{13}$CCN]:[HCC$^{13}$CN].
Table \ref{tab:HC3N} summarizes correspondences of the possible main formation pathway of HC$_3$N and its expected $^{13}$C isotopic fractionation \citep{tani2017}.
Regarding the last one (the electron recombination reaction of HC$_3$NH$^{+}$), various formation pathways of the HC$_3$NH$^{+}$ ion should compete, and then clear $^{13}$C isotopic fractionation would not be seen in HC$_3$NH$^{+}$, as well as HC$_3$N.
The reaction between of ``C$_{2}$H$_{2}$ + CN $\rightarrow$ HC$_3$N +H'' can explain the observed $^{13}$C isotopic fractionation in TMC-1 CP.

\begin{center}
\begin{table*}[t]
\caption{Main formation mechanisms of HC$_3$N and expected $^{13}$C isotopic fractionation} \label{tab:HC3N}
\begin{tabular}{lll}
\hline
 & Expected  &  \\
Formation Route & fractionation & Sources \\
\hline
C$_{2}$H$_{2}$ + CN $\rightarrow$ HC$_3$N + H & $1:1:x$ & TMC-1, L1521B, L1527, G28.28-0.36 \\
CCH + HNC $\rightarrow$ HC$_3$N + H & $y:1:z$  & L134N \\
HC$_3$NH$^{+}$ + e$^{-}$ $\rightarrow$ HC$_3$N + H & $\approx 1:1:1$ & G12.89+0.49, G16.86-2.16 \\
\hline
\end{tabular}
Note ``Expected fractionation pattern'' means the [H$^{13}$CCCN] : [HC$^{13}$CCN] : [HCC$^{13}$CN] ratio. $x$, $y$, and $z$ are arbitrary values.
\end{table*}
\end{center}

At the time five $^{13}$C isotopologues of HC$_5$N were detected \citep{takano1990}, there was no evidence for the $^{13}$C isotopic fractionation due to low S/N ratios.
More than 25 years later, \citet{tani2016} successfully detected the lines of the five $^{13}$C isotopologues of HC$_5$N ($J=9-8$ and $16-15$ at 23 GHz and 42 GHz bands, respectively) with sufficient S/N ratios of 12--20.
The derived abundance ratios among the five $^{13}$C isotopologues of HC$_5$N are $1.00:0.97:1.03:1.05:1.16$ ($\pm0.19$) ($1\sigma$) for [H$^{13}$CCCCCN] : [HC$^{13}$CCCCN] : [HCC$^{13}$CCCN] : [HCCC$^{13}$CCN] : [HCCCC$^{13}$CN].
Hence, even if the S/N ratios increase, there is no clear difference in abundance among the five $^{13}$C isotopologues of HC$_5$N, unlike HC$_3$N.
\citet{tani2016} proposed that the reactions between hydrocarbon ions (C$_5$H$_m$$^{+}$; $m=3-5$) and nitrogen atoms, followed by electron recombination reactions are the most plausible main formation mechanism of HC$_5$N at TMC-1 CP.
This partly agrees with its general formation route in early starless cores indicated in Fig.\,\ref{fig:route}.

\citet{burkhardt2018} detected six $^{13}$C isotopologues of HC$_7$N and five $^{13}$C isotopomers of HC$_5$N using the Green Bank 100m telescope.
H$^{13}$CC$_{6}$N could not be detected in their observations.
They found no significant difference among the $^{13}$C isotopomers of HC$_7$N, as similar to the case of HC$_5$N.
They concluded that the significant formation route for HC$_7$N is the reaction of hydrocarbon ions and nitrogen atoms, which is the same conclusion for HC$_5$N by \citet{tani2016}.

In addition to cyanopolyynes, C$_{n}$S ($n=2,3$) and C$_{2n}$H ($n=1,2$) have been studied for $^{13}$C fractionation in TMC-1 CP \citep{sakai2007,sakai2010,sakai2013JPCA}.
\citet{sakai2007} detected the lines of $^{13}$CCS and C$^{13}$CS ($J_N=2_1-1_0$, $F=5/2-3/2$).
The abundance ratio of C$^{13}$CS/$^{13}$CCS was derived to be $4.2\pm2.3$ ($3\sigma$).
They proposed that the reaction between CH and CS is the main formation route of CCS in TMC-1 CP.
The abundance ratio of C$^{13}$CH/$^{13}$CCH was derived to be $1.6\pm0.4$ ($3\sigma$) \citep{sakai2010}. 
To explain the abundance difference between the two $^{13}$C isotopomers of CCH, \citet{sakai2010} proposed that the reaction of CH + C is the main formation mechanism of CCH at TMC-1 CP.

Unlike cyanopolyynes, the $^{13}$C isotopic fractionation of CCS and CCH does not necessarily provide information on their main formation mechanisms.
\citet{furuya2011} ran chemical simulations including $^{13}$C and investigated effects of the isotopomer-exchange reactions.
They considered the following two isotopomer-exchange reactions for CCH and CCS, respectively:
\begin{equation} \label{equ:CCH}
^{13}{\rm {CCH}} + {\rm {H}} \rightleftharpoons {\rm {C}}^{13}{\rm {CH}} + {\rm {H}} + 8.1 {\rm {K}};
\end{equation}
and 
\begin{equation} \label{equ:CCS}
^{13}{\rm {CCS}} + {\rm {S}} \rightleftharpoons {\rm {C}}^{13}{\rm {CS}} + {\rm {S}} + 17.4 {\rm {K}}.
\end{equation}
They also included the following neutral-neutral exchange reaction of CCS to reproduce the observed isotopomer ratio of CCS:
\begin{equation} \label{equ:CCS2}
^{13}{\rm {CCS}} + {\rm {H}} \rightleftharpoons {\rm {C}}^{13}{\rm {CS}} + {\rm {H}} + 17.4 {\rm {K}}.
\end{equation}
This reaction is regarded as a catalytic reaction by the hydrogen atom.
At low temperature conditions ($T\approx10$ K), C$^{13}$CH and C$^{13}$CS should be more abundant than the other $^{13}$C isotopomers by reactions (\ref{equ:CCH}) -- (\ref{equ:CCS2}).
Their model results can explain the observed abundance differences between the two $^{13}$C isotopomers of CCH and CCS in TMC-1 CP \citep{sakai2007,sakai2010}.
It was found that C$^{13}$CH is more abundant than $^{13}$CCH in the other starless cores \citep[L1521B and L134N;][]{tani2019}, and such a character may be common for cold dark clouds.
Such exchange reactions may contribute to larger species such as C$_3$S and C$_4$H \citep{sakai2013JPCA}.
These studies suggest that it is difficult to establish the main formation routes directly from $^{13}$C isotopic fractionation, especially radical-type species.
Hence, the main formation routes of CCH and CCS remain uncertain.

The $^{13}$C isotopic fractionation of HC$_3$N has been investigated toward various types of sources; low-mass starless cores L1521B and L134N \citep{tani2017}, the low-mass YSO L1527 \citep{tani2016HC3N,araki2016}, and three MYSOs \citep[G12.89+0.49 G16.86-2.16, and G28.28-0.36;][]{tani2016HC3N,tani2021carbon}.
Table \ref{tab:13C} summarizes these results categorized into three types of fractionation patterns.

By comparing the results among the three starless cores (TMC-1 CP, L1521B, L134N), it was found that the $^{13}$C isotopic fractionation in L1521B and TMC-1 CP is similar to each other; HCC$^{13}$CN is more abundant than the others, and the other two $^{13}$C isotopologues have similar abundances.
On the other hand, the results in L134N are different from the other two starless cores; the abundance ratios in L134N are $1.5$ ($\pm0.2$)$:1.0:2.1$ ($\pm0.4$) ($1\sigma$) for [H$^{13}$CCCN]:[HC$^{13}$CCN]:[HCC$^{13}$CN].
Based on the classifications (Table \ref{tab:HC3N}), the main formation mechanisms of HC$_3$N are determined as the reactions of C$_{2}$H$_{2}$ + CN in L1521B and CCH + HNC in L134N.
The C$^{13}$CH/$^{13}$CCH abundance ratio was found to be $>1.4$ in L134N \citep{tani2019}, which agrees with the abundance ratio of $1.5$ ($\pm0.2$)$:1.0$ for [H$^{13}$CCCN]:[HC$^{13}$CCN].
This is further supporting evidence for the conclusion that the main formation pathway of HC$_3$N includes CCH in L134N.
The difference between L134N and TMC-1CP/L1521B is probably brought about by different HNC/CN abundance ratios (HNC/CN = 35.6 and 54.2 in TMC-1 CP and L134N, respectively).
The HNC/CN abundance ratio depends on the cloud age, and then the main formation mechanism of cyanopolyynes likely changes throughout the cloud evolution.

In the case of star-forming cores (the low-mass YSO L1527 and three MYSOs; G12.89+0.49, G16.86-2.16, G28.28-0.36), L1527 and the MYSO G28.28-0.36 show the same feature as TMC-1 CP.
This pattern agrees with the main formation route of the reaction between C$_2$H$_2$ and CN.
This is consistent with the chemical simulations confirming WCCC \citep{hassel2008} and formation reactions of cyanopolyynes during the warm-up stage \citep{tani2019model}. 
On the other hand, there were no significant differences among the three $^{13}$C isotopologues of HC$_3$N in the other two MYSOs (G12.89+0.49 and G16.86-2.16).
Such results suggest that the main formation route of HC$_3$N is the electron recombination reaction of the HC$_3$NH$^+$ ion.
The differences among the MYSOs may come from different stellar luminosities or different environments where the target MYSOs were born.
For example, strong stellar UV radiation 
may have changed the initial $^{13}$C isotopic fractionation of HC$_3$N in G12.89+0.49 and G16.86-2.16, because the following reactions destroy and reform HC$_3$N in harsh environments \citep{tani2019model}:
\begin{equation} \label{equ:HC3N1}
{\rm {HC}}_3{\rm {N}} + {\rm {H}}_3^+/{\rm {HCO}}^+ \rightarrow {\rm {HC}}_3{\rm {NH}}^+ + {\rm {H}}_2/{\rm {CO}},
\end{equation}
and 
\begin{equation} \label{equ:HC3N2}
{\rm {HC}}_3{\rm {NH}}^+ + {\rm {e}}^- \rightarrow {\rm {HC}}_3{\rm {N}} + {\rm {H}}.
\end{equation}
Even if HC$_3$N is mainly formed by the reaction between C$_2$H$_2$ and CN initially and shows clear $^{13}$C isotopic fractionation, the above reaction cycle could erase the $^{13}$C isotopic fractionation.
Another possibility is that the initial main formation route is reaction (\ref{equ:HC3N2}) in G12.89+0.49 and G16.86-2.16 due to strong stellar feedback from other stars which had been formed earlier than the MYSOs G12.89+0.49 and G16.86-2.16.
To solve these questions, we need to increase source samples in various environments and obtain maps of the fractionation.

\subsection{Carbon-Chain Species in Disks} \label{sec:3_5}

Before the ALMA era, there were only a few reported detections of carbon-chain species in the protoplanetary disks around T Tauri stars and Herbig Ae stars.
\citet{henning2010} detected CCH from two T Tauri stars, DM Tau and LkCa\,15, with the IRAM Plateau de Bure Interferometer (PdBI). 
The first detection of HC$_3$N from protoplanetary disks was achieved using the IRAM 30m telescope and PdBI \citep{chapillon2012}.
They detected the HC$_3$N lines ($J=12-11$ and $16-15$) from protoplanetary disks around two T Tauri stars, GO Tau and LkCa\,15, and the Herbig Ae star MWC\,480. 
Studies of disk chemistry have dramatically progressed, thanks to ALMA observations.
In this subsection, we summarize studies related to carbon-chain species in protoplanetary disks.

\citet{qi2013} reported the first detection of $c$-C$_{3}$H$_{2}$ in a disk around the Herbig Ae star HD\,163296 using the ALMA Science Verification data. Its emission is consistent with the Keplerian rotating disk and traces a ring structure from an inner radius of $\sim 30$ au to an outer radius of $\sim 165$ au.
The HC$_3$N line ($J=27-26$; $E_{\rm {up}}=165$ K) has been detected from the protoplanetary disk of MWC\,480, which is a Herbig Ae star in the Taurus region, using ALMA \citep{oberg2015}.
Angular resolutions are $0.4^{\prime \prime}-0.6^{\prime \prime}$, corresponding to 50--70 au.
The data can spatially resolve the molecular emission, and show a velocity gradient caused by Keplerian rotation of the protoplanetary disk.
\citet{oberg2015} also detected CH$_3$CN and H$^{13}$CN in the same observation and found that the abundance ratios among the three species in the protoplanetary disk of MWC\,480 are different from those in the solar-type protostellar binary system IRAS\,16293-2422.
Thus, they suggested that varying conditions among protoplanetary disks can lead to chemical diversity in terms of carbon-chain species.

\citet{bergner2018} conducted survey observations of CH$_3$CN and HC$_3$N toward four T Tauri stars (AS\,209, IM Lup, LkCa\,15, and V4046 Sgr) and two Herbig Ae stars (MWC\,480 and HD\,163296) with ALMA.
Typical angular resolutions are from $\sim0.5^{\prime\prime}$ to $\sim1.5^{\prime\prime}$.
They detected the HC$_3$N ($J=27-26$) line from all of their target sources.
Besides, the $J=31-30$ and $J=32-31$ lines have been detected from MWC\,480.
The spatial distributions of HC$_3$N and CH$_3$CN show similarity; compact and typically well within the bounds of the dust continuum.
HC$_3$N is considered to be formed by only the gas-phase reactions: C$_2$H$_2$ + CN and CCH + HNC (see also Table \ref{tab:HC3N}). 

The Molecules with ALMA at Planet-forming Scales (MAPS) ALMA Large Program has studied disk chemistry around five target sources (IM Lup, GM Aur, AS\,209, HD\,163296, and MWC\,480) in Bands 3 and 6 \citep{oberg2021maps}.
Typical beam sizes are around $0.3^{\prime \prime}$ and $0.1^{\prime \prime}$ in Band 3 and Band 6, respectively.
\citet{ilee2021} presented the results for HC$_3$N, CH$_3$CN, and $c$-C$_3$H$_2$.
The HC$_3$N and $c$-C$_3$H$_2$ lines have been clearly detected from four of the target sources, with the exception being IM Lup, where only one $c$-C$_3$H$_2$ line has been tentatively detected.
The $c$-C$_3$H$_2$ emission shows clear ring-like features in AS\,209, HD\,163296, and MWC\,480, suggestive of an association with the outer dust rings.
Two HC$_3$N lines  ($J=11-10$ and $29-28$) show ring-like distributions in AS\,209 and HD\,163296, whereas the $J=29-28$ line appears centrally peaked in MWC\,480.
The HC$_3$N emission of the $J=11-10$ line is similarly extended to that of $c$-C$_3$H$_2$, but the $J=29-28$ line seems to be more compact.
CH$_3$CN, on the contrary, appears to have a ring-like feature only in AS\,209, while more centrally peaked structures are seen in the other sources.
\citet{ilee2021} demonstrated that the observed HC$_3$N emission traces upper layers ($z/r=0.1-0.4$) of the protoplanetary disks compared to that of CH$_3$CN ($z/r\leq0.1-0.2$).
They also found that the HC$_3$N/HCN and CH$_3$CN/HCN abundance ratios of the outer regions (50--100 au) in the target disks are consistent with the composition of cometary materials.
The warmer disks, HD\,163296 and MWC\,480, likely have comet formation zones at correspondingly larger radii.

\citet{guzman2021} presented distributions of CCH toward the five protoplanetary disks of the MAPS program.
They proposed that the CCH emission comes from relatively warmer (20--60 K) layers.
In HD\,163296, there is a decrease in the column density of CCH and HCN inside of the dust gaps near $\sim 83$ au, at which a planet has been considered to be located.
The similar spatial distributions of CCH and HCN suggest that they are produced by the same chemical processes, and photochemistry is the most probable one.

ALMA observations have revealed the presence of disks around not only T Tauri and Herbig Ae stars, but also around more massive, O-/B-type stars.
\citet{csengeri2018} detected the vibrationally-excited HC$_3$N line ($J=38-37$, $v_{7}=1e$) around the O-type star G328.2551-0.5321 (O5--O4 type star) with ALMA.
This source is a high-mass protostar in the main accretion phase.
Their data have a spatial resolution of around 400 au.
The position-velocity (PV) diagram of this HC$_3$N vibrationally-excited line is consistent with a Keplerian disk rotation profile, and they proposed that such HC$_3$N vibrationally-excited emission could be a new tracer for compact accretion disks around high-mass protostars.

\begin{figure*}[ht]
\begin{center}
\includegraphics[bb = 0 0 700 250, scale=0.5]{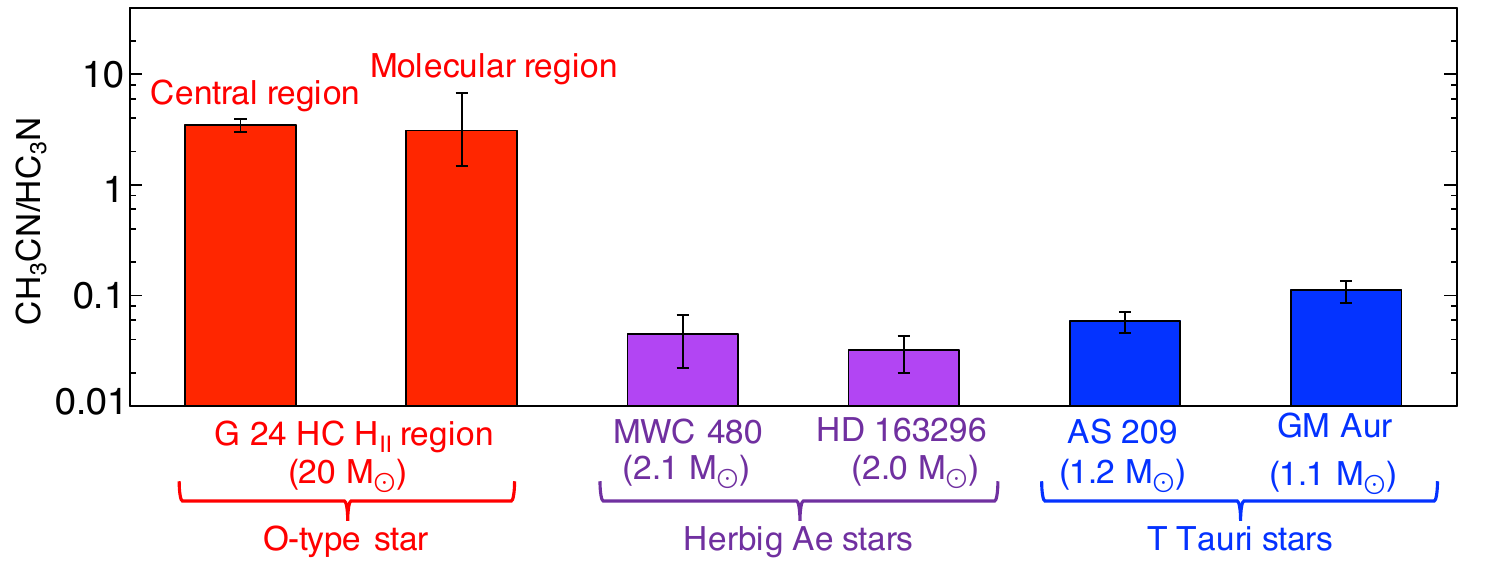}
\end{center}
\caption{Comparison of the CH$_3$CN/HC$_3$N abundance ratios in disks around various stellar masses, which is modified from \citet{tani22}. Results of Herbig Ae and T Tauri stars are from \citet{ilee2021}.}
\label{fig:diskchem}
\end{figure*}

\citet{tani22} detected the HC$_3$N vibrationally-excited lines ($J=24-23$, $v_{7}=2, l=0$ and $2e$) from the hypercompact H$_{\rm {II}}$ (HC\,H$_{\rm {II}}$) region G24.47-0.08 A1 using ALMA Band 6 data.
Their emission morphologies are largely similar to those of CH$_3$CN, which was suggested to trace Keplerian disk rotation around a central mass of $20\:M_{\odot}$ in the previous study of \citet{moscadelli2021}.
The column densities of HC$_3$N and CH$_3$CN were derived using lines of their $^{13}$C isotopologues, and the CH$_3$CN/HC$_3$N abundance ratios were compared with those in protoplanetary disks around the lower-mass stars obtained by the MAPS program \citep{ilee2021}.
Fig. \ref{fig:diskchem} shows the comparisons of the CH$_3$CN/HC$_3$N abundance ratios in disks.
It is clear that the ratio in the disk around the G24 HC\,H$_{\rm {II}}$ region is higher than those around the lower-mass stars by more than one order of magnitude.

Such a difference in the CH$_3$CN/HC$_3$N abundance ratio was explained by the HC$_3$N and CH$_3$CN chemistry in the disk: efficient thermal sublimation of CH$_3$CN from ice mantles and rapid destruction of HC$_3$N by the UV photodissociation and/or reactions with ions (H$^+$, H$_{3}^{+}$, HCO$^+$).
In the protoplanetary disks, CH$_3$CN is considered to be efficiently formed by dust-surface reactions: (1) the successive hydrogenation reactions of C$_2$N; and (2) a radical-radical reaction between CH$_3$ and CN \citep{loomis2018disk}.
The derived excitation temperature of CH$_{3}$CN in the G24 HC\,H$_{\rm {II}}$ region ($T_{\rm {ex}}\approx 335$ K) is much higher than its sublimation temperature ($\sim 95$ K), which suggests that CH$_3$CN formed on dust surfaces efficiently sublimates by the thermal desorption mechanism.
On the other hand, its excitation temperatures around the Herbig Ae and T Tauri stars were derived to be 30--60 K, which is suggestive of the non-thermal desorption mechanisms such as photodesorption \citep{loomis2018disk,ilee2021}.
This means that CH$_3$CN sublimation is not efficient in disks around the Herbig Ae and T Tauri stars, leading to low gas-phase abundances of CH$_3$CN.
Both HC$_3$N and CH$_3$CN could be destroyed by the UV radiation, and the UV photodissociation rate of HC$_3$N is higher than that of CH$_3$CN by a factor of $\sim2.4$ \citep{legal2019}.
Thus, HC$_3$N could be more rapidly destroyed by UV photodissociation.
In addition, HC$_3$N is destroyed by reactions with ions, which are expected to be abundant in the H$_{\rm {II}}$ region.
In summary, HC$_3$N is likely destroyed rapidly in the G24 HC\,H$_{\rm {II}}$ region.

Until now, there is only one O-type star disk in which the CH$_3$CN/HC$_3$N abundance ratio has been derived.
We need similar data for an increased sample of sources, including T Tauri, Herbig Ae, and O/B type disks, including probing various evolutionary stages, to further test the apparent tentative trends that have been so far revealed.
In addition, for such studies of disk structures, especially around the more distant massive protostars, it is important to conduct unbiased line survey observations with high angular resolution.

\subsection{Carbon-Chain Species in Other Environments} \label{sec:3_6}

Carbon-chain molecules have been detected not only from star-forming regions in our Galaxy, but also other environments of the ISM. We briefly summarize the carbon-chain species detected in these regions.

Small hydrocarbons have been known to be present in photodissociation regions (PDRs).
\citet{cuadrado2015} observed small hydrocarbons toward the Orion Bar PDR using the IRAM 30m telescope.
They detected various small hydrocarbons (CCH, C$_4$H, $c$-C$_3$H$_2$, $c$-C$_3$H, C$^{13}$CH, $^{13}$CCH, $l$-C$_3$H, and $l$-H$_2$C$_3$) and the $l$-C$_3$H$^+$ ion.
They found that the spatial distributions of CCH and $c$-C$_3$H$_2$ are similar but do not follow the PAH emission, and suggested that photo-destruction of PAHs is not a necessary requirement for the observed abundances of the smallest hydrocarbons. 
Instead, the gas-phase reactions between C$^+$ and H$_2$ produce the small hydrocarbons.
\citet{guzman2015} observed small hydrocarbons (CCH, $c$-C$_3$H$_2$), $l$-C$_3$H$^+$, and DCO$^+$ toward the Horsehead PDR with the PdBI.
They demonstrated that top-down chemistry, in which large polyatomic molecules or small carbonaceous grains are photo-destroyed into smaller hydrocarbon molecules or precursors, works in this PDR.
Suggestions by these two studies \citep{cuadrado2015, guzman2015} seem in contradiction but may imply that the carbon-chain chemistry in PDRs differs among regions. 
Thus, further study is needed of carbon-chain chemistry in PDRs.

The envelopes of the carbon-rich Asymptotic Giant Branched (AGB) star IRC+10216 are known as a carbon-chain-rich site.
Several carbon-chain molecules have been discovered for the first time in this source in both radio and infrared regimes (see also Table \ref{tab:quant}). 
For example, diacetylene (C$_4$H$_2$) has been detected in mid-infrared observations with the 3m Infrared Telescope Facility (IRTF) from this source \citep{fonfria2018}.
This molecule is distributed in two populations with different excitation conditions; cold and hot rotational temperatures were derived to be $47\pm7$ K and $420\pm120$K, respectively.
\citet{pardo2022} conducted deep line survey observations in the Q band with the Yebes 40m telescope and summarized the detected carbon-chain species in this source.
The rotational temperatures of the carbon-chain species are around 5--25 K, suggesting that carbon-chain species may exist in different regions.
The species detected with the radio telescope exist in colder regions compared to the species detected with the infrared telescope.
There remain a lot of unidentified lines (U lines), and future laboratory spectroscopic experiments are necessary for line identifications.

\begin{figure}
    \centering
    \includegraphics[bb = 0 50 850 500, scale=0.5]{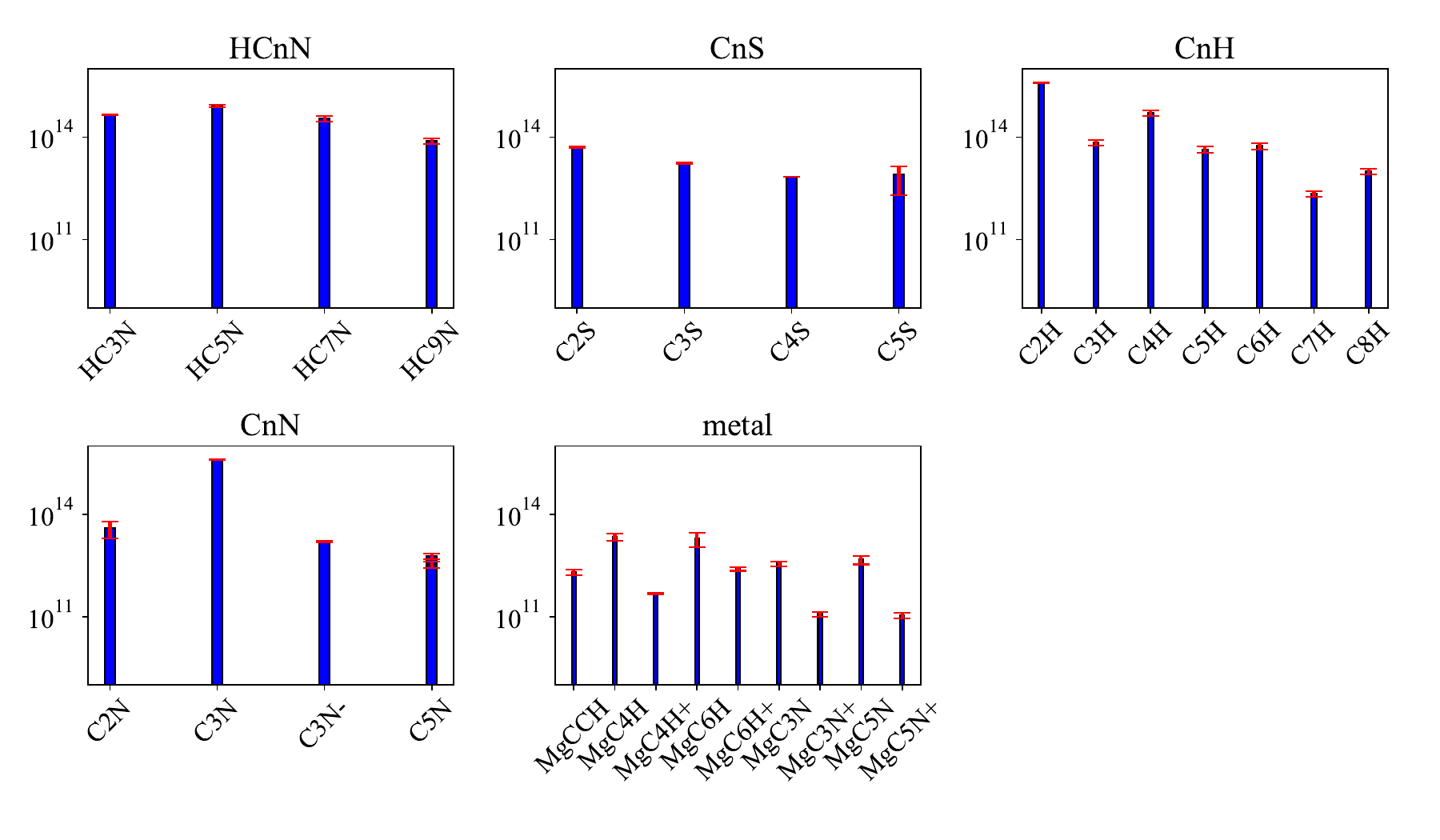}
    \caption{Column density variation of carbon chains toward IRC+10216.}
    \label{fig:irc_column}
\end{figure}

Fig.\,\ref{fig:irc_column} shows column densities of several carbon-chain series in IRC+10216.
Their column densities are provided in Table \ref{tab:colden-irc} in Appendix \ref{sec:appendix2}.
As discussed in section \ref{sec:3_11}, we can compare column densities straightforwardly if they were derived by the same observing conditions and analytical methods.
Otherwise, one needs to carefully assess the original studies and the differences in their methods.
In the case of IRC+10216, the spatial distributions of carbon-chain species are not uniform \citep[e.g.,][]{Agundez2017}, and so different beam sizes and beam positions can cause significant differences in the derived column density.

Although column densities of C$_n$H were derived by different authors (see Table \ref{tab:colden-irc} in Appendix \ref{sec:appendix2}), a similar zigzag feature as TMC-1 CP (Fig.\,\ref{fig:tmc_column} in section \ref{sec:3_11}) can be seen in this series.
One of the unique points of its envelope is that metal-bearing carbon-chain species including ions have been discovered.
The abundance ratios of MgC$_4$H/MgC$_4$H$^+$ and MgC$_6$H/MgC$_6$H$^+$ are calculated at $\sim105$ and $\sim20$, and  MgC$_3$N/MgC$_3$N$^+$ and MgC$_5$N/MgC$_5$N$^+$ are $\sim1029$ and $\sim10$, respectively.
Hence, larger species tend to be in the ionic forms.
These ionic forms are considered to be produced by radiative association between Mg$^+$ and C$_n$H or C$_n$N, respectively.
Longer species of C$_{2n}$H and C$_n$N show larger dipole moments (see Table \ref{tab:quant}), which could enhance reactions with the Mg$^+$ ion. 
The difference in the radiative association coefficient between C$_4$H and C$_6$H is one order of magnitude, whereas that between C$_3$N and C$_5$N is more than three orders of magnitude (Table D.2. in \citet{Cernicharo2023metal}), and this seems to be able to explain the observational results.
However, \citet{Cernicharo2023metal} could not reproduce all of their observed abundances; the modeled column density of MgC$_3$N$^+$ was lower than the observed one by a factor of $\sim30$.
Experimental data on these radiative association reactions and other formation and destruction reactions are needed to reduce modeling uncertaintites to better understand these potential discrepancies.
The longer cyanopolyynes may be lower in abundance compared to TMC-1 CP; [HC$_5$N]:[HC$_7$N]:[HC$_9$N] ratios in IRC+10216 are derived to be 1:$\sim0.4$:$\sim0.09$, while these ratios in TMC-1 CP are approximately 1:0.6:0.2, respectively.
These differences likely reflect different carbon-chain growth mechanisms between starless cores and envelopes around the carbon-rich AGB stars.

The planetary nebula CRL\,618, another carbon-chain-rich source, has been studied by radio and infrared observations. Polyacetylenic chains (C$_4$H$_2$ and C$_6$H$_2$) and benzene (C$_6$H$_6$) have been detected here with the Infrared Space Observatory (ISO) \citep{cernicharo2001}.
The abundances of C$_4$H$_2$ and C$_6$H$_2$ are lower than that of C$_2$H$_2$ by only a factor of 2--4, while benzene is less abundant than acetylene by a factor of $\sim40$.
These authors suggested that UV photons from the hot central star and shocks associated with its high-velocity winds affect the chemistry in CRL\,618: i.e., the UV photons and shocks drive the polymerization of acetylene and the formation of benzene.
These hydrocarbons likely survive in harsh regions compared to star-forming regions. \citet{pardo2005} observed cyanopolyynes up to HC$_7$N, and proposed rapid transformation from small cyanide to longer cyanopolyynes in this source.

\citet{berne2012} investigated formation process of C$_{60}$ by the infrared observations with {\it {Spitzer}} and {\it {Herschel}} toward the NGC\,7023 reflection nebula.
They found that C$_{60}$ is efficiently formed in cold environments of an interstellar cloud irradiated by the strong UV radiation field. The most plausible formation route is the photochemical processing of large PAHs. 

Carbon-rich AGB stars or planetary nebulae, like IRC+10216 and CRL\,618, appear to possess unique carbon-chain chemistry differing from that in star-forming regions and PDRs, and thus be important laboratories to study carbon chemistry, including PAHs and benzene.
Future observations with infrared telescopes, such as James Webb Space Telescope (JWST), may give us new insights into carbon chemistry, in particular carbonaceous dust grains, PAHs, and fullerenes.
In fact, recent observations with JWST have revealed the presence of dust shells around the Wolf-Rayet binary WR\,140, which likely implies an episodic dust formation process \citep{Lau2022}.
Using JWST, we can also reveal the spatial distributions of PAHs and fullerenes, as well as dust grains. 
Similar observations will be able to be applied for C-rich AGB stars or planetary nebulae and reveal the carbon chemistry around these harsh environments.
Furthermore, we can investigate relationships among carbon-chain species, hydrocarbons, fullerenes, and carbonaceous dust by a combination of ALMA and JWST.
These studies are essential to reveal what forms of carbon are ejected into the ISM and how they are incorporated into future star formation events.
This point will be essential to reveal the origin of large carbon-chain molecules and species including benzene rings that have been recently detected at TMC-1 CP (section \ref{sec:3_11}).

ALMA observations have detected carbon-chain species in extragalactic sources.
The ALMA Comprehensive High-resolution Extragalactic Molecular Inventory (ALCHEMI) large program has conducted line survey observations from 84.2 GHz to 373.2 GHz toward the starburst galaxy NGC\,253.
Several carbon-chain species (e.g., CCH, $c$-C$_3$H$_2$, HC$_3$N, HC$_5$N, HC$_7$N, CCS) have been detected from this galaxy \citep{martin2021}. 
The detection of more complex carbon-chain species will be reported (Dr. Sergio Martin, ESO/JAO, private comm.).
\citet{shimonishi2020} detected CCH in a hot molecular core in the Large Magellanic Cloud (LMC), and found that the CCH emission traces outflow cavity, as also seen in the low-mass YSOs in our Galaxy (see section \ref{sec:3_2}).
Such observations toward extragalactic sources with different metallicities will be important for a comprehensive understanding of carbon chemistry in the ISM.

Summaries of this section are as follows: 
\begin{enumerate}
\item Recent line survey observations toward TMC-1 by the Green Bank 100m and Yebes 40m telescopes discovered various and unexpected carbon-chain molecules. However, abundances of some of them cannot be explained by chemical simulations, meaning that our current knowledge about carbon chemistry in the ISM lacks important processes.
\item Survey observations toward low-mass YSOs revealed that both carbon-chain species and COMs are present around most of the low-mass YSOs. 
Hot corino and WCCC states are likely to be extreme ends of a continuous distribution.
\item Since carbon-chain molecules trace chemically young gas, their lines can be powerful tracers of streamers, which are important structures to understand star formation and disk evolution.
\item Carbon-chain species are formed around MYSOs. 
ALMA observations have shown that they exist in hot core regions with temperatures above 100 K. Thus, the carbon-chain chemistry is not the WCCC mechanism found around low-mass YSOs, but rather indicates the presence of ``Hot Carbon-Chain Chemistry (HCCC)''.
\item The vibrationally-excited lines of HC$_3$N can be used as a disk tracer around massive stars. 
The disk chemistry around massive stars may be different from that around lower-mass stars (i.e., T Tauri and Herbig Ae stars), although there is the need to increase the source samples to confirm this.
\item Infrared observations toward carbon-rich AGB stars and planetary nebulae have detected polyacetylene chains, benzene, and fullerenes in their envelopes. 
These sites are unique laboratories to study carbon chemistry which is different from that in star-forming regions.
\item Beyond our Galaxy, several carbon-chain species have also been detected in other galaxies enabled by high-sensitivity ALMA observations.
\end{enumerate}

\section{Chemical Simulations} \label{sec:4}

Modeling studies of carbon-chain chemistry in starless cores have tried to obtain good agreement with the observed abundances in the dark cloud TMC-1. 
Recent work has focused on particular species that were newly detected in TMC-1 CP (section \ref{sec:3_11}).
Here, we review modeling studies covering various types of carbon-chain molecules in dark clouds.

\citet{loison2014} studied several carbon-chain groups (C$_{n}$, C$_{n}$H, C$_{n}$H$_{2}$, C$_{2n+1}$O, C$_{n}$N, HC$_{2n+1}$N, C$_{2n}$H$^{-}$, C$_{3}$N$^{-}$) with gas-grain chemical models including updated reaction rate constants and branching ratios assuming two different C/O ratios (0.7 and 0.95).
They added a total of 8 new species and 41 new reactions, and modified 122 rate coefficients taken from the KInetic Database for Astrochemistry (KIDA, kida.uva.2011).
Their results clearly show that some carbon-chain molecules depend on the C/O elemental abundance ratio (e.g., C$_n$H where $n=4,5,6,8$).
Their models with new parameters can obtain good agreement with the observed abundances in the dark cloud TMC-1 CP, and the models with two C/O ratios (0.7 and 0.95) obtain a similar agreement at different times.
Two ages show better agreement between observations and models; $10^5$ yr and around ($1-2$)$\times 10^6$ yr.
The gas-phase chemistry is dominated at the earlier phase, whereas the grain surface chemistry and gas-grain interaction become more efficient at the later stages because of more frequent collisions between gas-phase particles and dust grains.
These authors also compared the modeled results to the observed abundances in another starless core, L134N. Here the models with a C/O ratio of 0.7 are in better agreement with the observed abundances compared to the case with the higher C/O ratio.
Ages when the models agree with the observed abundances in L134N best are ($3-5$)$\times10^4$ yr and $\sim6 \times 10^5$ yr.
Large amounts of free C, N, and O are available in the gas phase at the first age, while strong depletion effects are predicted at the later stage.
They also suggested that experimental work for the determination of rate constants for the reactions of O + C$_{n}$H and N + C$_{n}$H, especially at low-temperature conditions, is necessary.

The best-fitting ages differ by about one order of magnitude in the two different C/O elemental abundance ratios.
These results imply that it is not straightforward to compare chemical ages among starless cores located in different parent molecular clouds that have different elemental abundances.
Thus, we need caution when treating carbon-chain molecules as chemical evolutionary indicators (\citet{suzuki92,benson98} and section \ref{sec:3_31}).

Most recent studies have focused on particular molecules that were newly detected at TMC-1 CP (section \ref{sec:3_11}).
Classical models for dark clouds consider bottom-up chemistry starting from C$^+$, with carbon-chain molecules considered to form mainly by gas-phase reactions (see Fig.\,\ref{fig:route} in section \ref{sec:1_2}).
However, such classical views need to be revisited. 
For example, it has been revealed that both gas-phase and grain-surface formation routes are important for the reproduction of the observed abundance of H$_2$CCCHC$_3$N \citep{shingledecker2021}. 
Some molecules detected by the GOTHAM project, especially cyclic molecules, have not been explained by the chemical simulations yet \citep[e.g.,$c$-C$_9$H$_8$;][]{burkhardt2021}.
These results suggest that small rings are formed through the destruction of PAHs or other unknown processes.
Our knowledge about connections among different categories (e.g., linear, cyclic, PAHs) likely lacks important processes.
Besides, we need to reveal the initial conditions of carbon-bearing species in molecular clouds.
Further observations and chemical simulations are necessary to understand carbon-chain chemistry including the newly detected molecules.
To achieve this science goal, this is a need to expand such studies to the diffuse ISM and evolved stars, where carbon is ejected into the ISM. 
This will allow better constraints on the initial chemical composition of star-forming regions and the potential impact on carbon-chain chemistry in starless cores.

Some recent studies with sophisticated chemical simulations focused on the origin of the chemical diversity around YSOs.
\citet{aikawa2020} demonstrated their results with two phases (the static phase and the collapse phase) and a multilayered ice mantle model, and investigated how the WCCC and hot corino chemistry depend on the physical conditions during the static phase.
They found:
\begin{enumerate}
    \item The lower temperatures ($T\leq15$ K) during the static phase can produce the WCCC sources more efficiently. 
    \item The lower visual extinction during the static phase can form CH$_4$ and carbon-chain molecules become more abundant. 
    \item A longer static phase is preferable for producing the WCCC sources.  
    \item It is difficult to produce the prototypical WCCC sources, where carbon-chain species are rich but COMs are deficient. On the contrary, the hot corino sources and hybrid-type sources where both COMs and carbon-chain species are reasonably abundant could be reproduced.
\end{enumerate}
In warm conditions, grain-surface formation and freeze out of CH$_4$, which is a key species for WCCC, become less effective. 
Moreover, the conversion of CO to CO$_2$ on grain surfaces becomes important, and the gaseous CO abundance decreases. 
These lead to a low abundance of C$^+$, which is formed by the destruction of CO by He$^+$.
The C$^+$ ion is another key species for WCCC.
Therefore, warm conditions are not suitable for the production of WCCC sources.
In the model with a longer static phase, CH$_4$ accumulates during the static phase, leading to a more favorable condition for WCCC.
These results disfavor the scenario of a short timescale of prestellar collapse to explain observed WCCC \citep{sakai2008} (see section \ref{sec:2_2}).

\citet{kalvans2021} investigated the effects of the UV radiation field and cosmic rays on the formation of WCCC sources.
They concluded that WCCC can be caused by exposure of a star-forming core to the interstellar radiation field (ISRF) or just to cosmic rays (with $\zeta \geq 10^{-16}$ s$^{-1}$).
Such a conclusion agrees with the observational results that hot corino type sources are located inside dense filamentary clouds, while the WCCC sources are located outside such structures \citep[e.g.,][]{lefloch2018}.
These two model studies show that various factors, including conditions before the onset of core collapse, are related to carbon-chain chemistry around low-mass YSOs.
These factors likely are entangled in most sources.

\begin{figure*}
\begin{center}
\includegraphics[bb = 0 10 550 400, scale=0.5]{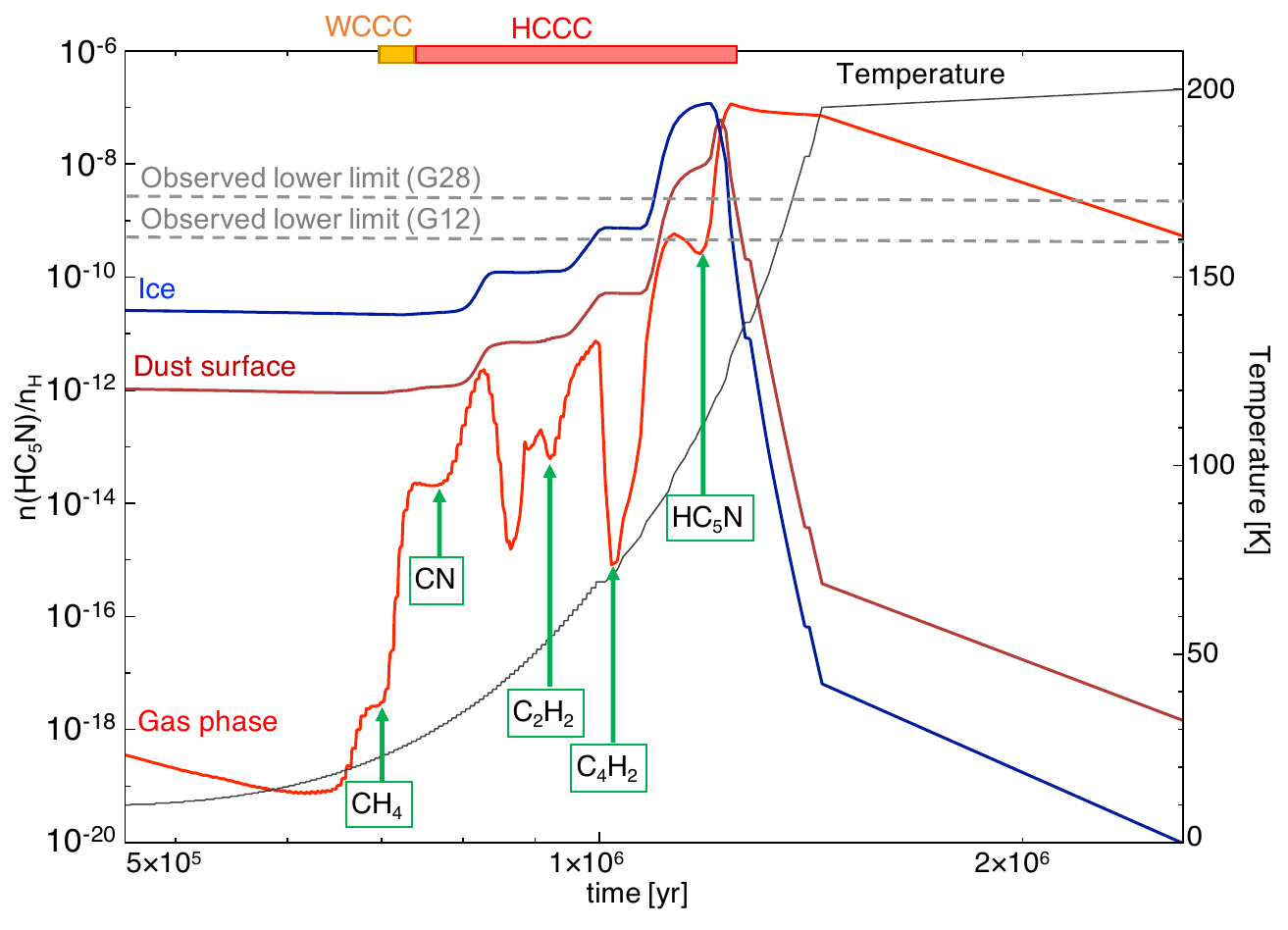}
\end{center}
\caption{The modeled HC$_5$N abundances in the gas phase (red), dust surface (brown), and ice mantle (blue) during the warm-up and hot-core periods \citep{tani2019model}. The black line indicates the temperature evolution. Indicated molecules mark the times of efficient sublimation from dust grains. The gray horizontal dashed lines indicate the observed lower limits of the gas-phase HC$_5$N abundances toward two MYSOs; G12.89+0.49 ($5\times10^{-10}$) and G28.28-0.36 ($2.1\times10^{-9}$).}
\label{fig:modelHC5N}
\end{figure*}

As with observational studies, chemical simulations focusing on the carbon-chain chemistry around MYSOs are still relatively poorly explored.
\citet{tani2019model} tried to reproduce the observed abundances of HC$_5$N around the three MYSOs observed with the single-dish telescopes (section \ref{sec:3_32}) with chemical simulations of hot-core models with a warm-up period.
They utilized three different heating timescales ($t_{h}$); $5\times10^4$ yr, $2\times10^5$ yr, and $1\times10^6$ yr, approximating high-mass, intermediate-mass, and low-mass star-formation, respectively \citep{garrod2006}.
Fig.\,\ref{fig:modelHC5N} shows the model results of HC$_5$N (gas phase, dust surface, and ice mantle) during the warm-up and hot-core periods with a heating timescale of $1\times10^6$ yr. 
In Fig.\,\ref{fig:modelHC5N}, we indicate the lower limits of the gas-phase HC$_5$N abundance in two MYSOs; G12.89+0.49 and G28.28-0.36 that show the minimum and maximum abundances, respectively \citep{tani2017MYSO}.
The observed abundances derived by single-dish telescopes are considered to be lower limits due to beam dilution. 
The observed lower limits of HC$_5$N around the MYSOs can be reproduced when the temperature reaches its sublimation temperature ($\sim115$ K) or the hot-core phase ($T=200$ K).

These authors also investigated cyanopolyyne chemistry in detail during the warm-up and hot-core periods. 
Basically, the formation and destruction reactions of HC$_3$N, HC$_5$N, and HC$_7$N are similar.
We give an explanation of HC$_5$N as an example (Fig.\,\ref{fig:modelHC5N}). 
The gas-phase HC$_5$N abundance (red curves) shows a drastic change with time (or temperature) evolution.
During the warm-up period, HC$_5$N is mainly formed by the reaction between C$_4$H$_2$ and CN.
In addition to this, the reaction between CCH and HC$_3$N partly contributes to the HC$_5$N formation.
This reaction (CCH+HC$_3$N) is important around $t\approx8.5\times10^5$ yr, when the gas-phase HC$_3$N abundance increases. 
At that time, the HC$_3$N production is enhanced by the reaction between CCH and HNC.
We indicate some molecules with green arrows in Fig.\,\ref{fig:modelHC5N}.
These indicate that each molecule directly sublimates from dust grains at these various ages.
Methane (CH$_{4}$) sublimates from dust grains around 25 K ($t\approx7.2\times10^5$ yr), and carbon-chain formation starts, namely WCCC.
After that, CN and C$_2$H$_2$ sublimate from dust grain at $t=7.7\times10^5$ yr ($T\approx31$ K) and $t=9.3\times10^5$ yr ($T\approx55$ K), respectively.
The C$_4$H$_2$ species is formed by the gas-phase reaction of ``CCH + C$_2$H$_2$ $\rightarrow$ C$_4$H$_2$ + H''.
When the temperature reaches around 73 K ($t=1.0\times10^6$ yr), C$_4$H$_2$ directly sublimates from dust grains.
The enhancement of the gas-phase abundances of CN and C$_4$H$_2$ boosts the formation of HC$_5$N in the gas phase.

We can see that the HC$_5$N abundances in dust surface and ice mantles increase when the gas-phase HC$_5$N abundance decreases.
This means that the HC$_5$N molecules, which are formed in the gas phase, adsorb onto dust grains and accumulate in ice mantles before the temperature reaches its sublimation temperature ($T\approx115$ K, corresponding to $t=1.2\times10^6$ yr).
\citet{tani2023} have advocated the concept of HCCC (Fig.\,\ref{fig:HCCCconcept}) based on these results.

This chemical simulation is supported by observations of the $^{13}$C isotopic fractionation (c.f., section \ref{sec:3_12}) of HC$_3$N toward the carbon-chain-rich MYSO G28.28-0.36 \citep{tani2016HC3N}.
The proposed main formation pathway of HC$_3$N is the reaction between C$_2$H$_2$ and CN in this MYSO.
This is consistent with the formation process seen in the chemical simulations during the warm-up stage.

\citet{tani2019model} proposed that longer heating timescales of the warm-up stage ($t_h$) could produce the carbon-chain-rich conditions by comparisons of six modeled results, focusing on cyanopolyynes.
In the HCCC mechanism, cyanopolyynes are formed in the gas phase, adsorb onto dust grains, and accumulate in the ice mantle.
Their ice-mantle abundances of just-before sublimation determine the gas-phase peak abundances.
Thus, longer heating timescales allow cyanopolyynes to accumulate in the ice mantle abundantly, leading to their higher gas-phase abundances in hot core regions.
The long heating timescale of the warm-up stage ($t_h$) does not necessarily reflect the timescale of stellar evolution or accretion.
It depends on the relationships between the size of the warm region ($R_{\rm{warm}}$) and the infall velocity ($v_{\rm{infall}}$) as suggested by \citet{aikawa2008}:
\begin{equation}
 t_{h}\propto\frac{R_{\rm{warm}}}{v_{\rm {infall}}}.
\end{equation}
If $R_{\rm{warm}}$ becomes larger or $v_{\rm{infall}}$ becomes smaller, $t_h$ will be longer.
The $R_{\rm{warm}}$ and $v_{\rm{infall}}$ values should be related to various physical parameters (e.g., protostellar luminosity, density structure, magnetic field strength).
The conditions of $R_{\rm{warm}}$ and $v_{\rm{infall}}$ can be investigated by observations.
Combined observations to derive chemical composition and the values of $R_{\rm{warm}}/v_{\rm {infall}}$ are needed for a more comprehensive understanding of the chemical evolution and diversity around YSOs.

Although the chemical simulations by \citet{tani2019model} were able to reproduce the observational results around MYSOs, we need to carefully consider the physical parameters for each MYSO. 
As shown in \citet{tani2019model}, the cosmic-ray ionization rate is a key parameter for the abundance of cyanopolyynes during the warm-up and hot-core stages. 
Moreover, different timescales of the warm-up stage affect the HC$_5$N abundance.
Effects from both factors cannot be disentangled easily. 
The cosmic-ray ionization rate is expected to be higher than the standard value ($\sim 10^{-17}$ s$^{-1}$) due to stellar feedback from the MYSO or nearby stars, and then one should potentially vary the cosmic-ray ionization rate for each source.
Since $R_{\rm{warm}}$ and $v_{\rm{infall}}$ are observable parameters, we can carry out survey observations of cyanopolyynes toward MYSOs with various $\frac{R_{\rm{warm}}}{v_{\rm {infall}}}$ and compare the results with the chemical simulations.
Such studies may be able to evaluate the effect of the warm-up stage on the cyanopolyyne abundances around MYSOs.
In addition, more sophisticated chemical simulations are needed, including a systematic exploration of the relevant parameter space to fully understand the chemical diversity around MYSOs.  

More recently, \citet{Lee2021Apml} demonstrated a machine-learning method.
They could reproduce the observed column densities of 87 molecules in TMC-1 within an order of magnitude without prior knowledge about the physical conditions of the source.
One of the merits of machine learning is that it can predict the presence of undiscovered species.
We can expect further astronomical detections with predictions by machine-learning approaches.

\section{Theoretical Studies} \label{sec:5}

\subsection{Role of Quantum Chemical Studies} \label{sec:5_1}

Quantum chemistry is an indispensable tool to study structures and spectral properties of a given molecule, regardless of laboratory stability, and makes this a necessary component of astrochemical analyses. It is probably the best tool to explore and interpret chemical structures, properties, and, most importantly, detectable spectra of unusual molecules detected in space. In the 1960s and 1970s, when radio telescopes came into action with better sensitivity, many species were identified based on the laboratory data of their rotational spectra. However, there were also many unidentified signatures. Closed shell molecules can be easily handled in the laboratory and generate rotational spectra. The problem arises in the case of radical and charged species because of their highly unstable and reactive nature. Initial detection of HCO$^+$ and N$_2$H$^+$ was in the 1970s and based on quantum chemistry \citep[see][and references therein]{fort15}. Following these, Patrick Thaddeus (Columbia University and then Harvard University) was a part of teams that detected nearly three dozen new molecular species \citep{mccarthy2001}. Thaddeus' group standardized the use of quantum chemistry in astrochemical detection. 

Several carbon-chain species such C$_2$H, C$_3$N, and C$_4$H were identified based on SCF (self-consistent field) computation. The usage of quantum chemistry in astrochemistry became popular and common if species had not been synthesized in a laboratory. Several carbon-chain species, such as $l$-C$_3$H$^+$ and C$_6$H$^-$, were identified based on the insight of quantum chemistry. This has continued until recent times with the detection of the kentenyl radical (HCCO) towards various dark clouds in 2015. In another example, the detection of C$_5$N$^-$ in the circumstellar envelope of the carbon-rich star IRC+10216 was based solely on quantum chemical data.

Modern quantum computations consider a lot of improvements with a wide range of quantum mechanical methods, basis sets, especially coupled cluster theory, at the single, double, and perturbative triples [CCSD(T)] level. The CCSD(T) method is exceptionally good at providing molecular structures and rotational constants, which provide reference rotational spectra.  The CCSD(T) method is considered the gold standard in quantum chemical calculations to obtain accurate bond energies and molecular properties \citep[e.g.,][]{ramabhadran2013,fort15,barone2015,fortenberry2017quantum}.
Over the years, many groups have made remarkable contributions to astrochemistry through quantum chemistry for astrochemical detection, understanding of the formation/destruction, and collisional excitation of different interstellar molecules (e.g., groups led by L. Allamandola, C. Bauschlicher, N. Balucani, V. Barone, M. Biczysko, P. Botschwina,  R. C. Fortenberry, J. Gauss, J. Kästner, T. J. Lee, J-C Loison, K. Peterson,  C. Puzzarini, A. Rimola, H. F. Schaefer, J. Stanton, P. Ugliengo, D. Woon). A comprehensive strategy for treating larger molecules with the required accuracy has been presented by \citet{barone2015}.

\subsubsection{Ground State and Stability} \label{sec:5_2}

The ground states of carbon-chain species are crucial because they decide their stability and eventually help us find the true state of the species either in laboratory experiments or in astrophysical environments via a spectral search. Table \ref{tab:quant} summarizes the ground state of all the carbon-chain species included in this review article. The ground state, enthalpy of formation, rotational constants, dipole moment, and polarizability of many carbon-chain species can be found in KIDA database (\url{https://kida.astrochem-tools.org/}) as well as in several works in the literature \citep[e.g.,][]{woon09,etim16,etim2017,baldea2019,etim20}. For any particular carbon-chain species, the ground state can be found with the help of quantum chemical study.  

\subsubsection{Dipole moment and Polarizability} \label{sec:5_3}

The dipole moment is an important parameter that decides whether a molecule is polar or non-polar. 
Non-polar molecules do not have a permanent electric dipole moment and do not show rotational transitions.
On the other hand, polar molecules have a permanent electric dipole moment and they show rotational transitions. A higher dipole moment value means a higher intensity of rotational transitions. It is crucial because one can say whether it is detectable or not through their rotational transitions. In addition to the symmetry of molecules, several other factors are important for their detectability in interstellar environments. These factors include the density of spectral lines, the partition functions, and the intrinsic line strengths of the transitions. The dipole moment values of all the carbon-chain species are summarized in Table \ref{tab:quant}. These can be measured theoretically with various quantum chemical methods and basis set by inclusion of molecular structures. 

The polarizabilities of all carbon-chain species, if available, are summarized in Table \ref{tab:quant}. 
The dipole moment and polarizability are crucial for the estimation of ion-neutral reaction rates. 
Ion-neutral reactions play a crucial role in the ISM, especially in cold dark clouds, for the formation and destruction of various species. For non-polar neutrals, the rate coefficient is given by the so-called Langevin expression:
\begin{equation}
k_{L} = 2\pi e \sqrt{\frac{\alpha_{\rm pola}}{\mu}},
\end{equation}
where $e$ is the electronic charge, $\alpha_{\rm pola}$ is the polarizability,  $\mu$ is the reduced mass of reactants, and cgs-esu units are utilized so that the units for the rate coefficient are cm$^{3}$ s$^{-1}$. 
This reaction rate coefficient is independent of the temperature and activation energy, and then the ion-molecule reaction can proceed even in cold molecular clouds ($T\approx 10$ K). These lead efficient formation of carbon-chain species in the early molecular cloud stage (section \ref{sec:1_2}).
For polar-neutral species, trajectory scaling relation is usually used. 
The best-known formula for the ion-dipolar rate coefficient $k_{D}$ in the classical regime and for linear neutrals is the Su-Chesnavich expression \citep{su82}. 
The Su-Chesnavich formula (see \cite{wake10,Maergoiz2009}) is based on the parameter $x$, defined by
\begin{equation}
x = \frac{\mu_{D}}{\sqrt{2\alpha_{\rm pola} k_BT}},
\end{equation}
where $\mu_{D}$ is an effective dipole moment of the neutral reactant, which is generally very close to the true dipole moment, and $k_B$ is the Boltzmann constant. 
As we can see in Table \ref{tab:quant}, the dipole moment and polarizability become larger for the longer species among the same series. The parameter $x$ is proportional to the dipole moment and inversely proportional to the square root of polarizability. Thus, it is important to obtain accurate dipole moment and polarizability by quantum calculation for chemical simulations that reproduce the observed abundances (section \ref{sec:4}).

Apart from the estimation of the rate coefficient of ion-neutral reactions, the dipole moment is one of the key parameters to determine the column density of observed species. Accurate estimation of the dipole moments is essential to derive realistic values of column density. 
For instance, the dipole moment of C$_4$H was used $\sim$ 0.87 Debye in previous literature, which was the value of mixed states, i.e., ground and excited states. 
The recent result suggests the dipole moment value is 2.10 Debye, which is 2.4 times larger than the values used before \citep{oya20}. As a result, its column densities have been overestimated by a factor of $\sim6$.

\subsubsection{Binding Energy of Carbon-Chain Species} \label{sec:5_4}

A major portion of carbon-chain species is primarily formed in the gas phase (Sections \ref{sec:1_2}, \ref{sec:2_2}, \ref{sec:2_3}). 
In addition, gas-grain exchange occurs, and many reactions occur on the grain surface.
The binding energy (BE) plays a pivotal role in interstellar chemistry, especially grain surface chemistry, which eventually enriches the gas-phase chemistry. 
Here, we describe the role of BE in interstellar chemical processes.

Binding energy controls several key processes on the grain surface. In particular, two of these are ``desorption'' and ``diffusion'', which are directly related to the BE of the species. Gas-phase species accrete onto the grain surface depending on their energy barriers with the surfaces. The species may bind to the grain surface via physisorption or chemisorption processes. $E_{\rm {des}}$ expresses the binding energy (or desorption energy) of the species. Species attached to grain surfaces can migrate from one site to another depending on the barrier energy of migration ($E_{\rm {dif}}$). This is known as a surface diffusion barrier. The surface diffusion barrier is related to the binding energy of the species following the formula $E_{\rm {dif}} = x E_{\rm {des}}$. Even though $x$ is a key parameter to describe the mobility of species on grain surfaces, the exact value has not historically been well constrained, broadly ranging from 0.3 to 0.8 \citep{Jin2020}. Desorption involves the release of species from the grain surface back into the gas phase. There are various desorption mechanisms, with thermal desorption and non-thermal desorption being the most significant. Both mechanisms directly depend on the binding energy (BE) of the species. Hence, the binding energy is a key parameter in understanding desorption processes and final gas phase abundances of carbon chains in astrochemical models, especially for HCCC (Section \ref{sec:4}).

Table \ref{tab:Ebind} in Appendix \ref{sec:appendix1} summarizes the BE values of all carbon-chain species (if available).
Most of the BE values are mainly taken from KIDA (\url{https://kida.astrochem-tools.org/}). BE estimation of a species heavily depends on the methods and surface used for the calculations \citep[e.g.,][]{pent17,das18,ferrero2020,villadsen2022,minissale2022}. The BE values of all carbon-chain species, especially higher-order linear chains, are estimated based on the addition method that are provided in KIDA. In this method, for instance, the BE of HC$_7$N is estimated by the addition of binding energies of HC$_6$N and C. For such calculations 800 K is usually adopted as the BE of a carbon atom. 
However, we see a huge difference in BE of a carbon atom between the old value and newly measured ones based on quantum chemical study \citep[$\ge 10,000$ K;][]{wake17,minissale2022}. 
Thus, the addition method of BE may lead to large uncertainties in estimating BEs of carbon-chain molecules, especially longer ones. If BE of a carbon atom is $\ge$10,000 K is correct, the addition method of BE may not be valid, because long carbon-chain species will have huge BE values. A recent experimental study suggests that the BE of carbon atoms is $\sim 2900$ K \citep{tsuge2023} in shallower binding sites.  A lower value ($<$5000 K) of carbon atom BE has also been reported in \cite{das18}. Nonetheless, to overcome this issue, dedicated quantum chemical calculations and/or experiments are required to estimate the BE values of carbon-chain species with higher accuracy. 
This problem becomes more serious for HCCC (sections \ref{sec:2_3} and \ref{sec:3_32}). In the HCCC mechanism, cyanopolyynes absorb dust grains during the warm-up stage and evaporate into the gas phase in the hot-core stage. If their estimated BE changes, their sublimation temperatures could change leading to different results in the chemical simulations and an impact on the interpretation of observed abundances.

\section{Experimental Studies}\label{sec:Experiment}

Experimental studies are essential to measure the rotational, vibrational and electronic spectra of molecules with high resolution and accuracy. They can provide accurate spectra, which are used for their precise identification in data obtained from various radio and infrared telescopes (both ground and space based). Laboratory spectroscopy plays a pivotal role in the detection of molecules in space via such telescopes. Numerous experiments have been conducted over the years to analyze the laboratory spectra of various molecular groups. In this context, we provide a concise overview of experiments focused on carbon-chain species. Cyanopolyynes stand out as one of the most intriguing and significant species in recent decades, primarily due to their profound implications for understanding the physical and chemical properties of star-forming environments and carbon-chain chemistry. 

The first carbon-chain molecule, the simplest cyanopolyyne HC$_3$N, was detected with the NRAO 140-foot radio telescope in 1971 based on the laboratory rotational spectra, as documented by \cite{tyler1963} and \cite{westenberg1950microwave}.
Subsequently, the detection of HC$_5$N, the second member of this series, relied on measured rotational spectra in a laboratory setting, as reported by \cite{alexander1976microwave}. Furthermore, the existence of higher-order cyanopolyynes, including HC$_7$N, HC$_9$N, and HC$_{11}$N, in the ISM was confirmed through various laboratory rotational spectra measurements \citep{iida1991, travers1996,mccarthy2000experimental}. In 1998, Thaddeus and colleagues compiled a comprehensive review of numerous studies focusing on the detection of carbon chains in space, alongside corresponding laboratory investigations \citep{thaddeus1998}. Henning and Schnaiter's 1998 review article delves into both astronomical spectroscopy and laboratory investigations pertaining to cosmic carbon \citep{henning1998}. It specifically explores the diverse structural forms of carbon-containing species and their characterization through experimental studies. Laboratory measurements of the gas-phase electronic spectra of various carbon-chain anions have been conducted and subsequently compared to the diffuse interstellar bands \citep{tulej1998gas}. These comparative analyses yield the initial and persuasive indications of the identity of certain carriers responsible for the diffuse interstellar bands.
\cite{mccarthy2000} conducted a comprehensive laboratory study of the rotational spectra of 11 carbon-chain molecules, encompassing cyanopolyynes, isocyanopolynes, methylcyanopolyynes, and methylpolyynes.

Laboratory experiments to investigate reactions have been developed too.
Experimental investigations of carbon-bearing molecule formation via neutral-neutral reactions are summarized in \citet{Kaiser2002}. 
Subsequently, \cite{savic2004} discussed the formation of small hydrocarbons and ions that were measured under the interstellar and circumstellar conditions by using ion trap conditions. Laboratory and astronomical identification of the first interstellar anion, $\rm{C_6H^{-}}$, was carried out by \cite{McCarthy06}. In the following year, McCarthy and colleagues measured the rotational spectra of $\rm{C_4H^{-}}$ and $\rm{C_8H^{-}}$ \citep{gupta2007}. The ion chemistry of the ISM has been extensively reviewed, featuring detailed discussions of observations and experimental results in the work of \cite{snow2008} and the associated references. Cosmic rays have been utilized to measure the formation of unsaturated hydrocarbons in interstellar ice analogs \citep{pilling2012}. This was achieved through irradiation of a mixture comprising H$_2$O, NH$_3$, CO (or CO$_2$), simple alkanes, and CH$_3$OH. As a result, various compounds were identified, including hexene, cyclohexene, benzene, OCN$^{-}$, CO, CO$_2$, as well as a range of aliphatic and aromatic alkenes and alkynes. In a review article by \cite{zack2014}, various laboratory spectroscopic techniques for studying carbon species of astrophysical significance, including methods like matrix isolation, cavity ringdown, resonance-enhanced multiphoton ionization, and ion trapping, were presented and discussed in detail. In a recent study, \cite{gatchell2021} found that during energetic collisions with heavy particles, defective PAHs can form and remain stable in the ISM under thermal equilibrium conditions. These defective PAHs exhibit enhanced reactivity compared to intact or photo-fragmented PAHs, potentially playing a crucial role in interstellar chemistry. Very recently rotational spectra of unsaturated carbon chains such as propadienone, cyanovinylacetylene, and allenylacetylene have been measured in the laboratory \citep{melli2022}.

Here we briefly describe a few studies of carbon-chain species, mainly focusing on molecules containing a benzene ring. The first benzene ring (C$_6$H$_6$) and its cyano derivative, benzonitrile ($c$-C$_6$H$_5$CN) were detected toward CRL 618 and TMC-1, respectively, based on their laboratory infrared and rotational spectra \citep[e.g.,][]{McGuire2018}. 
The most striking and groundbreaking results, the detection of fullerenes (C$_{60}$, C$_{70}$) and its protonated form C$_{60}^+$, was also based on their laboratory data \citep[e.g.,][]{martin93, nemes94,kato91}. 
In the 2020s, several PAHs have been identified in the ISM with the aid of their laboratory spectra. 
For instance, two isomers of cyanonapthalene \citep{mcguire2021,McNaughton2018}, two isomers of ethynyl cyclopentadiene \citep{mccarthy2021,cernicharo2021cyclopentadiene,McCarthy2020}, 2-cyanoindene \citep{sita2022}, fulvenallene \citep{McCarthy2020,sakaizumi1993}, ethynyl cyclopropenylidene, cyclopentadiene, and indene \citep[][and references therein]{cernicharo2021hydrocarboncycle,burkhardt2021}. 
Recently, \citet{Martinez2020} investigated the growth of carbon-containing species from C and H$_2$ in conditions analogous of circumstellar envelopes around carbon-rich AGB stars. 
They found that nanometer-sized carbon particles, pure carbon clusters, and aliphatic carbon species are formed efficiently, whereas aromatics are generated at trace levels and no fullerenes are detected.
They also suggested that the formation of aromatic species must occur via other processes, such as the thermal processing of aliphatic material on the surface of dust grains. The astronomical detection of most of the carbon-chain species was based on their laboratory rotational spectra.  A more comprehensive list of references to laboratory experiments of rotational, vibrational, and electronic spectra of different molecules can be found in the individual papers reporting their interstellar detection.

\section{Summary and Open Questions of This Review} \label{sec:7}

\subsection{Summary}

We have reviewed carbon-chain chemistry in the ISM, mainly focusing on recent updates.
A summary of our main points is as follows.

\begin{enumerate}
\item By the end of 2023, 132 carbon-chain species have been detected in the ISM. This accounts for almost 43\% of 305 interstellar molecules detected in the ISM or circumstellar shells. These include various families of carbon-chain species, involving elements of O, N, S, P, and Mg. 

\item Two line survey projects toward TMC-1 CP (GOTHAM and QUIJOTE) have recently reported detections of many new carbon-chain species. Abundances of some of these species are not yet reproducible in chemical simulations indicating a need for improved models.

\item In addition to the cold gas conditions of early-phase molecular clouds, carbon-chain formation also occurs around low-, intermediate- and high-mass YSOs. Warm Carbon-Chain Chemistry (WCCC) was found in 2008 around low-mass YSOs, while Hot Carbon-Chain Chemistry (HCCC) has been proposed based on recent observations around high-mass YSOs. 

\item Chemical simulations aim to explain conditions forming hot corino and WCCC sources. There are several possible parameters to reproduce the chemical differentiation: e.g., UV radiation field and temperature of the prestellar phase. The most plausible effect is the ISRF rather than the timescale of the prestellar phase.

\item Thanks to high-angular resolution and high-sensitivity observations with ALMA, several carbon-chain species (e.g., CCH, $c$-C$_3$H$_2$, HC$_3$N) have been detected from protoplanetary disks around Herbig Ae and T Tauri stars. Vibrationally-excited lines of HC$_3$N have been found to trace disk structures around massive stars.

\item Circumstellar envelopes around carbon-rich AGB stars and planetary nebulae are unique factories of carbon chemistry. Infrared observations have revealed the presence of PAHs and fullerenes in such environments. Recent laboratory experiments investigated chemistry in such regions.

\item Carbon-chain species have been detected even in extragalactic environments, such as the starburst galaxy NGC253 via the ALCHEMI project. In the Large Magellanic Cloud (LMC), CCH emission has been found to trace outflow cavities, as seen in low-mass YSOs in our Galaxy. 

\item Theoretical and experimental studies are important for the observational detection of carbon-chain species and for obtaining an understanding of their formation and destruction processes. Developments of these techniques are important to reveal carbon-chain chemistry in various physical conditions in the ISM.

\item Recent new approaches with machine learning will provide us with predictions of possible detectable species in the ISM, which is expected to help accelerate the discoveries of new interstellar molecules.
\end{enumerate}

The presence of carbon-chain species in the ISM has been known since the early 1970s, and many researchers have investigated their features through observations, chemical simulations, laboratory experiments, and quantum calculations. Recent findings raise new questions about carbon-chain chemistry, and it is an exciting time of progress.
To solve the newly raised questions, collaborative research involving observations, laboratory experiments, and chemical simulations is crucial.

To understand the carbon-chain chemistry better by observational methods, we need more dedicated low-frequency, high-sensitivity, and high angular resolution observations towards dark clouds, low-, intermediate-, and high-mass YSOs, and other environments. 
ALMA Band 1 is now available, and the Square Kilometer Array (SKA) and ngVLA will become available in the near future.
Observations using these facilities will be essential to reveal links between ISM physics and carbon-chain chemistry, the origin of chemical differentiation around YSOs, and relationships between WCCC and HCCC. In addition, future observational studies combining infrared data (e.g., from JWST, Thirty Meter Telescope (TMT), European Extremely Large Telescope (E-ELT)) and radio (ALMA, ngVLA, SKA, and future single-dish) telescopes have the potential for breakthrough results. For instance, relationships between PAHs/small dust grains and common carbon-chain species, which can be observed by infrared and radio regimes, respectively, can be studied by such a combination. 
These studies will help open up the astrochemical field of ``the lifecycle of carbon in the ISM''.

\subsection{Open Key Questions}

Recent new discoveries of carbon-chain molecules in the ISM have raised new questions, and we have realized that our knowledge of carbon-chain chemistry is far from complete.
We highlight the following open questions:

\begin{enumerate}
\item How do large carbon-chain species, which have been found in TMC-1 CP, form? Is there a role for both bottom-up and top-down processes?
\item How are PAHs and fullerenes related to other carbon-chain species in the ISM?
\item In what form is the main carbon reservoir ejected into the ISM {\bf {from evolved stars}}? How does it propagate into the ISM and become incorporated into molecular clouds where the next-generation of star formation occurs?
\item Which, if any, important formation/destruction processes of carbon-chain species are missing from current chemical models? 
\item Can we estimate more accurate branching ratios for different species (e.g., isomers) in electron recombination reactions?
\item How can we obtain accurate binding energy of carbon-chain species?
\item Are there reactions forming and/or destroying carbon-chain species on dust surfaces and within ice mantles?
\end{enumerate}

Answers to these questions likely require combined efforts of multi-wavelength observational, theoretical, and experimental study.


\begin{appendices}

\section{Binding Energies} \label{sec:appendix1}

Table \ref{tab:Ebind} summarizes the binding energies (BE) of carbon-chain species.

\input{table4}

\clearpage

\section{Observed Column Densities in TMC-1 CP and IRC+10216} \label{sec:appendix2}

Tables \ref{tab:colden-tmc} and \ref{tab:colden-irc} provide information on column densities at TMC-1 CP and IRC+10216.
These values are used for Figs.\,\ref{fig:tmc_column} and \ref{fig:irc_column} in sections \ref{sec:3_11} and \ref{sec:3_6}, respectively.

\input{table2}

\input{table3}

\end{appendices}

\backmatter

\bmhead{Acknowledgments}
K.T. is grateful to Professor Eric Herbst (University of Virginia) for leading me to the astrochemical field, working with me, and giving a lot of comments on studies of carbon-chain molecules that are presented in this article. 
K.T. appreciates Professor Masao Saito (National Astronomical Observatory of Japan) for giving his advice and continuous encouragement.
K.T. is supported by JSPS KAKENHI grant No.JP20K14523.
P.G acknowledges the support from the Chalmers Initiative of Cosmic Origins Postdoctoral Fellowship.
We thank Professor Fumitaka Nakamura (National Astronomical Observatory of Japan) and Professor Kazuhito Dobashi (Tokyo Gakugei University) for providing original data of mapping observations of carbon-chain species toward TMC-1 obtained by the Nobeyama 45m radio telescope. 
We would also like to thank Dr. Emmanuel E. Etim for his comments and suggestions. J.C.T. acknowledges support from ERC Advanced Grant MSTAR.
We appreciate Dr. Ryan Lau for his comments and suggestions.


\bibliographystyle{aa}

\input{main.bbl}
\end{document}

%% file: table1.tex
\begin{longtable}[!h]{p{2.3cm}p{0.8cm}cp{0.8cm}p{4.3cm}p{1.5cm}p{1.5cm}}
\caption{Ground state, polarizability ($\alpha$), dipole moment ($\mu$) and present astronomical status of different carbon chain species}
\label{tab:quant}
\\ 
\hline
Species & Ground State & $\alpha$ & $\mu$ & \multicolumn{3}{c}{Detection Status} \\
&&($\mathring{A}^{3}$) & (D) & ISM or CSE & TMC-1 & IRC+10216\\
\hline\hline
\endfirsthead
\multicolumn{7}{c}{Cont.} \\
\hline
\endhead
C$_2$ &singlet &5.07$^{a}$  &0.0 &\cite{souz77c2} & & $\surd$\\
C$_3$ &singlet&5.18$^{a}$ &0.0 &\cite{hink88c3} &   & $\surd$ \\
C$_4$ &singlet&7.51$^{a}$  &0.0& & & \\
C$_5$ &singlet &11.16$^{a}$ &0.0&\cite{bern89c5} &   & $\surd$ \\
C$_6$ &singlet &14.32$^{a}$ &0.0 &-- & & \\
C$_7$ &singlet &20.50$^{a}$ &0.0 &-- & & \\
C$_8$ &singlet &23.96$^{a}$ &0.0 &-- & & \\
C$_9$ &singlet &33.36$^{a}$ &0.0 & --& & \\ 
C$_{10}$&singlet &37.70 $^{a}$ &0.0 &-- & & \\ 
\hline
\multicolumn{5}{c}{$\rm{C_{n}H}$}\\
\hline
C$_2$H &doublet &4.42$^{a}$ &0.81$^{a}$ & \cite{tuck74c2h}& $\surd$ & $\surd$ \\
$l$-C$_3$H & doublet&5.36$^{a}$ &3.52$^{a}$ &\cite{thad85c3h} & $\surd$ & $\surd$ \\
C$_3$H$^+$ &singlet&3.69$^{*}$ &3.00$^{b}$& \cite{pety12c3hp} & $\surd$ & \\
C$_4$H &doublet & 7.15$^{a}$ &2.40$^{c}$& \cite{guelin78c4h} & $\surd$ & $\surd$\\
C$_4$H$^-$ &singlet &9.57$^{*}$  &5.9$^{d}$& \cite{Cernicharo2007}& $\surd$ & $\surd$ \\
C$_5$H &doublet &10.50$^{a}$ &4.84$^{a}$ & \cite{Cernicharo86} & $\surd$ & $\surd$ \\
C$_5$H$^+$ &singlet &8.44$^{*}$ &2.88$^{e}$ &\cite{cernicharo2022C5H} & $\surd$ & \\
c-C$_5$H &doublet &8.77$^{*}$& 3.39$^{f}$ &\cite{Cabezas2022cyclicC5H}& $\surd$ & \\
C$_6$H &doublet &13.68$^{a}$ &5.6$^{a}$ &\cite{suzu86c6h} & $\surd$ & $\surd$ \\
C$_6$H$^-$ &singlet &15.74$^{*}$&8.2$^{d}$ &\cite{McCarthy06}& $\surd$ & $\surd$ \\
C$_7$H &doublet &17.37$^{a}$ &5.83$^{a}$ & \cite{Guelin97} &  & $\surd$ \\
C$_8$H &doublet &21.85$^{a}$ &6.43$^{a}$ & \cite{Cernicharo96} & $\surd$ & $\surd$ \\
C$_8$H$^-$ &singlet &24.39$^{*}$&11.9$^{5}$& \cite{Brunken2007,Remijan2007} & $\surd$ & $\surd$ \\
C$_9$H &doublet &26.38$^{a}$ &6.49$^{a}$ &-- &  & \\
C$_{10}$H &doublet &31.05$^{a}$ &7.13$^{a}$ &\cite{Remijan2023} & $\surd$ & \\
C$_{10}$H$^-$ &singlet &-- &--&\cite{Remijan2023} & $\surd$ & $\surd$ \\
\hline
\multicolumn{5}{c}{$\rm{HC_{n}H}$}\\
\hline
HC$_2$H &singlet&3.38$^{a}$ &0.0$^{a}$ &\cite{Ridgway1976}& & $\surd$ \\
HC$_3$H &triplet&2.58$^{a}$ &0.51$^{g}$ &-- & &\\
HC$_4$H &singlet&7.05$^{a}$ &0.0$^{a}$&\cite{cernicharo2001}& & \\
HC$_5$H &triplet & -- &--& --& & \\
HC$_6$H &singlet & 11.95$^{a}$ &0.0$^{a}$&\cite{cernicharo2001}& &  \\
HC$_7$H &triplet & -- &--&-- & & \\
HC$_8$H &singlet& 18.59$^{a}$ &0.0&-- & & \\
\hline
\multicolumn{5}{c}{$\rm{H_2C_n}$}\\
\hline
$l$-H$_2$C$_3$ &singlet &5.61$^{a}$  &4.16$^{a}$ &\cite{cernicharo1991c3h2} & $\surd$ & $\surd$ \\
$l$-H$_2$C$_4$ &singlet&8.12$^{a}$ &4.43$^{a}$&\cite{Cernicharo1991c4h2}& & $\surd$ \\
$l$-H$_2$C$_5$ &singlet &11.32$^{a}$  &5.89$^{a}$ &\cite{cabezas2021cumulene}  & $\surd$ & \\
$l$-H$_2$C$_6$ &singlet & 15.15$^{a}$ &6.15&\cite{Langer97} & $\surd$ &  \\
\hline
\multicolumn{5}{c}{$\rm{C_{n}O}$}\\
\hline
C$_2$O &triplet&4.09$^{a}$  &1.43$^{a}$ & \cite{Ohishi91} & $\surd$ & $\surd$ \\
C$_3$O &singlet &6.03$^{a}$ &2.39$^{h}$ & \cite{Matthews84} & $\surd$ & $\surd$ \\
HC$_3$O$^+$ &singlet&-- &3.41$^{i}$& \cite{cernicharo2020HC3O} & $\surd$ &\\
C$_4$O &triplet &9.21$^{a}$  &3.01$^{j}$&-- & & \\
C$_5$O &singlet &10.93$^{*}$&4.06$^{k}$& \cite{Cernicharo2021C5O} & $\surd$ & $\surd$ \\
C$_6$O &triplet &15.41$^{*}$ &4.88$^{j}$&-- & & \\
C$_7$O &singlet &19.51$^{*}$ &4.67$^{l}$& --& & \\
C$_8$O &triplet &25.99$^{*}$ &4.80$^{l}$&-- & & \\
\hline
\multicolumn{5}{c}{$\rm{C_{n}S}$}\\
\hline
C$_2$S &triplet&6.87$^{a}$ &3.12$^{a}$& \cite{saitoc3s1987} & $\surd$ & $\surd$ \\
HC$_2$S$^+$ &triplet&&2.29$^{l1}$& \cite{cabezas2022} & $\surd$ & \\
C$_3$S &singlet &9.65$^{a}$ &3.939$^{a}$& \cite{Kaifu87}& $\surd$ & $\surd$ \\
HC$_3$S$^+$ &singlet &--&1.73$^{d}$& \cite{cernicharo2021HC3S}  & $\surd$ & \\
C$_4$S &triplet &13.70$^{a}$ &4.62$^{a}$& \cite{Cernicharo2021S} & $\surd$ & $\surd$ \\
C$_5$S &singlet &15.79$^{*}$ &4.65$^{m}$& \cite{Agundez2014} & $\surd$ & $\surd$ \\
C$_6$S &triplet &21.53$^{*}$ &5.40$^{l}$&-- & & \\
C$_7$S &singlet &26.44$^{*}$&6.17$^{l}$&-- & & \\
C$_8$S &triplet &34.54$^{*}$&6.50$^{l}$& --& & \\
\hline
\multicolumn{5}{c}{$\rm{C_{n}N}$}\\
\hline
C$_2$N & doublet&4.27$^{a}$  & 0.60$^{a}$ & \cite{anderson14c2n} & & $\surd$ \\
C$_3$N &doublet &5.68$^{a}$ & 2.86$^{a}$ & \cite{guel77c3n} & $\surd$ & $\surd$ \\
C$_3$N$^-$ &singlet&7.58$^{*}$&3.1$^{n}$ & \cite{Thaddeus08} & $\surd$ & $\surd$ \\
C$_4$N &doublet &8.75$^{a}$  &0.06$^{a}$ & --& & \\
C$_5$N &doublet &9.43$^{a}$ &3.33$^{a}$ & \cite{Guelin98}& $\surd$ & $\surd$ \\
C$_5$N$^-$ & singlet&13.26$^{*}$&5.20$^{o}$& \cite{Cernicharo2008C5Nm}& $\surd$ & $\surd$ \\
C$_6$N &doublet &14.91$^{*}$& 0.21$^{p}$& --& & \\
C$_7$N &doublet &18.95$^{a}$ &0.87$^{a}$ & --& & \\
C$_7$N$^-$ &singlet&--&7.5 &\cite{cernicharo2023c7n} & $\surd$ & $\surd$ \\
C$_8$N &doublet &24.16$^{*}$&--&-- & & \\
NC$_4$NH$^+$ &singlet &--&9.1&\cite{agundez2023}& $\surd$ & \\
\hline
\multicolumn{5}{c}{$\rm{C_{n}P}$}\\
\hline
C$_2$P &doublet &7.52$^{a}$  &3.24$^{a}$ & \cite{Halfen08} & & $\surd$ \\
C$_3$P &doublet &10.50$^{a}$ &3.89$^{a}$&-- & &\\
C$_4$P &doublet &12.76$^{a}$  &4.19$^{a}$&-- & &\\
C$_5$P &doublet &17.22$^{*}$&-- & --& & \\
C$_6$P &doublet &22.13$^{*}$&-- &-- & & \\
C$_7$P &doublet &27.90$^{*}$&-- &-- & &\\
C$_8$P &doublet &34.41$^{*}$&-- & --& &\\
\hline
\multicolumn{5}{c}{$\rm{HC_{2n}N}$}\\
\hline
HC$_2$N &triplet& &3.30$^{q}$ &  \cite{Guelin91} & & $\surd$ \\
HC$_4$N &triplet &8.84$^{a}$&4.30$^{a}$ &  \cite{Cernicharo04} & & $\surd$ \\
HC$_6$N &triplet&15.07$^{a}$  &4.89$^{a}$&-- & &\\
HC$_8$N &triplet &22.96$^{a}$ &5.57$^{a}$ &-- & &\\
H$_2$CCCN &doublet&&3.60&\cite{cabezas2023h2c3n}&  $\surd$ & \\

\hline
\multicolumn{5}{c}{$\rm{HC_{2n+1}N}$}\\
\hline
HC$_3$N &singlet &5.85$^{a}$ &3.78$^{a}$ &\cite{Turner71} & $\surd$ & $\surd$ \\
HNC$_3$ &singlet &6.41$^{*}$ &6.46$^{p}$ & \cite{kawaguchi1992}& $\surd$ & $\surd$ \\
HC$_3$NH$^+$ &singlet&4.90$^{*}$ &1.87$^{r}$ &\cite{Kawaguchi94} & $\surd$ & \\
HC$_5$N &singlet &10.42$^{a}$ &4.41$^{a}$& \cite{Avery76} & $\surd$ & $\surd$ \\
HC$_5$NH$^+$ &singlet &10.67$^{*}$ &3.26$^{n1}$& \cite{Marcelino20} & $\surd$ & \\
HC$_7$N &singlet &16.69$^{a}$ &4.90$^{a}$& \cite{Kroto78} & $\surd$ & $\surd$ \\
HC$_7$NH$^+$ &singlet &-- &6.40$^{o1}$& \cite{cabezas2022}& $\surd$ & \\
HC$_9$N &singlet &23.89$^{a}$ &5.29$^{a}$& \cite{Broten78} & $\surd$ & $\surd$ \\
HC$_{11}$N &singlet &--&5.47& \cite{loomis2021} & $\surd$ & \\
\hline
\multicolumn{5}{c}{$\rm{HC_{n}O}$}\\
\hline
HC$_2$O &doublet&4.20$^{c}$&1.8$^{s}$ &\cite{Agundez2015hc2o} & $\surd$ & \\
HC$_3$O &doublet &5.20$^{c}$&2.74$^{s}$&\cite{cernicharo2020HC3O} & $\surd$ & \\
HC$_4$O &doublet &7.83$^{*}$&2.64$^{m1}$& & &\\
HC$_5$O &doublet &10.87$^{*}$& 2.16$^{m1}$&  \cite{McGuire2017} & $\surd$ & \\
HC$_6$O &doublet &14.46$^{*}$&2.11$^{m1}$ & & &\\
HC$_7$O &doublet &--&2.17$^{m1}$ &\cite{Cordiner2017}& $\surd$ &\\
HC$_8$O &doublet &--&2.19$^{m1}$ & & &\\
\hline
\multicolumn{5}{c}{$\rm{HC_{n}S}$}\\
\hline
HC$_2$S &doublet&6.92$^{s}$&1.36$^{s}$&\cite{cernicharo2021HC3S} & $\surd$ & \\
HC$_3$S &doublet &9.62$^{s}$&1.28$^{s}$ & & &\\
HC$_4$S &doublet &11.60$^{*}$ &1.45$^{p1}$&\cite{fuentetaja2022} & $\surd$ & \\
HC$_5$S &doublet &--&1.92$^{q1}$& & &\\
HC$_6$S &doublet &19.86$^{*}$&2.75$^{r1}$& & &\\
HC$_7$S &doublet &--&2.10$^{q1}$& & &\\
HC$_8$S &doublet &--&3.21$^{r1}$& & &\\
\hline
\multicolumn{5}{c}{Metal Containing}\\
\hline
MgC$_{2}$H &doublet &10.44$^{*}$&1.68$^{t}$& \cite{Agundez2014} & & $\surd$ \\
MgC$_{4}$H &doublet &15.01$^{*}$&2.12$^{u}$ & \cite{Cernicharo2019Mg} & & $\surd$ \\
MgC$_{3}$N &doublet &13.16$^{*}$&6.30$^{u}$ & \cite{Cernicharo2019Mg} & & $\surd$ \\
MgC$_{5}$N &doublet &19.22&7.30$^{v}$ &\cite{pardo2021}& $\surd$  & $\surd$ \\
MgC$_{6}$H &doublet &21.54$^{*}$&2.50$^{v}$ &\cite{pardo2021} & &  $\surd$ \\
$c$-C$_2$Si &doublet &6.79$^{a}$ &2.4$^{a}$& \cite{thaddeus1984sic2} & & $\surd$ \\
$c$-C$_3$Si &singlet&11.90$^{a}$ &4.1$^{a}$ &  \cite{Apponi99}& & $\surd$ \\
C$_4$Si &singlet &--& 6.3$^{w}$& \cite{ohishi1989sic4} & & $\surd$ \\
HMgCCN&singlet&&4.5&\cite{cabezas2023}&& $\surd$\\
NaCCCN &singlet&&12.9&\cite{cabezas2023}&& $\surd$\\
MgC$_4$H$^+$ &singlet&&13.5&\cite{Cernicharo2023metal}&& $\surd$\\
MgC$_3$N$^+$ &singlet&&18.5&\cite{Cernicharo2023metal}&& $\surd$\\
MgC$_6$H$^+$ &singlet&&18.2&\cite{Cernicharo2023metal}&& $\surd$\\
MgC$_5$N$^+$ &singlet&&23.7&\cite{Cernicharo2023metal}&& $\surd$\\
\hline
\multicolumn{5}{c}{Cyclic carbon chains}\\
\hline
$c$-C$_3$H & doublet&4.80$^{a}$ &2.60$^{a}$ &\cite{Yamamoto87} & $\surd$ &\\
$c$-C$_3$H$_2$  &singlet &4.58$^{a}$  &3.41$^{a}$ & \cite{Thaddeus85C3H2} & $\surd$ & $\surd$ \\
$c$-C$_3$HCCH &singlet& -- &4.93$^{x}$ & \cite{cernicharo2021hydrocarboncycle} & $\surd$ & \\
$c$-H$_2$C$_3$O &singlet &5.20$^{s}$ &4.39$^{s}$&\cite{Hollis06} & $\surd$ & \\
\hline
\multicolumn{5}{c}{PAHs and benzene ring related species}\\
\hline
C$_6$H$_6$&singlet &10.35$^{a}$ &0.0 & \cite{cernicharo2001} & & \\
$\rm{C_{60}^{+}}$&doublet &&0.0 &\cite{Foing1994}& & \\
C$_{60}$&singlet&79.0$^{y}$ &0.0 & \cite{Cami2010} & & \\
C$_{70}$&singlet &10$^{z}$&0.0 & \cite{Cami2010} & & \\
$c$-$\rm{C_6H_5CN}$& singlet &11.91$^{y}$&4.51$^{k1}$ &\cite{McGuire2018} & $\surd$ & \\
$c$-$\rm{C_9H_8}$&singlet &121.20$^{a1}$&0.87$^{b1}$&\cite{cernicharo2021hydrocarboncycle}& $\surd$ & \\
$c$-$\rm{C_5H_4CCH_2}$& singlet&--&0.69$^{c1}$ & \cite{cernicharo2022fulvenallene} & $\surd$ & \\
$c$-$\rm{C_5H_6}$& singlet&--& 0.42$^{d1}$& \cite{cernicharo2021hydrocarboncycle}& $\surd$ & \\
1-$c$-$\rm{C_5H_5CN}$&singlet &10.93$^{*}$&4.42$^{e1}$ & \cite{mccarthy2021} & $\surd$ &\\
2-$c$-$\rm{C_5H_5CN}$&singlet &10.65$^{*}$&5.13$^{e1}$ & \cite{lee2021}& $\surd$ & \\
1-$\rm{C_{10}H_7CN}$&singlet &19.83$^{*}$&6.60$^{f1}$ & \cite{mcguire2021} & $\surd$ & \\
2-$\rm{C_{10}H_7CN}$&singlet &20.43$^{*}$&6.10$^{f1}$ & \cite{mcguire2021}& $\surd$ & \\
$o$-$\rm{C_6H_4}$&singlet&--&1.38$^{h1}$ & \cite{cernicharo2021benzyne} & $\surd$ & \\
1-$c$-$\rm{C_5H_5CCH}$&singlet &12.36$^{*}$&1.13$^{g1}$& \cite{cernicharo2021cyclopentadiene} & $\surd$ & \\
2-$c$-$\rm{C_5H_5CCH}$&singlet &11.93$^{*}$&1.48$^{g1}$& \cite{cernicharo2021cyclopentadiene} & $\surd$ & \\
$\rm{C_6H_5CCH}$& singlet&--&0.66$^{i1}$ & \cite{loru23} & $\surd$ & \\
$\rm{C_9H_7CN}$&singlet &--& 5.04$^{j1}$&\cite{sita2022} & $\surd$ & \\
CH$_3$C$_5$N &singlet&&5.75&\cite{Snyder06}& $\surd$ &\\
CH$_3$C$_6$H &singlet&&1.50&\cite{Remijan06}& $\surd$ &\\
{\it{E}}-1-C$_4$H$_5$CN &singlet&11.18&4.81&\cite{Cooke2023} & $\surd$ &\\
(CH$_3$)$_2$CCH$_2$ &singlet&&0.50&\cite{Fatima2023} & &\\
\hline
\end{longtable}
The $\surd$ mark in the last three rows means detection including tentative detection. \\
References: $^{a}$\cite{woon09},
$^{b}$\cite{pety12c3hp}, $^{c}$\cite{oya20},
$^{d}$\cite{blanksby2001},
$^{e}$\cite{bots91},
$^{f}$\cite{crawford1999},
$^{g}$\cite{nguyen2001},
$^{h}$\cite{brown1983},
$^{i}$\cite{cernicharo2020HC3O},
$^{j}$\cite{ewing1989},
$^{k}$\cite{bots93},
$^{l}$\cite{etim20},
$^{m}$\cite{pascoli1998},
$^{n}$\cite{Thaddeus08},
$^{o}$\cite{Cernicharo2008C5Nm},
$^{p}$\cite{kawaguchi1992},
$^{q}$\cite{hirano1989},
$^{r}$\cite{bots1987},
$^{s}$KIDA (\url{https://kida.astrochem-tools.org/}), 
$^{t}$\cite{woon1996},
$^{u}$\cite{Cernicharo2019Mg},
$^{v}$\cite{pardo2022},
$^{w}$\cite{ohishi1989},
$^{x}$\cite{trav97},
$^{y}$ \url{https://cccbdb.nist.gov/pollistx.asp},
$^{z}$\cite{compa2001},
$^{a1}$\cite{ghiasi2006},
$^{b1}$\cite{caminati1993},
$^{c1}$\cite{sakaizumi1993},
$^{d1}$\cite{laurie1956},
$^{e1}$\cite{sakaizumi1987},
$^{f1}$\cite{McNaughton2018},
$^{g1}$\cite{cernicharo2021cyclopentadiene},
$^{h1}$\cite{kraka1993},
$^{i1}$\cite{cox1975},
$^{j1}$\cite{sita2022},
$^{k1}$\cite{WOHLFART2008},
$^{l1}$\cite{puzzarini2008},
$^{m1}$\cite{mohamed2005},
$^{n1}$\cite{Marcelino20},
$^{o1}$\cite{cabezas2022HC7NH},
$^{p1}$\cite{Fuentetaja2022HCCCHCCC},
$^{q1}$\cite{gordon2002},
$^{r1}$\cite{wang2009}
$^{*}$This work (estimated with B3LYP/6-311++G(d,p) level of theory using Gaussian software \citep{fris13})\\

%% file: table4.tex
\begin{table*}[h]
\tabcolsep7.5pt
\caption{Binding energy (BE) of carbon-chain species}
\label{tab:Ebind}
\begin{center}
\begin{tabular}{cc|cc|cc}
\hline
Species & BE (K) & Species& BE (K) & Species& BE (K) \\
\hline
C$_2$& 10000& C$_8$N&7200&C$_3$O&2750 \\
C$_3$&2500&C$_9$N&8000&C$_5$O&4350 \\
C$_{4}$&3200&C$_{10}$N&8800 & C$_7$O&5950  \\ 
C$_5$&4000&C$_2$H$_2$&2587 &C$_9$O&7550 \\
C$_6$&4800&C$_2$H$_4$&2500 & HC$_2$O&2400 \\
C$_7$&5600&C$_2$H$_5$&3100 &SiC$_2$&4300 \\
C$_{8}$&6400&C$_2$H$_6$&1600 &SiC$_3$&5100 \\
C$_{9}$&7200&C$_4$H$_2$&4187& SiC$_4$&5900 \\
C$_{10}$&8000&C$_5$H$_2$&4987 \\
C$_{11}$&9600&C$_6$H$_2$&5787\\
C$_2$H&3000&C$_7$H$_2$&6587\\
$l$-C$_3$H&4000&C$_2$P&4300 & \\
$c$-C$_3$H&5200&C$_3$P&5900\\
C$_4$H&3737&C$_4$P&7500\\
C$_5$H&4537&C$_2$S&2700\\
C$_6$H&5337&C$_3$S&3500\\
C$_7$H&6137&C$_4$S&4300\\
C$_8$H&6937&HC$_3$N&4580\\
$c$-$\rm{C_3H_2}$&5900&HC$_4$N&5380\\
C$_2$N&2400&HC$_5$N&6180\\
C$_3$N&3200&HC$_6$N&7780\\
C$_4$N&4000&HC$_7$N&7780\\
C$_5$N&4800&HC$_8$N&9380\\
C$_6$N&5600&HC$_9$N&9380\\
C$_7$N&6400&C$_2$O&1950\\


\hline
\end{tabular}
\end{center}
{Taken from the KIDA (\url{https://kida.astrochem-tools.org/}), and also see \citet{wake17}, \citet{pent17}, \citet{das18}.}
\end{table*}

%% file: table2.tex
\begin{longtable}[h]{p{2.2cm} p{3.5cm} p{2.3cm}p{4cm}}
\caption{Observed column density of carbon chain species toward TMC-1 CP}
\label{tab:colden-tmc}
\\ 
\hline
Species & $N$ (cm$^{-2}$) & Telescope  & Ref.  \\ 
\hline
        HC$_3$N & (2.3 $\pm$ 0.2)$\times 10^{14}$  & RT40m & \cite{cernicharo2020HC4NC}\\ 
        HC$_5$N & (1.08$\pm$0.2)$\times 10^{14}$ & RT40m &  \cite{cernicharo2020HC4NC} \\ 
        HC$_7$N & (6.4$\pm$0.4)$\times 10^{13}$  & RT40m &  \cite{cernicharo2020HC4NC} \\ 
        HC$_9$N & $2.15_{-0.20}^{+0.23}\times 10^{13}$ & GBT100m &\cite{loomis2021}  \\ 
        HC$_{11}$N & $7.8_{-5.08}^{+21.27}\times 10^{11}$ & GBT100m &\cite{loomis2021} \\ 
        ~ & ~ & ~ & ~ \\ 
        C$_2$H & ($6.5\pm2.7)\times10^{14}$ & ~ & \cite{sakai2010} \\ 
        C$_4$H & $1.62_{-0.22}^{+0.25}\times10^{14}$ & GBT100m & \cite{Remijan2023} \\ 
        C$_4$H$^-$ & $6.79_{-2.59}^{+4.68}\times10^{10}$ & GBT100m & \cite{Remijan2023} \\ 
        C$_6$H & $5.17_{-1.83}^{+0.62}\times10^{12}$ & GBT100m & \cite{Remijan2023} \\ 
        C$_6$H$^-$ & $2.84_{-0.44}^{+0.58}\times10^{11}$ & GBT100m & \cite{Remijan2023} \\ 
        C$_8$H & $7.25_{-1.23}^{+2.06}\times10^{11}$ & GBT100m & \cite{Remijan2023}\\ 
        C$_8$H$^-$ & $8.00_{-2.59}^{+7.79}\times10^{10}$ & GBT100m & \cite{Remijan2023} \\ 
        C$_{10}$H & $2.02_{-0.82}^{+2.68}\times10^{11}$ & GBT100m & \cite{Remijan2023} \\  
        C$_{10}$H$^-$ & $4.04_{-2.23}^{+10.67}\times10^{11}$ & GBT100m & \cite{Remijan2023} \\ 
        ~ & ~ & ~ & ~ \\ 
        HC$_2$O & (7.7$\pm$0.7)$\times10^{11}$ & RT40m &\cite{Cernicharo2021C5O} \\ 
        HC$_3$O & (1.3$\pm$0.2)$\times10^{11}$ & RT40m & \cite{Cernicharo2021C5O} \\ 
        HC$_4$O & $\leq 9.0 \times10^{10}$ & RT40m & \cite{Cernicharo2021C5O} \\ 
        HC$_5$O & (1.4$\pm$0.2)$\times10^{12}$ & RT40m & \cite{Cernicharo2021C5O}\\ 
        HC$_6$O & $\leq 1.8 \times10^{11}$ & RT40m & \cite{Cernicharo2021C5O} \\ 
        HC$_7$O & (6.5$\pm$0.5)$\times10^{11}$ & RT40m & \cite{Cernicharo2021C5O} \\ 
        ~ & ~ & ~ & ~ \\  
        C$_2$O & (7.5$\pm$0.3)$\times10^{11}$ & RT40m &\cite{Cernicharo2021C5O} \\ 
        C$_3$O & (1.2$\pm$0.2)$\times10^{12}$ & RT40m & \cite{Cernicharo2021C5O}  \\ 
        C$_4$O & $\leq 9.0 \times 10^{10}$ & RT40m & \cite{Cernicharo2021C5O}  \\ 
        C$_5$O & (1.5$\pm$0.2)$\times10^{10}$ & RT40m &\cite{Cernicharo2021C5O}  \\ 
        C$_6$O & $\leq 1.1 \times 10^{11}$ & RT40m &\cite{Cernicharo2021C5O}  \\ 
        ~ & ~ & ~ & ~ \\ 
        C$_2$S & (5.5$\pm$0.65)$\times10^{13}$ & RT40m & \cite{cernicharo2021HC3S} \\ 
        C$_3$S & (1.3$\pm$0.13)$\times10^{13}$ & RT40m & \cite{cernicharo2021HC3S} \\ 
        C$_4$S & (3.8$\pm$0.4)$\times10^{10}$ & RT40m & \cite{Cernicharo2021S} \\ 
        C$_5$S & (5.0$\pm$1.0)$\times10^{10}$ & RT40m & \cite{Cernicharo2021S} \\ 
        ~ & ~ & ~ & ~ \\ 
        CH$_3$C$_3$N & (8.66$\pm$0.46)$\times10^{11}$ & GBT100m & \cite{siebert2022} \\ 
        CH$_3$C$_5$N & (2.86$\pm$0.30)$\times10^{11}$ & GBT100m & \cite{siebert2022} \\ 
        CH$_3$C$_7$N & (0.86$\pm$0.19)$\times10^{11}$ & GBT100m & \cite{siebert2022} \\ 
        CH$_3$C$_4$H & (100.8$\pm$5.7)$\times10^{11}$ & GBT100m & \cite{siebert2022} \\ 
        CH$_3$C$_6$H & (10.4$\pm$0.72)$\times10^{11}$ & GBT100m & \cite{siebert2022} \\ 
        ~ & ~ & ~ & ~ \\ 
        $c$-C$_6$H$_5$CN & $(4.0\pm1.6) \times10^{11}$ & GBT100m & \cite{McGuire2018} \\ 
        $c$-C$_9$H$_8$  & $9.04_{-0.96}^{+0.96}\times10^{12}$ & GBT100m  & \cite{sita2022}  \\ 
        $c$-C$_5$H$_4$CCH$_2$  & (2.7 $\pm$ 0.3) $\times10^{12}$  & RT40m & \cite{cernicharo2022fulvenallene} \\ 
        $c$-C$_5$H$_6$  & (1.2 $\pm$ 0.3) $\times10^{13}$ & RT40m  & \cite{cernicharo2021cyclopentadiene} \\ 
        1-$c$-C$_5$H$_5$CN  & $8.27_{-1.0}^{+0.9}\times10^{11}$ & GBT100m & \cite{lee2021} \\ 
        2-$c$-C$_5$H$_5$CN & $1.89_{-0.15}^{+0.18}\times10^{11}$ & GBT100m & \cite{lee2021}  \\ 
        1-C$_{10}$H$_7$CN  & $7.35_{-4.63}^{+3.33}\times10^{11}$ & GBT100m &\cite{mcguire2021} \\ 
        2-C$_{10}$H$_7$CN  & $7.05_{-4.50}^{+3.23}\times10^{11}$ & GBT100m & \cite{mcguire2021} \\ 
        1-$c$-C$_5$H$_5$CCH & (1.4 $\pm$ 0.2)$\times10^{12}$ & RT40m & \cite{cernicharo2021cyclopentadiene} \\ 
        2-$c$-C$_5$H$_5$CCH  & (2.0 $\pm$ 0.4)$\times10^{12}$ & RT40m & \cite{cernicharo2021cyclopentadiene} \\ 
        $c$-C$_6$H$_4$ & (5.0 $\pm$ 1.0)$\times10^{11}$ & RT40m & \cite{cernicharo2021benzyne} \\ 
        C$_6$H$_5$CCH & (3.00$\pm$0.5)$\times10^{12}$ & RT40m &\cite{loru23} \\ 
        C$_9$H$_7$CN  & $2.10_{-0.46}^{+0.60)}\times10^{11}$ & GBT100m & \cite{sita2022} \\ 
        $c$-C$_3$HCCH & (3.1 $\pm$ 0.8)$\times10^{11}$ & RT40m & \cite{cernicharo2021hydrocarboncycle} \\ 
        E-1-C$_4$H$_5$CN & $3.8_{-0.91}^{+1.0}\times10^{10}$ & GBT100m & \cite{Cooke2023} \\ 
        ~ & ~ & ~ & ~ \\ 
        $c$-C$_3$H & $6.12_{-3.06}^{+12.24}\times10^{12}$ & IRAM30m & \cite{loison2017} \\ 
        $l$-C$_3$H & $1.12_{-0.56}^{+2.24}\times10^{12}$ & IRAM30m & \cite{loison2017} \\ 
        $c$-C$_3$H$_2$ & $(1.22\pm0.61)\times10^{14}$ & IRAM30m & \cite{loison2017} \\ 
        $l$-C$_3$H$_2$ & $1.82_{-0.91}^{+3.64} \times10^{12}$ & IRAM30m &\cite{loison2017} \\ 
        $c$-C$_5$H & $(9.0\pm0.9)\times10^{10}$ & RT40m & \cite{Cabezas2022cyclicC5H} \\ 
        $l$-C$_5$H & (1.3 $\pm$ 0.3)$\times10^{12}$ & RT40m & \cite{Cabezas2022cyclicC5H}  \\ 
        $l$-C$_5$H$_2$ &  (1.80$\pm$0.5)$\times10^{10}$ & RT40m & \cite{cabezas2021cumulene} \\
        \hline
\end{longtable}

%% file: table3.tex
\begin{table}[!ht]
\caption{Column density of observed carbon chain species towards IRC+10216}
\label{tab:colden-irc}
    \centering
    \begin{tabular}{p{1.8cm} p{3.0cm} p{2.9cm} p{4cm}}
    \hline
        Species & $N$ (cm$^{-2}$) & Telescope  & Ref.  \\  
    \hline
        HC$_3$N & $(4.5\pm0.20)\times10^{14}$ & RT40m & \citet{cernicharo2020HC4NC} \\  
        HC$_5$N &  $(8.3\pm 0.70)\times10^{14}$ & RT40m & \citet{Pardo2020} \\  
        HC$_7$N & $(3.5\pm 0.70)\times10^{14}$  & RT40m & \citet{Pardo2020} \\  
        HC$_9$N & $(7.6\pm 1.40)\times10^{13}$ & RT40m & \citet{Pardo2020} \\  
        ~ & ~ & ~ & ~ \\  
        C$_2$S & $(5.0\pm0.30)\times10^{13}$ & IRAM30m &\citet{Agundez2014} \\  
        C$_3$S & $(1.7\pm0.10)\times10^{13}$ & IRAM30m & \citet{Agundez2014}  \\  
        C$_4$S & $<7.0\times10^{12}$ & IRAM30m & \citet{Agundez2014}  \\  
        C$_5$S & ($2-14$)$\times10^{12}$ & IRAM30m & \citet{Agundez2014}  \\  
        ~ & ~ & ~ & ~ \\  
        MgCCH & $(2.0\pm0.4)\times10^{12}$ & RT40m & \citet{Agundez2014}  \\  
        MgC$_4$H & $(2.2\pm0.50)\times10^{13}$ & RT40m & \citet{Cernicharo2019Mg} \\  
        MgC$_6$H & $(2.0\pm0.90)\times10^{13}$ & RT40m & \citet{pardo2021} \\  
        MgC$_4$H$^+$ &$(4.8\pm0.30)\times10^{11}$ & RT40m & \citet{Cernicharo2023metal} \\  
        MgC$_6$H$^+$ &$(2.5\pm0.30)\times10^{12}$ & RT40m & \citet{Cernicharo2023metal} \\  
        MgC$_3$N & $(3.6\pm0.60)\times10^{12}$  & RT40m & \citet{Cernicharo2019Mg} \\  
        MgC$_5$N & $(4.7\pm1.30)\times10^{12}$ & RT40m & \citet{pardo2021} \\  
        MgC$_3$N$^+$ &$(1.2\pm0.20)\times10^{11}$ & RT40m & \citet{Cernicharo2023metal} \\  
        MgC$_5$N$^+$ &$(1.1\pm0.20)\times10^{11}$ & RT40m & \citet{Cernicharo2023metal} \\  
        ~ & ~ & ~ & ~ \\ 
        C$_2$H & $(3.84\pm0.09)\times10^{15}$ & IRAM30m, Herschel & \citet{debeck2012} \\  
        C$_3$H & $(7.0\pm1.40)\times10^{13}$ & IRAM30m & \citet{cernicharo2000} \\  
        C$_4$H & $(5.1\pm1.02)\times10^{14}$ & IRAM30m & \citet{oya20} \\  
        C$_5$H & $(4.4\pm0.88)\times10^{13}$ & IRAM30m &\citet{cernicharo2000}  \\  
        C$_6$H & $(5.5\pm1.10)\times10^{13}$ & IRAM30m & \citet{cernicharo2000}\\  
        C$_7$H & $(2.2\pm0.44)\times10^{12}$ & IRAM30m & \citet{cernicharo2000}\\  
        C$_8$H & $(1.0\pm0.20)\times10^{13}$ & IRAM30m & \citet{cernicharo2000}\\  
        ~ & ~ & ~ & ~ \\  
        C$_2$N  & ($4.0\pm2.00)\times 10^{13}$  & ARO12m & \citet{anderson14c2n} \\  
        C$_3$N & ($4.0 \pm 0.10)\times10^{15}$ & ARO12m & \citet{anderson14c2n} \\  
        C$_3$N$^-$ &  (1.6 $\pm$0.60)$\times10^{12}$&IRAM30m& \cite{Thaddeus08}\\
        C$_5$N & ($6.0\pm1.20)\times10^{12}$ &  IRAM30m & \citet{Guelin98} \\   
        C$_5$N$^-$ & ($3.4\pm 0.68)\times10^{12}$ & IRAM30m & \citet{Cernicharo2008C5Nm}\\  
      \hline  
    \end{tabular}
When we cannot obtain errors from the original papers, we indicate 20\% errors.
\end{table}